\documentclass[sigconf,dvipsnames]{acmart}

\usepackage{microtype}
\usepackage{graphicx}
\usepackage{subcaption}
\usepackage{wrapfig}
\usepackage{amsmath}
\usepackage{commath}
\usepackage[mathscr]{euscript}
\usepackage{dsfont}
\usepackage{adjustbox}

\usepackage{float}

\usepackage{algorithmic}
\usepackage[ruled,vlined]{algorithm2e}

\usepackage{overpic}

\usepackage{paralist}
\usepackage{geometry}
\usepackage{booktabs} 

\usepackage{ulem}
\usepackage[margin=1cm]{caption}

\usepackage{hyperref}
\hypersetup{
    colorlinks=true,
    linkcolor=blue,
    filecolor=magenta,      
    urlcolor=cyan,
}

\newcommand{\Review}[1]{#1}

\usepackage{blindtext}
\usepackage{todonotes}
\usepackage{xfrac}

\usepackage{multirow}
	
\usepackage{color, colortbl}

\DeclareMathOperator*{\argmax}{arg\,max}

\newcommand{\Recover}{\texttt{RECOVER}}

\author[P. Bertin \textit{et al.}]{Paul Bertin$^1$, Jarrid Rector-Brooks$^1$, Deepak Sharma$^1$, Thomas Gaudelet$^2$, Andrew Anighoro$^2$, Torsten Gross$^2$, Francisco Mart\'{i}nez-Pe\~{n}a$^3$, Eileen L. Tang$^3$, Suraj M S$^2$, Cristian Regep$^2$, Jeremy B. R. Hayter$^2$, Maksym Korablyov$^1$, Nicholas Valiante$^4$, Almer van der Sloot$^5$, Mike Tyers$^5$, Charles Roberts$^2$,  Michael M. Bronstein$^{6,7}$, Luke L. Lairson$^3$, Jake P. Taylor-King$^{2}$, and Yoshua Bengio$^{1}$ \\[1ex]
\large{$^1$ Mila, the Quebec AI Institute, Canada $\quad$ $^2$ Relation Therapeutics, London, UK  \\ $^3$ Department of Chemistry, The Scripps Research Institute, USA $\quad$ $^4$ Glyde Bio, Inc, USA. \\
$^5$ IRIC, Institute for Research in Immunology and Cancer, Université de Montréal, Canada \\
$^6$ Department of Computer Science, University of Oxford, UK \quad $^7$ Twitter, UK } \\ \texttt{jake@relationrx.com, yoshua.bengio@mila.quebec}}

\begin{document}

\title[\bf RECOVER: sequential model optimization for combination drug repurposing]{{\bf RECOVER: sequential model optimization platform for combination drug repurposing identifies novel synergistic compounds \textit{in vitro}}}


\begin{abstract}

For large libraries of small molecules, exhaustive combinatorial chemical screens become infeasible to perform when considering a range of disease models, assay conditions, and dose ranges. Deep learning models have achieved state of the art results \textit{in silico} for the prediction of synergy scores. However, databases of drug combinations are biased towards synergistic agents and these results do not necessarily generalise out of distribution. We employ a sequential model optimization search utilising a deep learning model to quickly discover synergistic drug combinations active against a cancer cell line, requiring substantially less screening than an exhaustive evaluation. \Review{Our small scale wet lab experiments only account for evaluation of $\sim$5\% of the total search space}. After only $3$ rounds of ML-guided \textit{in vitro} experimentation (including a calibration round), we find that the set of drug pairs queried is enriched for highly synergistic combinations; two additional rounds of ML-guided experiments were performed to ensure reproducibility of trends. Remarkably, we rediscover drug combinations later confirmed to be under study within clinical trials. Moreover, we find that drug embeddings generated using only structural information begin to reflect mechanisms of action. \Review{Prior \textit{in silico} benchmarking suggests we can enrich search queries by a factor of $\sim$5-10$\times$ for highly synergistic drug combinations by using sequential rounds of evaluation when compared to random selection, or by a factor of $>$3$\times$ when using a pretrained model selecting all drug combinations at a single time point.}

Our method is available at: \url{https://github.com/RECOVERcoalition/Recover}.

\end{abstract}

\twocolumn[{%
\renewcommand\twocolumn[1][]{#1}%
\maketitle
}]

\section{Introduction}

\begin{figure*}[ht]
\centering
\includegraphics[width=\textwidth]{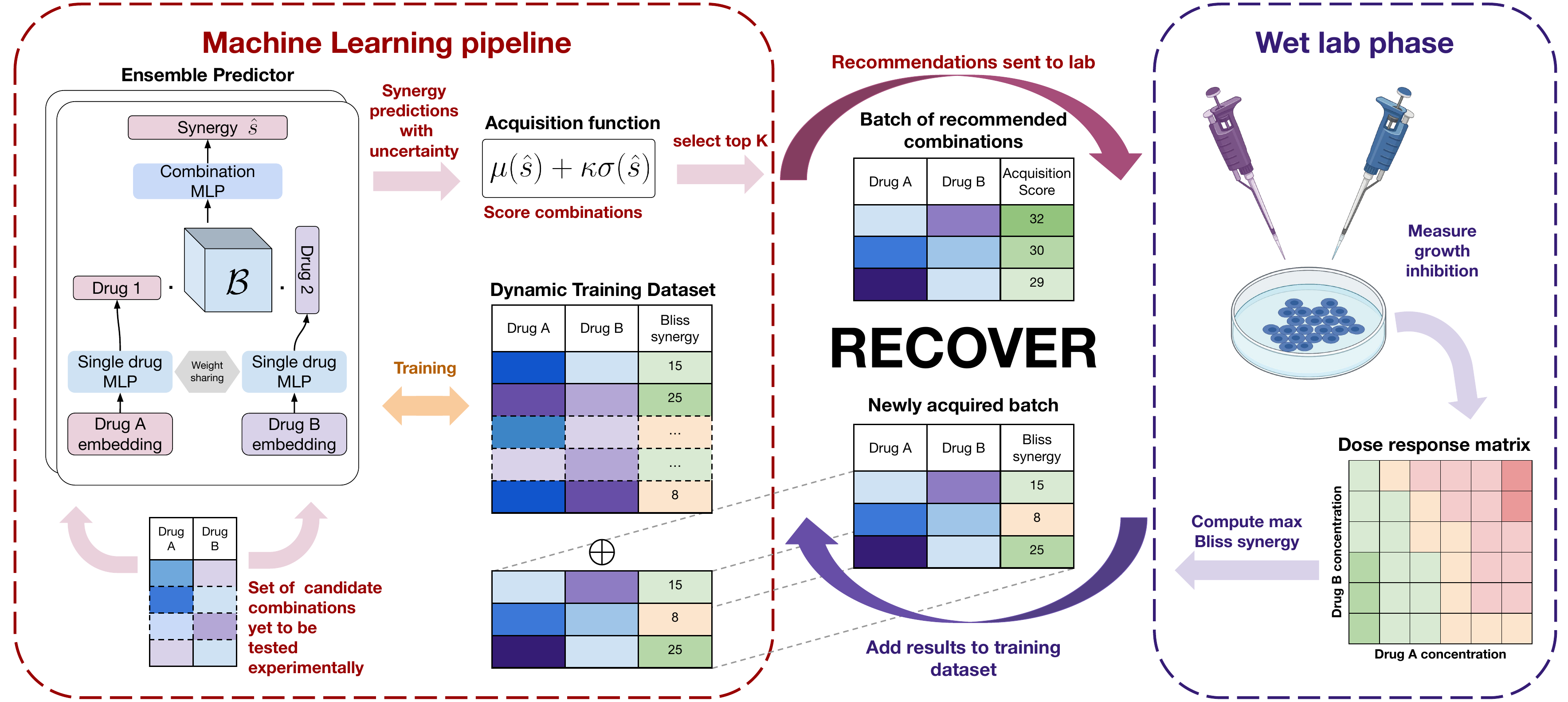}
\caption{Overview of the {\Recover} workflow integrating both a novel machine learning pipeline and iterated wet lab evaluation.}
\label{fig:pipeline_diagram}
\end{figure*}



Drug combinations are an important therapeutic strategy for treating diseases that are subject to evolutionary dynamics, in particular cancers and infectious disease \cite{tyers2019drug, mokhtari2017combination}. Conceptually, as tumours or pathogens are subject to change over time, they may develop resistance to a single agent \cite{delou2019highlights} --- motivating one to target multiple biological mechanisms simultaneously \citep{al-lazikaniCombinatorialDrugTherapy2012}. Discovering synergistic drug combinations is a key step towards developing robust therapies as they hold the potential for greater efficacy whilst reducing dose, and hopefully limiting the likelihood of adverse effects. For example, in a drug repurposing scenario (i.e., uncovering new indications for known drugs), the ReFRAME library of $\sim$12,000 clinical stage compounds \cite{janes2018reframe} leads to $\sim$72 million pairwise combinations; this does not appear tractable with standard high throughput screening (HTS) technology --- even at a single dose \cite{clare2019industrial}.


With the recent COVID-19 global health crisis, there has been the need for rapid drug repurposing that would allow for expedited and de-risked clinical trials. Due to the complexity of selecting drug combinations and minimal training data publicly available, studies have typically been limited towards monotherapy repurposing from a variety of angles --- often involving artificial intelligence (AI) techniques to provide recommendations \cite{zhou2020artificial}. The dearth of drug combination datasets is due to the large combinatorial space of possible experiments available --- ultimately limiting the quality of drug synergy predictions. Moreover, databases of drug combinations are biased towards suspected synergistic agents and thus making predictions outside the scope of the training dataset can be challenging. 

\Review{The goal of this work is to discover synergistic drug combinations whilst only requiring minimal wet lab experimentation. One cost-efficient tool at our disposal is sequential model optimization (SMO), whereby a machine learning (ML) model selects experiments (i.e., pairs of drugs) that it would like to be evaluated (in this case, for drug synergism). Both highly informative experiments (``\textit{exploration}'') and experiments that double-down on promising data-driven hypotheses (``\textit{exploitation}'') can be selected \cite{sverchkov2017review}. Between rounds of experimental evaluation, the model is iteratively adapted to new observations (via model training), which allows performance to gradually improve. This SMO process allows for queries that are more and more enriched with highly synergistic combinations, ultimately leading to reduced experimentation when compared to an exhaustive search.}


There have now been a number of approaches for predicting the effects of and subsequently prioritising drug combinations \cite{bulusuModellingCompoundCombination2016}. Classic bioinformatics approaches have focused on using machine learning and network statistics over specified features of drugs (e.g., molecular fingerprints \cite{cereto2015molecular}), cell lines (e.g., transcriptomics, copy number variations \cite{karczewski2018integrative}), and interactome topology between biomolecules (e.g., protein--protein interactions, chemical--genetic interactions \cite{wildenhain2015prediction}, or gene regulatory networks \cite{cheng2019network}). Initiatives such as the Dialogue on Reverse Engineering Assessment and Methods (DREAM) have led to a plethora of methods being benchmarked against one another in prospective challenges through the generation of novel datasets \cite{menden2019community}. Complex deep learning architectures, which have set state of the art performance across a number of domains \cite{lecun2015deep}, have been used to predict both adverse drug--drug interactions \cite{zitnik2018modeling, deac2019drug} and synergistic drug combinations \cite{preuer2018deepsynergy,jin2021deep, rozemberczki2021moomin}. Sequential approaches, wherein several rounds of selection are performed, have also been explored in the context of drug combinations; for example, Kashif \textit{et al.}  \cite{Kashif_Andersson_Hassan_Karlsson_Senkowski_Fryknas_Nygren_Larsson_Gustafsson_2015} have proposed a heuristic based (as opposed to model based) exploration strategy.


We present a sequential model optimization platform that can guide wet lab experiments: \Recover, a deep learning regression model that predicts synergy using molecular fingerprints as inputs. \Review{To motivate the use of {\Recover}, we demonstrate a real world use case whereby one observes both: a $\sim$5-10$\times$ estimate for the enrichment of synergistic drugs identified using SMO when compared to selecting drug combinations at random; and a $\sim$3$\times$ improvement when compared to selecting drugs in a single batch using a pretrained model.} We then perform a retrospective validation to benchmark the performance of our model and understand its generalization abilities using the DrugComb database --- largely pertaining to cancer cell line data \cite{zagidullin2019drugcomb}. Thereafter, we evaluate our SMO pipeline \textit{in silico}, which allows the model to select the most relevant data points to be labelled in order to discover most promising combinations while reducing model uncertainty. Finally, we test {\Recover} prospectively in an \textit{in vitro} experimental setting whereby we discover novel synergistic combinations active against a breast cancer model cell line, MCF7, which is also represented within our training dataset. 

\Review{With an SMO platform available in conjunction with an appropriate \textit{in vitro} assay, one has a powerful tool to rapidly respond to a future public health crisis. To encourage use by the scientific community, we detail a configuration that can be trained on a personal computer or laptop without requiring dedicated computational infrastructure. Remarkably, high predictive power is not a prerequisite for such an SMO system to be utilised effectively. In fact, as we are trying to identify pairs of drugs in prospective experiments that have more extreme synergy scores than those drug combinations evaluated within previous experiments (i.e., our training dataset), we cannot necessarily expect to have high predictive power. However, we achieve our ultimate goal: the identification of highly synergistic drugs --- not building highly accurate ML models. This work forms a proof of concept demonstration of {\Recover} --- which should then motivate greater community adoption of the method and extensions thereof.}

\section{Results}\label{sec:results}

\subsection{\Recover: sequential model optimization platform for rapid drug repurposing}
\label{sec:results:overview}

{\Recover} is an open-source SMO platform for the optimal suggestion of drug combinations, see Figure \ref{fig:pipeline_diagram}. Pairs of drug feature vectors are fed into a deep neural network which is used for the prediction of synergy scores. These feature vectors include molecular fingerprints as well as a one-hot encoding identifying a drug. For a full description of the model, see Appendix ~\ref{sec:model_description}.

Our core focus is the prediction of pairwise drug combination synergy scores. Whilst many mathematical descriptions of synergy have been proposed \cite{tyers2019drug}, in the following work, we utilise the Bliss synergy score due to its simplicity and numerical stability. In the context of cell viability, the Bliss independence model assumes that in the absence of synergistic effects, the expected fraction of viable cells after treatment with drugs $a$ and $b$ at doses $c_a$ and $c_b$, written $V(c_a,c_b)$, is identical to the product of the fractions of viable cells when utilising each drug independently, i.e., $V(c_a)V(c_b)$. We then define the Bliss synergy score as the difference between these quantities such that a fraction of surviving cells $V(c_a,c_b)$ smaller than the expected proportion $V(c_a)V(c_b)$ leads to a large Bliss synergy score
\begin{align}
\label{eqn:bliss}
\textnormal{s}_{Bliss}(c_a,c_b) 
&= V(c_a)V(c_b) - V(c_a,c_b)  \\ 
& = I(c_a,c_b) - I(c_a) - I(c_b) + I(c_a)I(c_b), \nonumber 
\end{align}
where $I(\cdot) = 1 - V(\cdot)$ is the experimentally measured growth inhibition induced by drug $a$, $b$, or both together at the associated doses. Given a dose-response matrix for the two drugs, a global synergy score can be obtained through a pooling strategy. In our case, we take the maximum value, i.e.,
\begin{align}
\label{eqn:maxpooling}
\hat{\textnormal{s}}_{Bliss} = \max_{c_a, c_b} \,  \textnormal{s}_{Bliss}(c_a,c_b) \, .
\end{align}
In many studies, the arithmetic mean is taken to calculate a global synergy score. Unfortunately, different laboratories use different dose intervals for each drug, and typically each drug combination shows a synergistic effect at a specific dose-pair interval. Therefore, the arithmetic mean is highly sensitive to the chosen dose interval, thus why we choose to prioritise a max-pooling strategy as in Eq.~\ref{eqn:maxpooling}. Unless explicitly stated otherwise, a synergy score refers to a global max-pooled Bliss score.

In addition to the prediction of synergy, {\Recover} estimates the uncertainty associated with the underlying prediction. More precisely, for a given combination of drugs, {\Recover} does not only provide a point estimate of the synergy, but estimates the distribution of possible synergy scores for each combination, which we refer to as the \textit{predictive distribution}. We define the \textit{model uncertainty} as the standard deviation of the predictive
distribution.

An \textit{acquisition function} is used to select the combinations that should be tested in subsequent experiments \citep{vzilinskas1972bayes}. This acquisition function is designed to balance between: \textit{exploration}, prioritising combinations with high model uncertainty whereby labelling said points should increase predictive accuracy in future experimental rounds; and \textit{exploitation}, selection of combinations believed to be synergistic with high confidence.

In summary, this SMO setting consists of generating recommendations of drug combinations that will be tested \textit{in vitro} at regular intervals. At each step, {\Recover} is trained on all the data acquired up to that point, and predictions are made for all combinations that could be hypothetically tested experimentally. The acquisition function is then used to provide recommendations for \textit{in vitro} testing. The results of the experiments are then added to the training data for the next round of experiments and the whole process repeats itself.

\Review{To illustrate the benefits of the SMO approach, we perform a preliminary simulation to mimic a scientist with a limited experimental budget of 300 drug combinations to be tested – with the aim to find synergistic drug combinations. We assume the experimentalist has access to a trained machine learning model and can use it in a number of ways.}

\Review{At a high level, we specify that there are two options: either to perform all 300 experiments in one go, or to perform experiments in 10 batches of 30. All models are pretrained on the O’Neil drug combination study \cite{o2016unbiased}, and validation by the experimentalist is simulated through uncovering specific examples from the NCI-ALMANAC drug combination study \cite{holbeck2017national}. In more detail, we test the following options: (Random) all 300 combinations are queried at random; (DeepSynergy) the synergies of all combinations in ALMANAC are predicted using the DeepSynergy model with the top 300 predictions queried; ({\Recover} without SMO)) the synergies of all combinations in ALMANAC are predicted using the {\Recover} model with the top 300 predictions queried; ({\Recover}) 30 combinations are queried at random followed by an SMO using batches of 30; and (DeepSynergy with SMO) which is the same SMO as before but using the DeepSynergy model.}

\begin{figure}[H]
\begin{overpic}[width=0.49\textwidth]{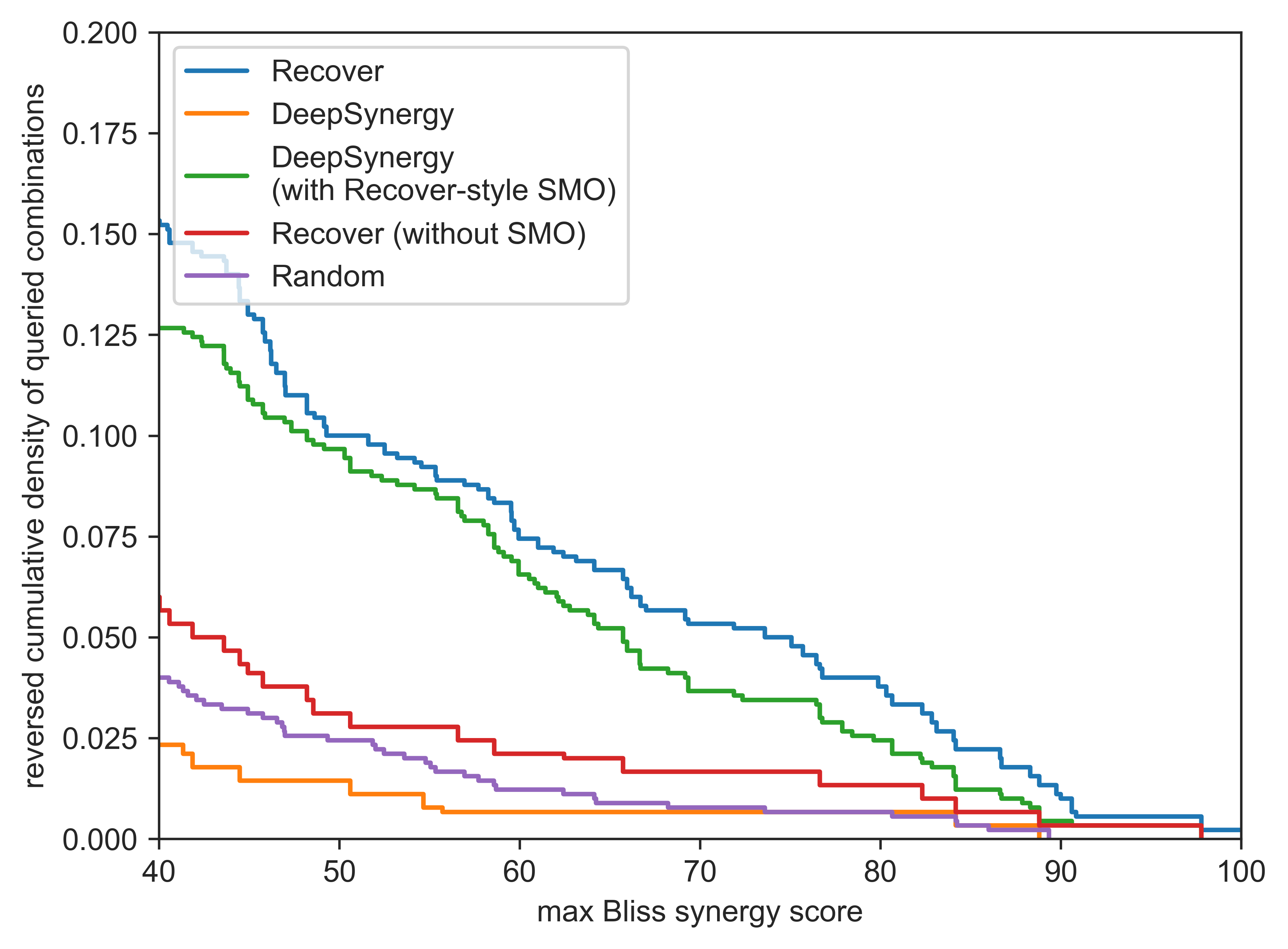}
\put(50,38){\includegraphics[clip,width=4cm]{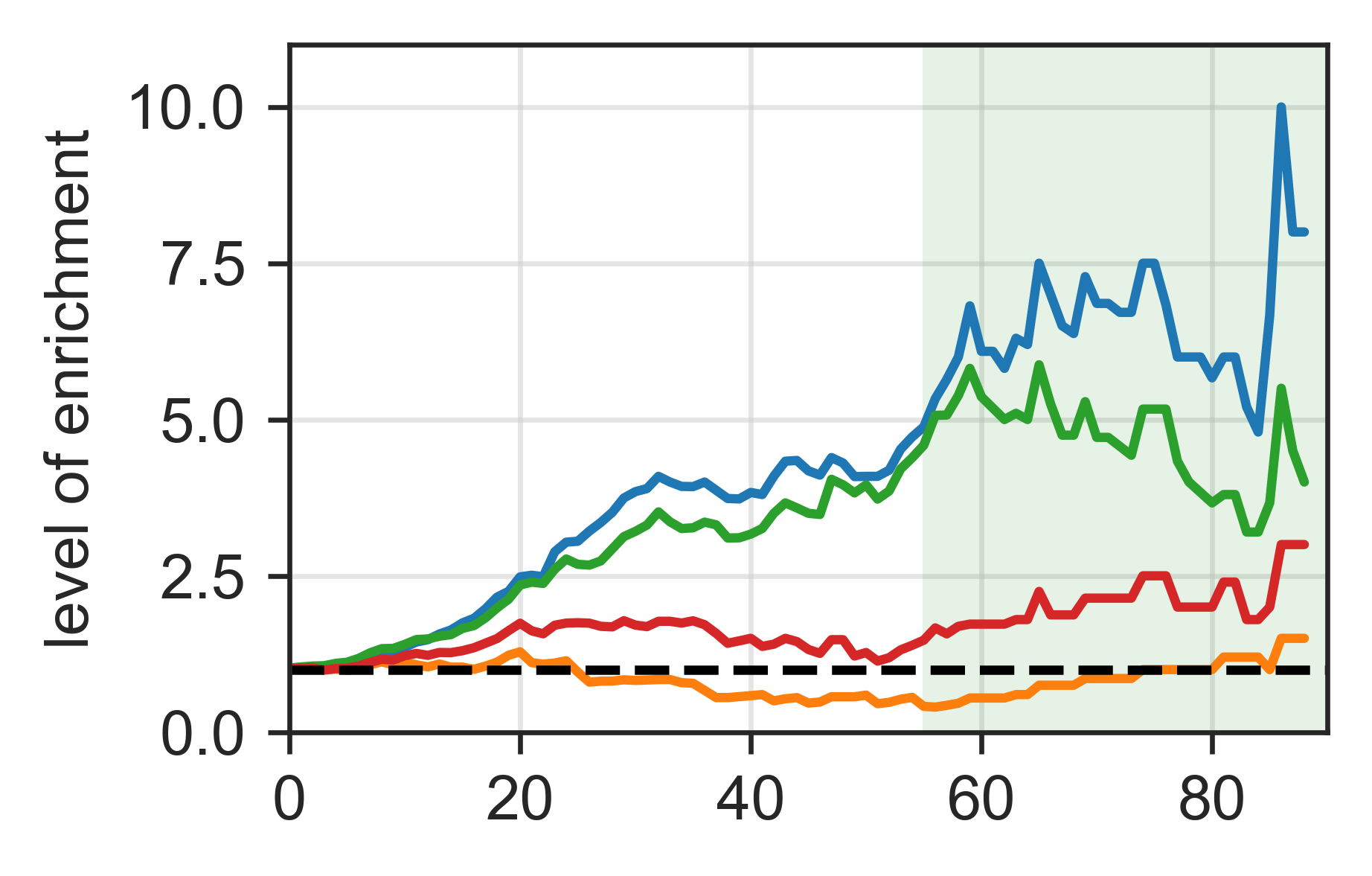}}
\end{overpic}
\caption{Reversed cumulative density of queried combinations, following different querying strategies. \textbf{(inset)} level of enrichment. Results are averaged over 3 seeds}
\label{fig:reversed_cumul_density}
\end{figure}

\Review{In Figure~\ref{fig:reversed_cumul_density}, we report the reversed cumulative density of the synergies of all 300 queried combinations (higher is better). We also report the level of enrichment defined as the ratio between the reversed cumulative density of a given strategy's queries and the reversed cumulative density of random queries. We first observe that DeepSynergy \cite{preuer2018deepsynergy} performs worse than random, while {\Recover} (without SMO) performs slightly above the level of randomness. Most importantly, the bulk of the performance gain comes from utilizing our SMO procedure. Finally, {\Recover} and DeepSynergy are compared head-to-head in the SMO setting, the {\Recover} model slightly outperforms the DeepSynergy model. The threshold for ``highly synergistic'' is challenging to specify, but we note that a drug combination in clinical trials has a max Bliss synergy score of $54.9$, see Section \ref{sec:ct_drug}. On this basis, these experiments suggest that our approach can reduce by a factor of $\sim$5--10$\times$ the amount of experiments needed to discover and validate highly synergistic drug combinations when compared to random selection, or by a factor of $>$3$\times$ when using a pretrained model selecting all drug combinations at a single time point.}

\subsection{Retrospective testing of {\Recover} informs the design of future experiments}\label{sec:backtesting}

\Review{In this section, we evaluate the \textit{in silico} performance of {\Recover} in multiple ways. In order to understand the scope of scenarios to which {\Recover} can be applied to, we benchmark {\Recover} against baseline models and test our ability to generalise in several out-of-distribution tasks without incorporating SMO. Thereafter, we perform backtesting through simulating mock SMO experiments (see Appendices \ref{sec:searching_space} and \ref{app:smo_evaluation}).}

Due to the limited size of most individual drug combination studies reported in the literature, we focus on the NCI-ALMANAC viability screen \cite{holbeck2017national} summarised in Figure \ref{fig:data_summary}. We refrain from combining multiple datasets because of the severe batch effects between studies; in Figure \ref{fig:data_summary}E, we show a scatter plot that demonstrates inconsistency between the O'Neil 2016 \cite{o2016unbiased} series of drug combination experiments against their NCI-ALMANAC counterpart. We note this may result from: variation in the readouts of these experiments, mutations in cell lines, or differences in harvest times. 

\Review{We investigate whether {\Recover} can generalize beyond the training (and validation) set in various ways: (i.) What is the performance on test cases drawn from the same distribution as the training set? Can {\Recover} generalize when (ii.) one of the drugs is unseen (during training), or (iii.) when both of the drugs are unseen? These tasks are illustrated graphically in Figure \ref{fig:out_of_dist_diagram}A. In Appendix~\ref{app:baseline_model_comparison}, we also present additional scenarios wherein we consider multiple cell lines and performance when our training and test sets come from different studies. For each task, we benchmark against several alternative models along with \Recover, including a linear support vector machine (SVM), Boosting Trees and DeepSynergy \cite{preuer2018deepsynergy}. In addition, we evaluate a version of {\Recover} without the invariance module, and another version for which the identities of the drugs (as well as cell lines) have been shuffled, see Appendix~\ref{app:baseline_model_comparison} for model details and their hyperparameter optimization. Through understanding the capacity for the generalisation of {\Recover}, we can design prospective experiments with greater confidence of success.}

In Figures \ref{fig:out_of_dist_diagram}B and \ref{fig:out_of_dist_diagram}C, we report the test performance metrics of {\Recover} across each of the first three tasks. Examining performance within task (a.), test statistics appear modest, however we demonstrate limits on achievable performance in Appendix \ref{app:subsection_theoretical_upper_bounds} --- resulting from experimental noise and non-uniformity of synergy scores. From task (a.) to task (c.), we note a drastic drop in performance for all models, but this effect is alleviated if only one of the drugs has not been seen before, see task (b.). 

\Review{We note that our benchmarking justifies various aspects of our deep learning architecture: the {\Recover} permutation invariance module can provide improvement in performance across some scenarios; moreover {\Recover} (shuffled labels) fails compared to other methods on task (ii.) with one unseen drug task, and is at the level of randomness on task (iii.) with two unseen drugs. In these cases, we demonstrate that drug structure is actually leveraged by the model in order to generalize (to some extent) to unseen drugs.}

From the above results, we can recommend that any prospective experiments should require that one of the two drugs in the combination have been seen in some context before, see task (iii.). \Review{Due to the severe batch experiments between studies in the public domain, as shown in Figure~\ref{fig:data_summary}E, models fail to generalize to data coming from a different study, as shown in Figure~\ref{fig:out_of_dist_diagram_appendix}A (Study Transfer task). As such, should we want to utilise publicly available resources, we will have to incorporate such data intelligently.} To this end we investigated using \textit{transfer learning}, wherein one trains a model on a large dataset (known as \textit{pretraining}), and thereafter refines the model on a smaller dataset (known as \textit{fine-tuning}) --- typically with some aspect of the task or the data changed between the two instances. We show that this is possible and beneficial (compared to not leveraging existing data) in an SMO setting between the O'Neil 2016 and NCI-ALMANAC studies in Appendix \ref{app:effect_pretraining}. Remarkably even with minimal correlation between studies, we are able to observe the benefits of transfer learning in this scenario. These findings inform design choices regarding our prospective experiments.

\begin{figure*}[ht]
    \centering
    \begin{minipage}{0.52\textwidth}
    \begin{overpic}[width=\textwidth] 
    {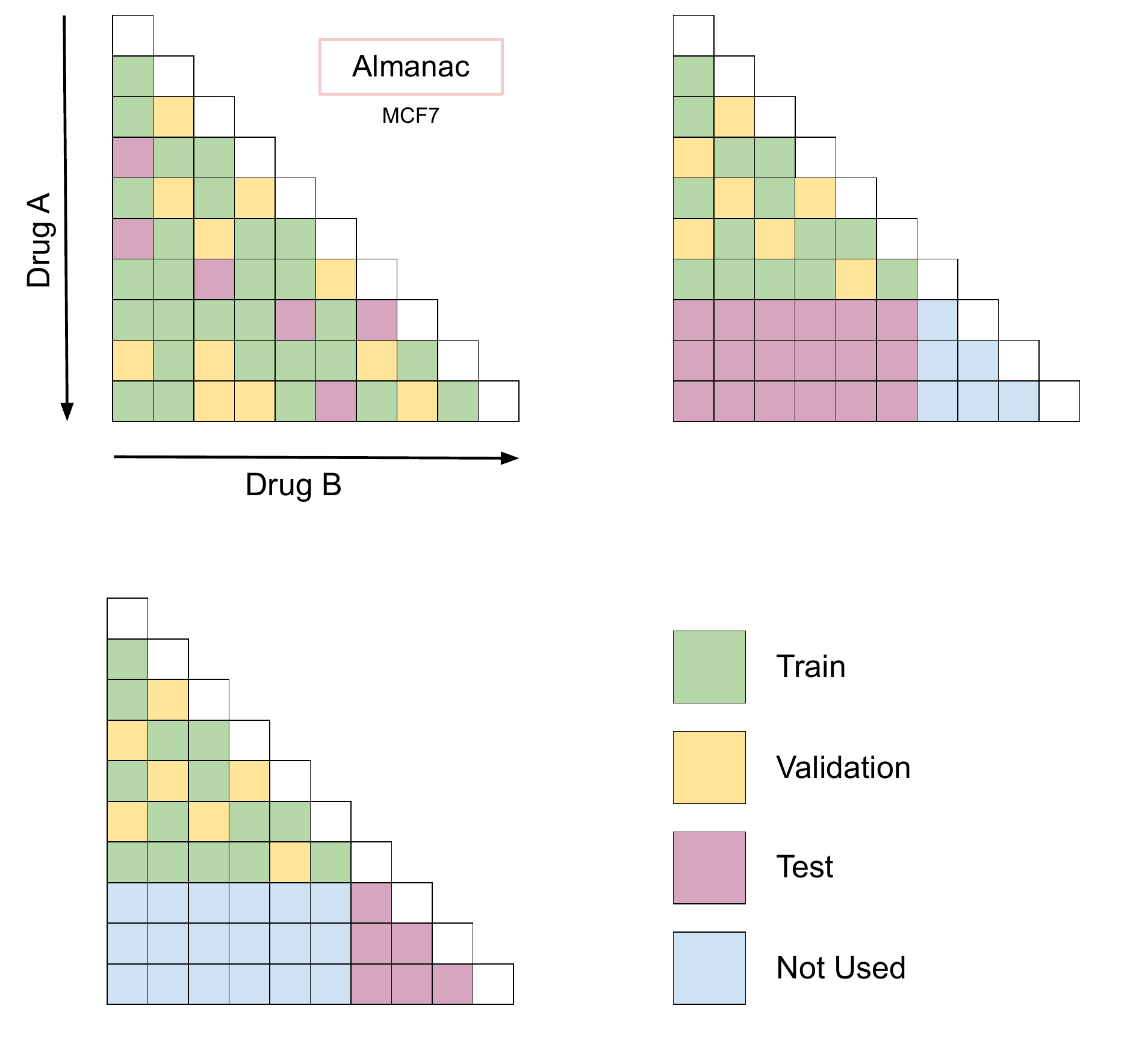}
    \put(0, 95){\Large A}
    \put(97, 95){\Large B}
    \put(97, 47){\Large C}
    \put(17, 43){(i.) Default}
    \put(62, 43){(ii.) One unseen drug}
    \put(12, -3){(iii.) Two unseen drugs}
    \end{overpic}
    \vspace{0.2cm}
    \end{minipage}
    \begin{minipage}{0.37\textwidth}
        \includegraphics[width=\textwidth]{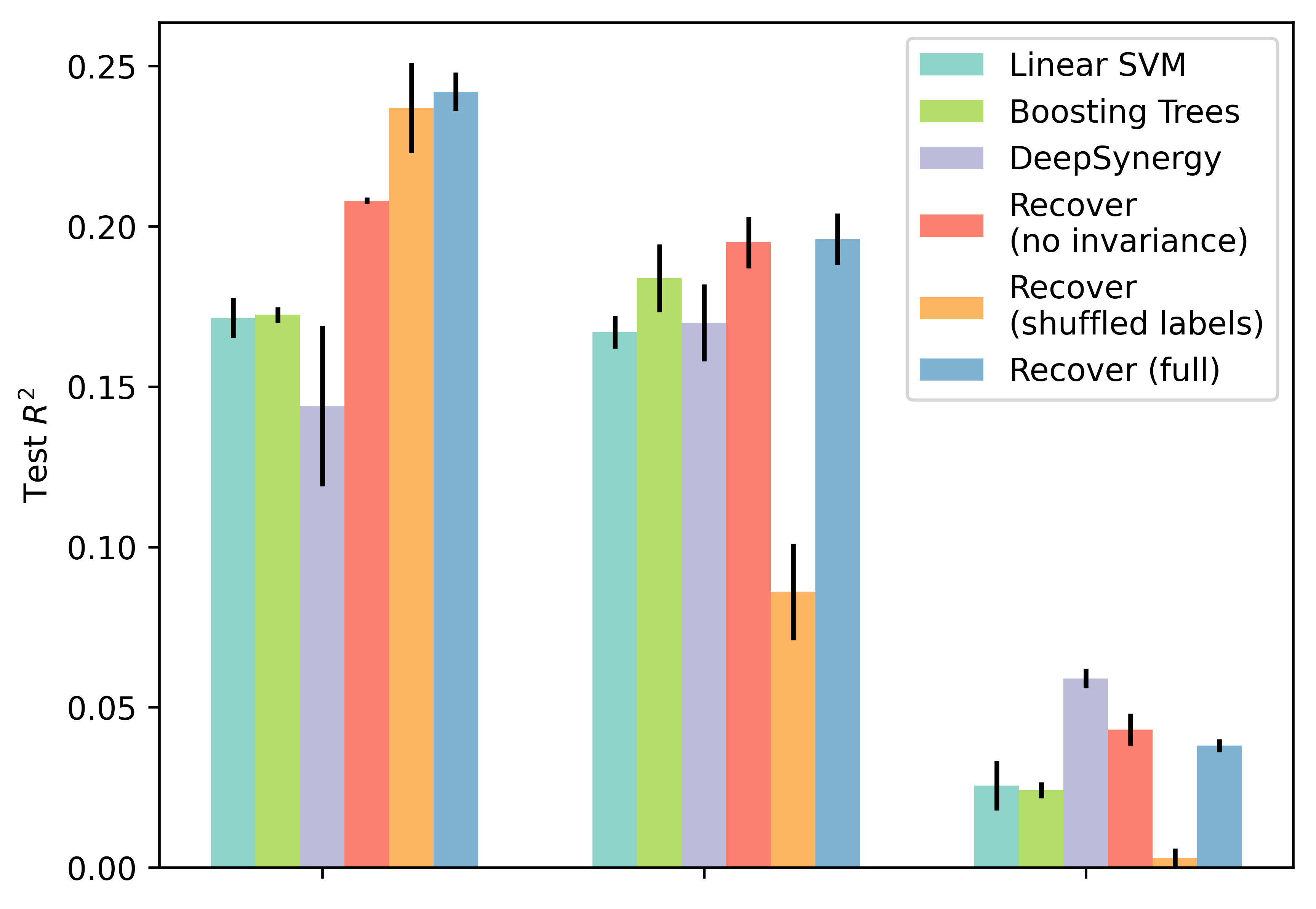}
        \hfill
    \includegraphics[width=\textwidth]{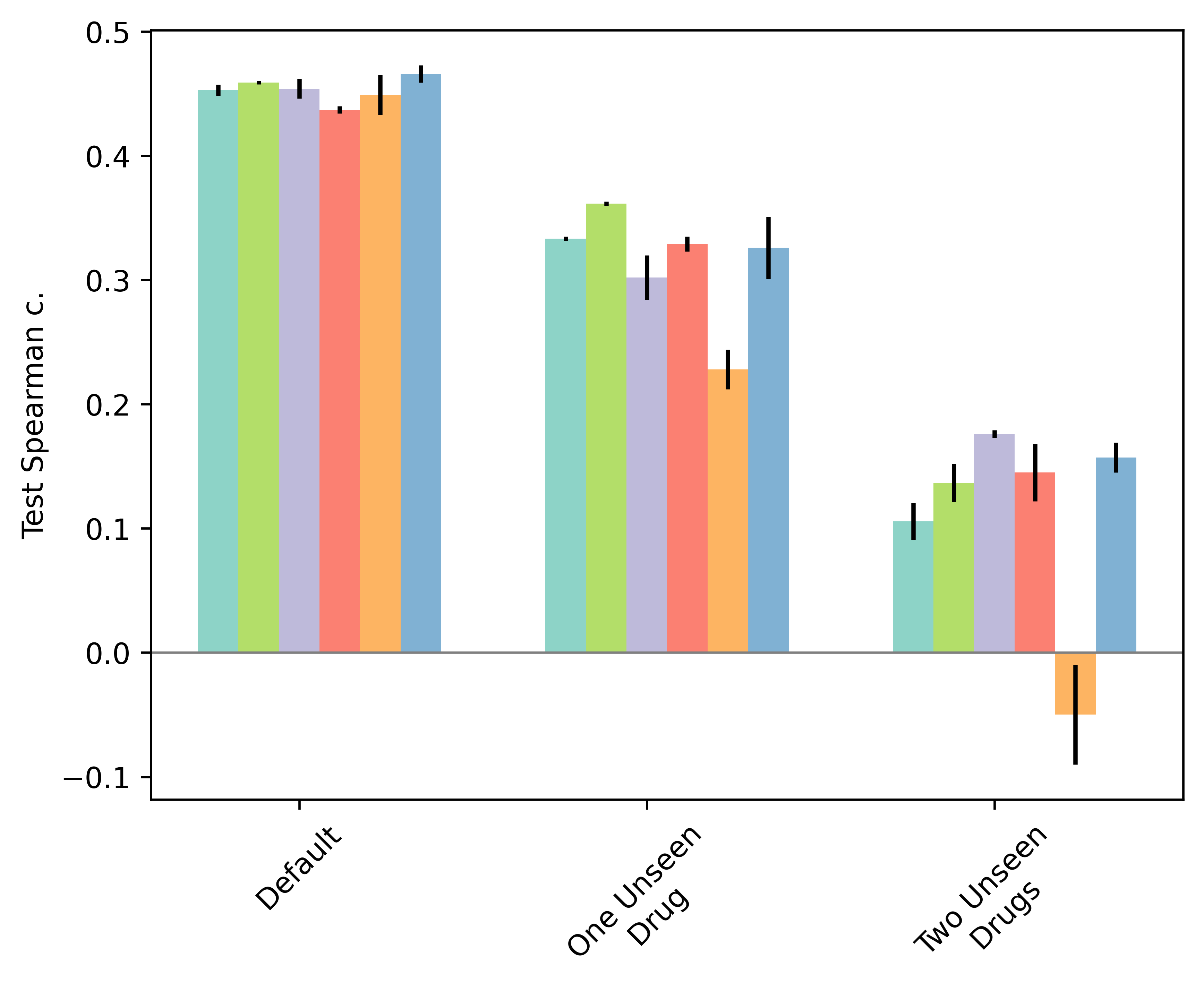}
    \end{minipage}
    \caption{ \textbf{(A.)} Overview of the different tasks on which {\Recover} has been evaluated. Each task corresponds to a different way to split the training, validation and test sets, and aims at evaluating a specific generalization ability of the model. 
    \textbf{(i.)} Default. Combinations are split randomly into training/validation/test (70\%/20\%/10\%). Only the MCF7 cell line is used.
    \textbf{(ii.)} One unseen drug. 30\% of available drugs are excluded from the training and validation sets. The test set consists of combinations between a drug seen during training and an unseen drug. Combinations among seen drugs are split into training and validation (80\%/20\%). Only the MCF7 cell line is used.
    \textbf{(iii.)} Two unseen drugs. Similar to Task (b.), but the test set consists of combinations of two unseen drugs. 
    \textbf{(B.) and (C.)} Performance of {\Recover} and other models for the three different tasks. Standard deviation computed over 3 seeds.}
    \label{fig:out_of_dist_diagram}
\end{figure*}
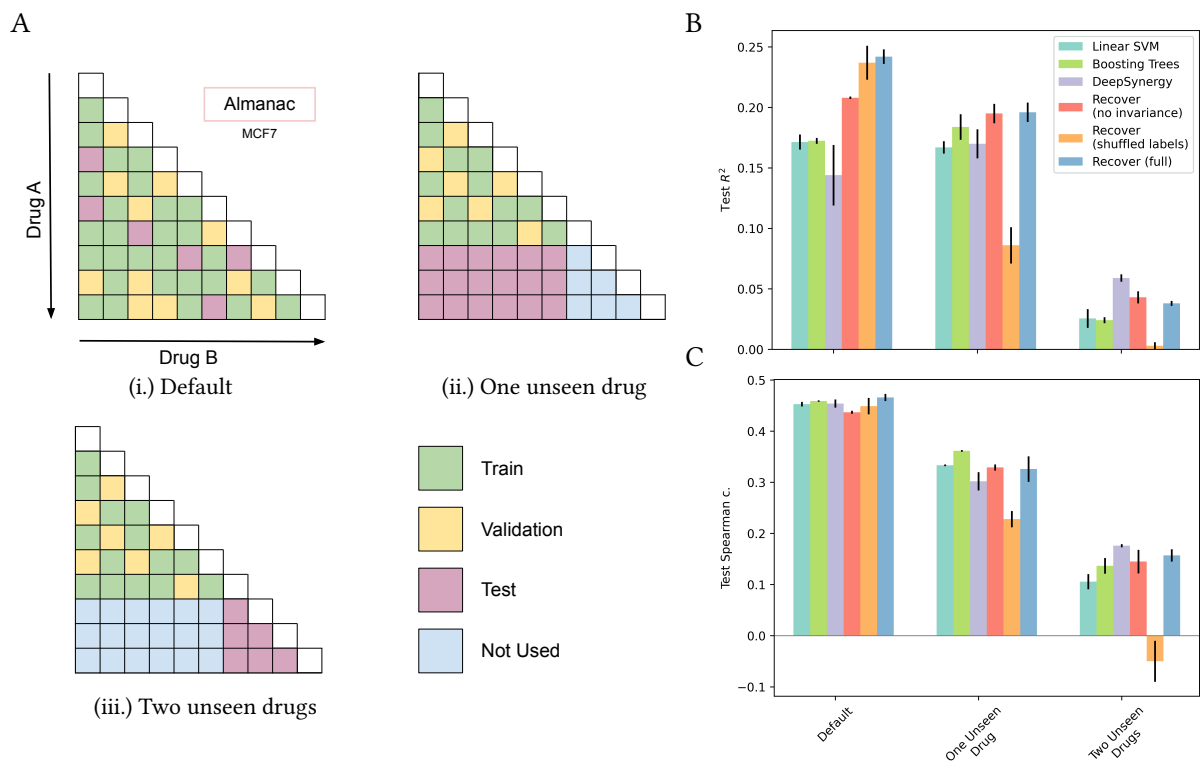



\subsection{Prospective use of {\Recover} enriches for selection of synergistic drug combinations}\label{subsec:prospective_use_of_recover}

\begin{figure*}[ht]

\begin{overpic}[clip, width=0.75\textwidth, trim={1cm 0.5cm 0 0.5cm}]{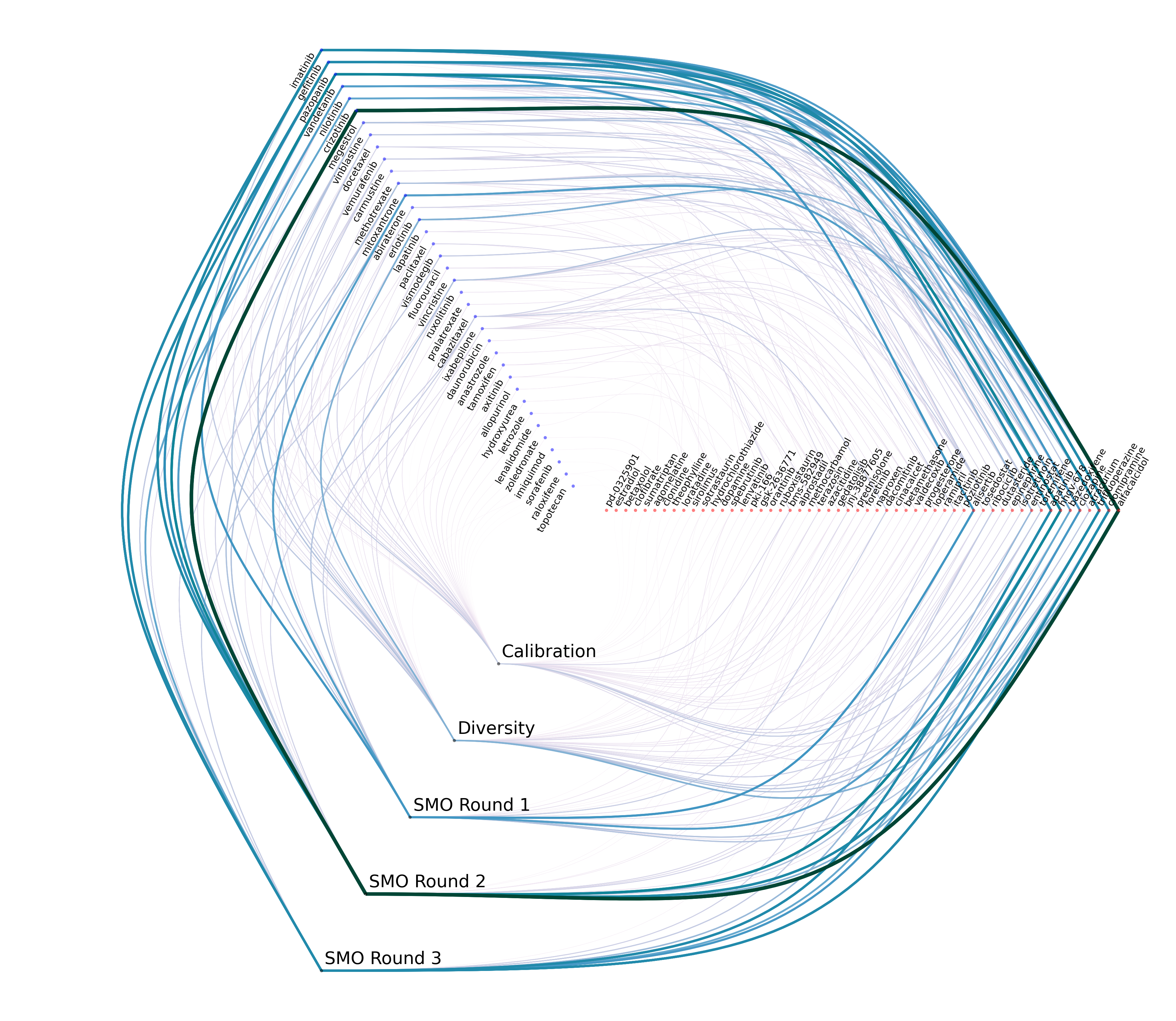}
\put(0,80){\Large A}
\put(60,33){\includegraphics[width=10cm]{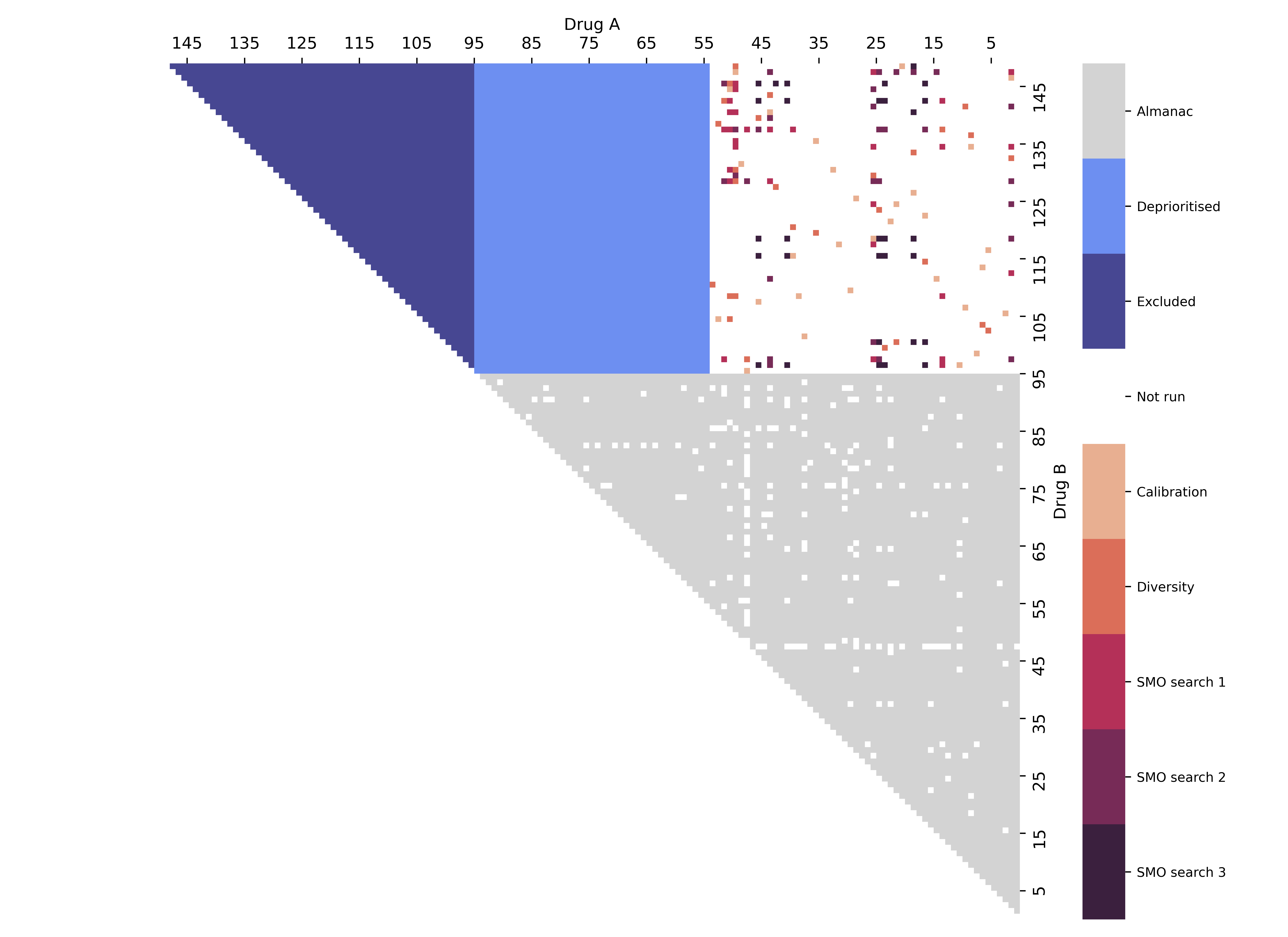}}
\put(75,80){\Large \textcolor{white}{B}}
\put(98,28){\footnotesize 146 drug combinations tested}
\put(98,24){\footnotesize 2,916 available combinations}
\put(98,20){\footnotesize $\sim$5\% of search space evaluated}
\end{overpic}\hspace{0.23\textwidth}
\vspace{0.3cm}
\begin{center}
\begin{overpic}[clip, width=0.495\textwidth]{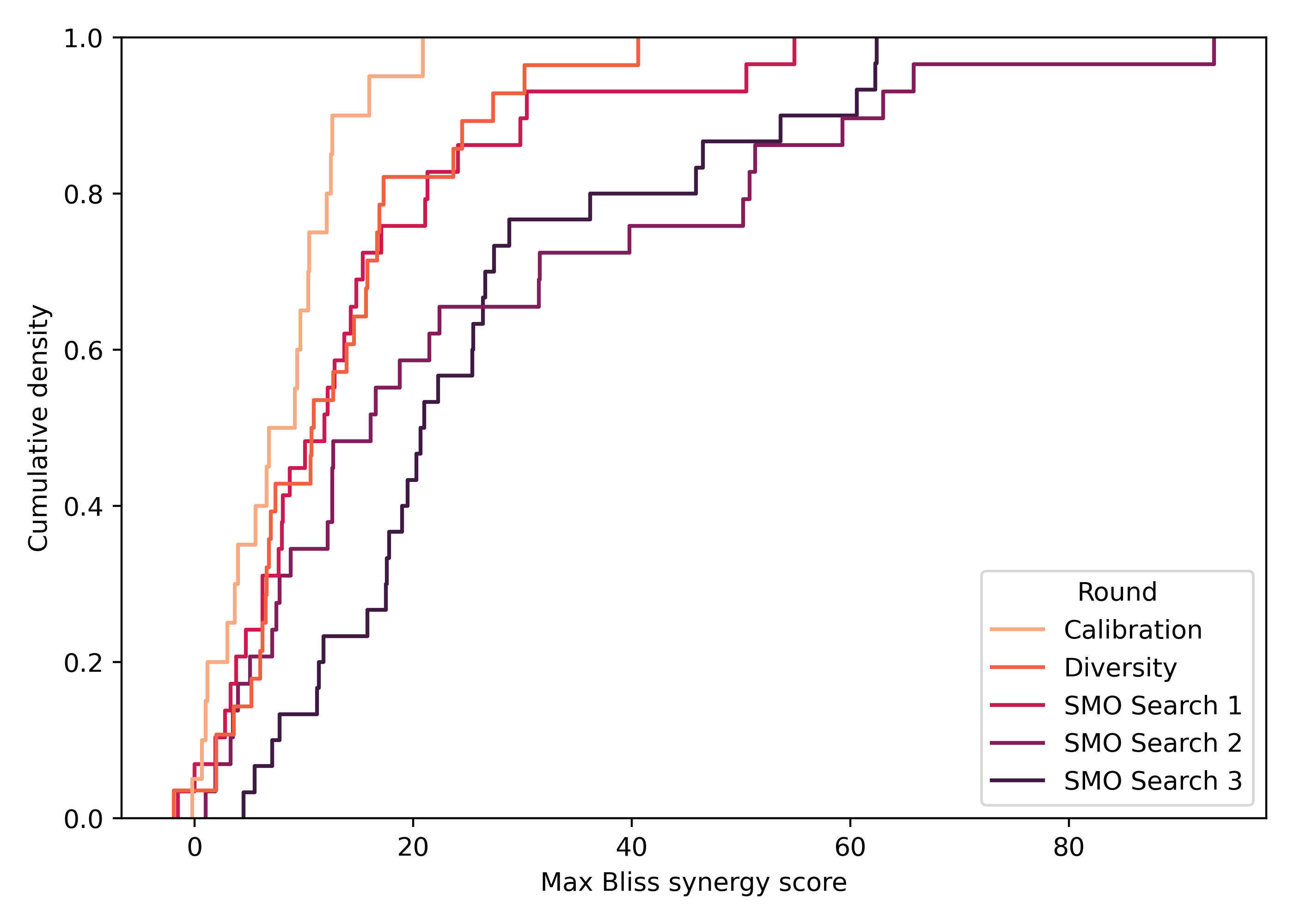}
\put(0,65){\Large C}
\put(56,27){\includegraphics[clip,width=3.75cm]{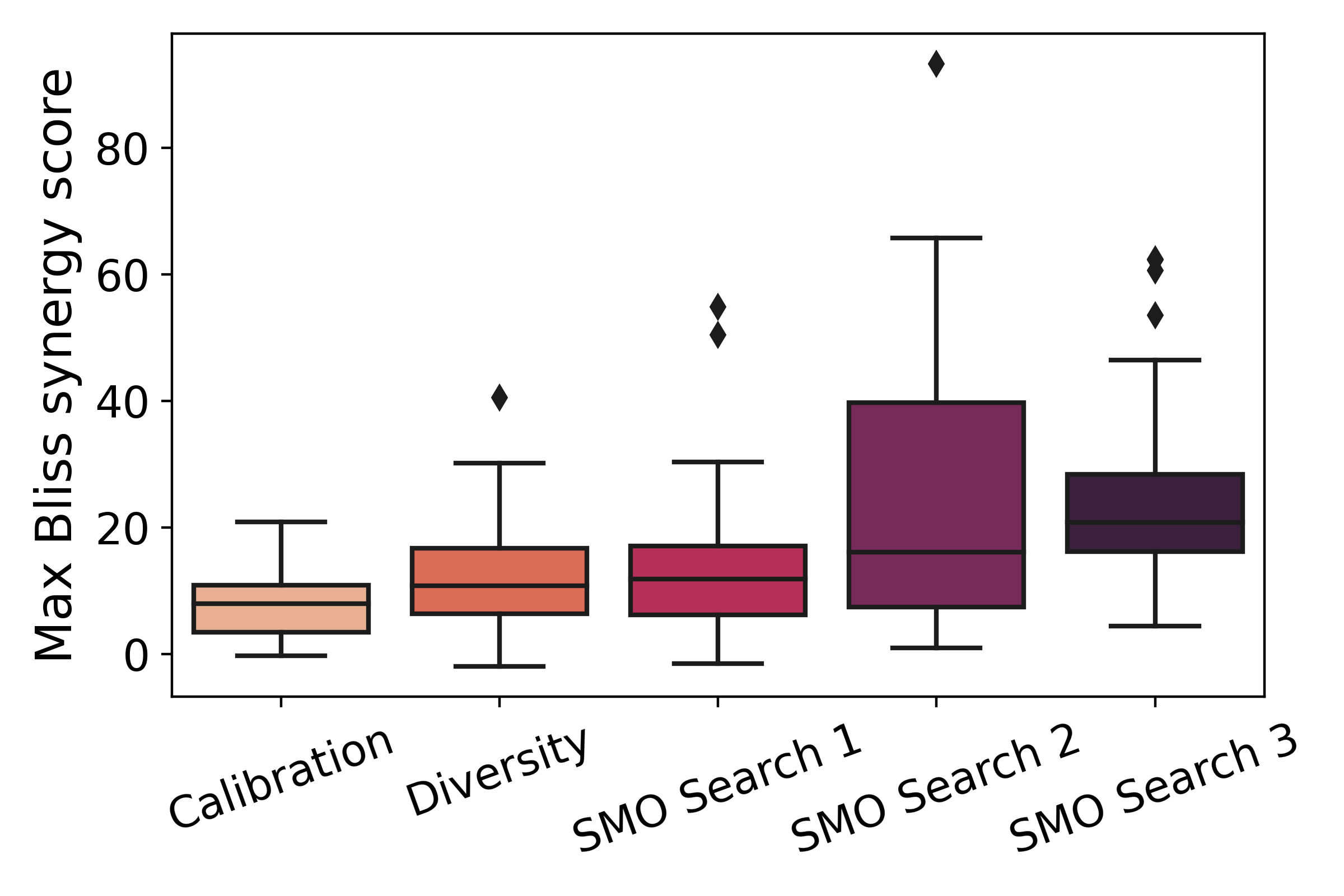}}
\put(45,10){\includegraphics[clip,width=2.75cm]{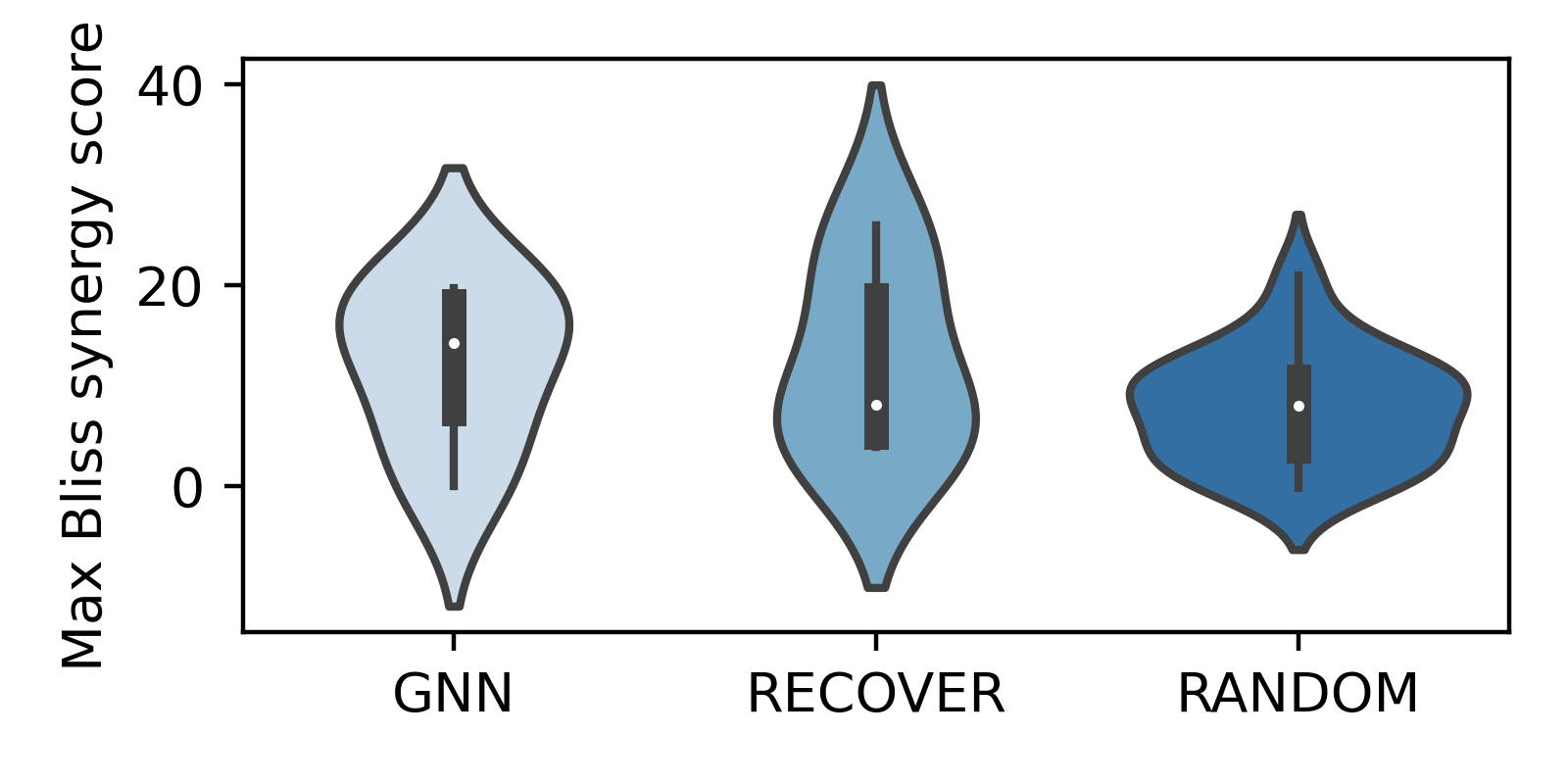}}
\put(63,26){\Large \textcolor{blue}{\textbf{/}}}
\end{overpic}
\begin{overpic}[clip, width=0.49\textwidth]{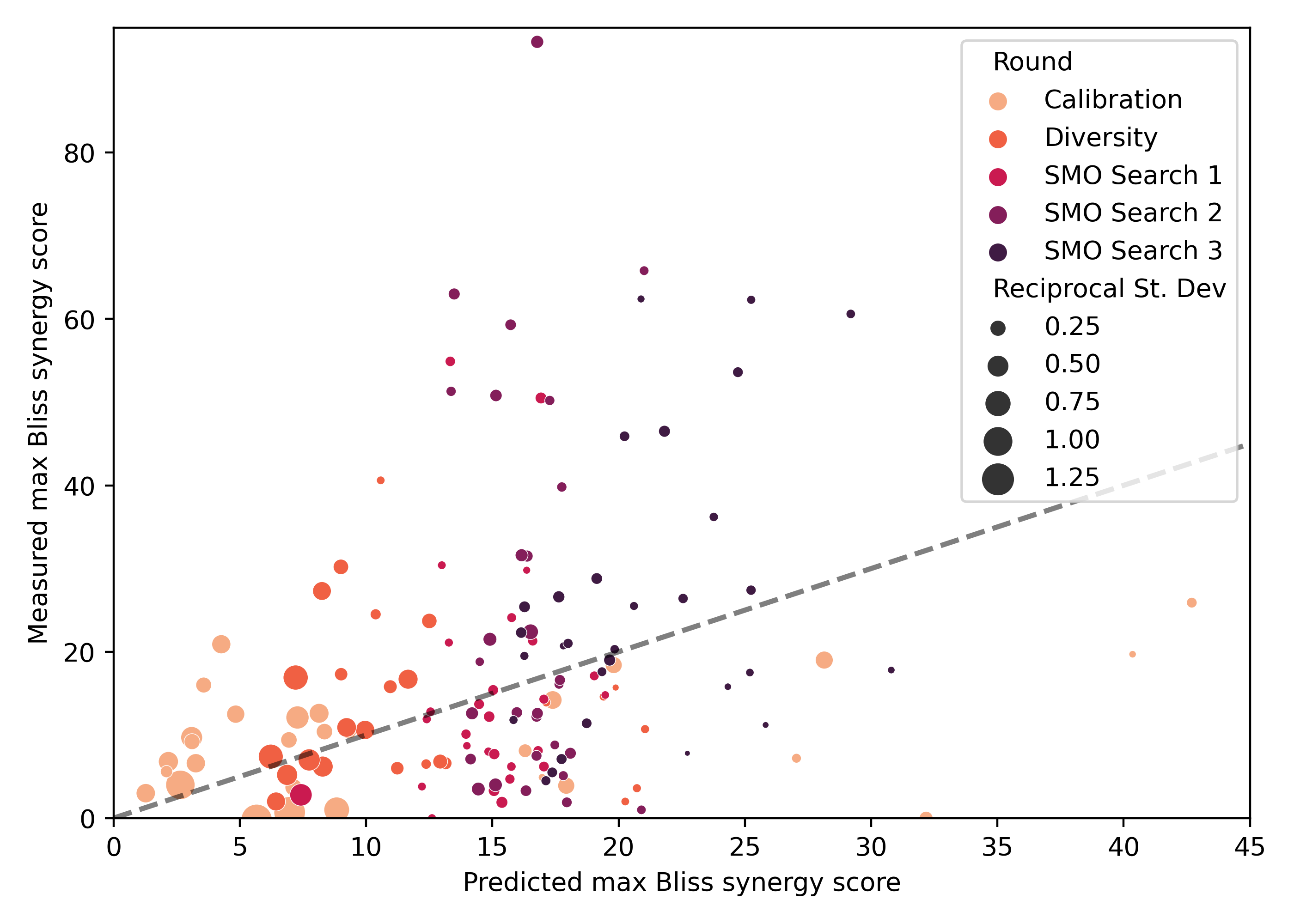}
\put(0,65){\Large D}
\put(8.25,45.5){\includegraphics[width=3cm]{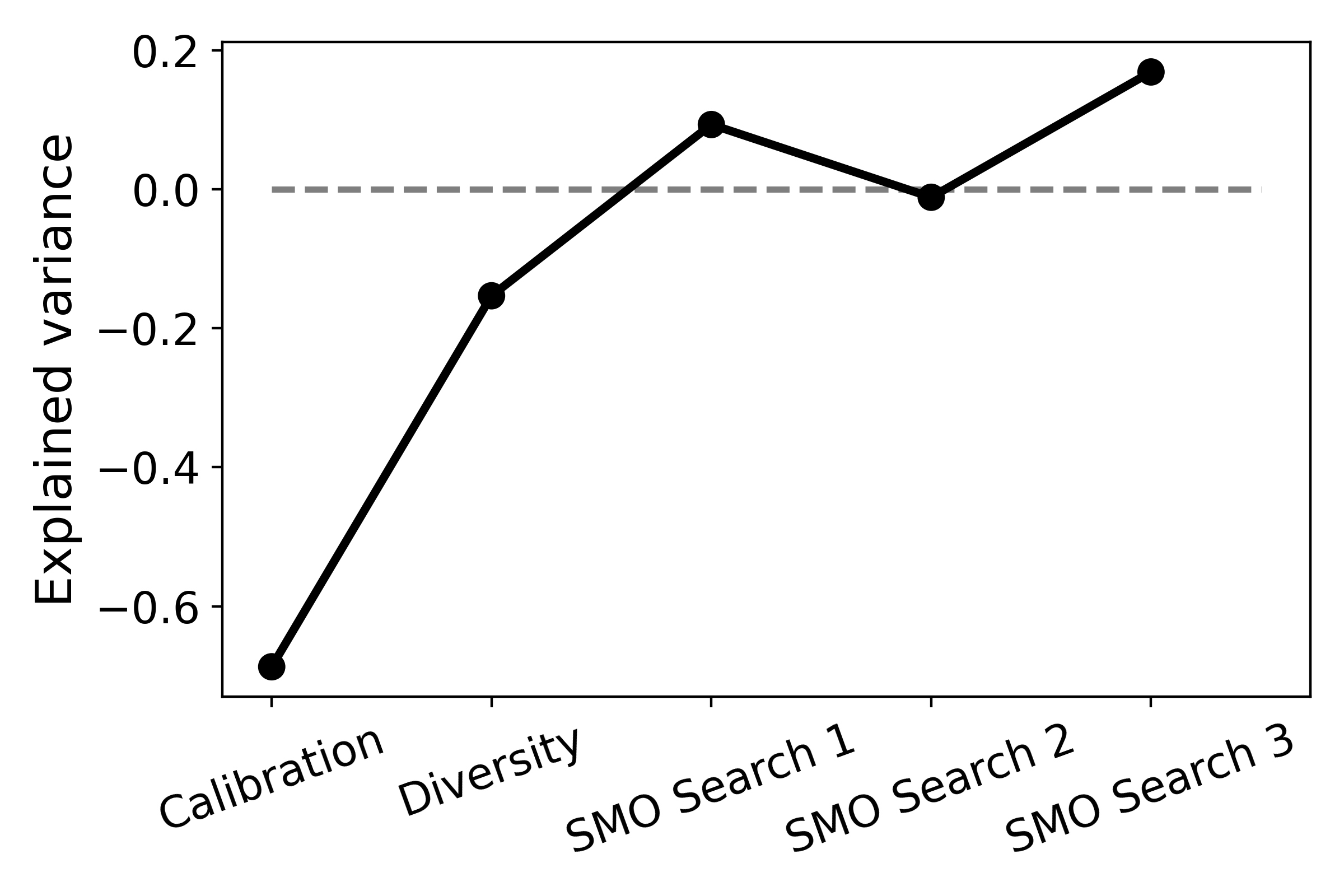}}
\end{overpic}
\end{center}

\caption{Prospective use of {\Recover} for \textit{in vitro} evaluation. \textbf{(A.)} Network plot indicating which pairs of drugs were identified at each round, line colour represents synergy. \textbf{(B.)} Heatmap representing drug combinations used during: pretraining (NCI-ALMANAC), in the five subsequent experiment rounds, and combinations excluded from the analysis. Drug combinations which were not available for pretraining or were not selected for experiments are represented in white. \textbf{(C.)} Cumulative density plot of max Bliss synergy score for each experimental round; (inset) box plot representation and calibration round details. \textbf{(D.)} Predicted versus actual plot for max Bliss synergy score and (inset) the explained variance is plotted for each experimental round.}
\label{fig:synergy_dist}
\end{figure*}

From the \textit{in silico} results, we now test {\Recover} prospectively using a cancer cell model, leveraging publicly available data for pretraining. Using the insights from Section \ref{sec:backtesting}, the queriable space of drug combinations was designed to include drug pairs where only one compound was already seen by the model during pretraining --- with a second compound not seen before, illustrated in Figure \ref{fig:data_summary}A. Details about the model used to generate recommendations are available in Appendix \ref{sec:recommendation_generation}. 

The MCF7 cell line was used to generate $6\times6$ dose-response matrices, see Appendix \ref{sec:experimental_protocol} for more details. We perform multiple rounds of {\Recover}-informed wet lab experiments and observe sequential improvements in performance. The rounds of experiments are described as follows:

\begin{itemize}

    \item[\textbf{1. Calibration}.] The initial round of experiments was performed to supplement publicly available data with 20 randomly selected unseen drug combinations. Furthermore, we confirmed the previous \textit{in silico} result that we could not predict synergy scores for unseen drugs \Review{(using a model trained on public data)} through selecting 5 highly synergistic combinations selected by {\Recover}. In addition, 5 more drug combinations were selected by a Graph Neural Network (GNN) model in the style of Zitnik \textit{et al.} \cite{zitnik2018modeling} that we did not develop further due to the computational overhead. It was also specified that each drug should appear in at most a single drug combination queried.
    
    \item[\textbf{2. Diversity}.] Thereafter, drug combinations are selected using model predictions in conjunction with the upper confidence bound (UCB) acquisition function. To ensure that we quickly observe all single drugs at least once (as we showed that the model cannot generalize well to combinations involving unseen drugs), we select our batch of experiments as follows. First we rank combinations according to their acquisition function score. We then find the first combination which involves a drug that has not yet been used (or involved in one of the combinations from the current batch), and add it to the batch. We repeat until we have 30 combinations in the batch.
    
    \item[\textbf{3. SMO Search}.] {\Recover} is now free to select any drug pairs of interest for testing, with the requirement that any single drug may be selected no more than 5 times (to avoid oversampling and depletion of chemical stock). Three such rounds have been performed in this manner.
    
\end{itemize}

The search space was constructed as follows. The NCI-ALMANAC includes 95 unique drugs that were employed in combinations tested on the MCF7 cell line, see grey area in Figure \ref{fig:synergy_dist}B. We chose to deprioritise drugs without a well-characterised mechanism of action (MoA) to facilitate biological interpretation and validation of the results, see light blue area in Figure \ref{fig:synergy_dist}B. To achieve this, drugs in NCI-ALMANAC were annotated with known targets extracted from the ChEMBL drug mechanism table: 54 drugs matched with at least one known target were thus selected. An additional 54 drugs were selected by clustering drugs with known MoA that are included in the DrugComb \citep{zagidullin2019drugcomb} database but not in NCI-ALMANAC, see Appendix \ref{sec:dataset_processing} for details. Hence, a search space including a total of 2,916 drug combinations was obtained, see the white area in Figure \ref{fig:synergy_dist}B. \Review{In Figure \ref{fig:synergy_dist}A, we illustrate the pairs of drugs selected in each round of experiments.}

\Review{We now evaluate both the synergy scores of the drug combinations selected and the underlying accuracy of the model.} In Figure \ref{fig:synergy_dist}C, we plot the cumulative density function of each experimental round. We note that the mean synergy score significantly increases between the first and the third round (\textit{t}-test, $\text{p}<0.05$); this trend further continues by the fifth round (\textit{t}-test, $\text{p}<10^{-5}$). Moreover, the distribution starts developing a heavier tail towards high max Bliss synergy scores. This emergent heavy tail already appears significant when comparing the distribution in the first SMO Search round to the background distribution of synergy scores in NCI-ALMANAC (Kolmogorov--Smirnov test, $\text{p}<0.025$), see Figure \ref{fig:data_summary}C. Finally, the highest max Bliss synergy score observed increases between rounds until the second SMO Search round, whereby the behaviour appears to have saturated.

\Review{In Figure \ref{fig:synergy_dist}D, we plot the predicted versus actual plot of the max Bliss synergy score. Here, the point size in the scatter plot is inversely proportional to the model uncertainty, therefore we display confident predictions as large points and vice versa. As expected, more confident predictions are closer to the $y=x$ line. Less confident predictions are associated with larger max Bliss synergy scores. Moreover, we systematically underestimate the measured max Bliss synergy score (more points far above $y=x$ line); this intuitively makes sense as we are trying to identify highly synergistic drug combinations that are not within our training dataset. Figure \ref{fig:synergy_dist}D (inset) displays the increase in (weighted) explained variance from one round to the next; weights are chosen to be the reciprocal of the model uncertainty. We find that initially the explained variance is negative, i.e., our model has no predictive power. However, as the experiments continue, a positive trend emerges such that we have a small amount of predictive power by the end of the experiments.}

\Review{This increase in the synergy of queried combinations from one round to the next demonstrated in Figure \ref{fig:synergy_dist}C is expected and can be attributed to two factors. First, we needed to adapt the model to predict in a new experimental setting. From the Study Transfer Task in Figure~\ref{fig:out_of_dist_diagram_appendix}, we know this would otherwise be an impossible task and thus motivates the Calibration round. After the Calibration round, one expects that the systematic biases learnt by the model during pretraining are minimised. At this point, the model is in a scenario akin to task (ii.) in Figure~\ref{fig:out_of_dist_diagram}. Second, we can improve performance further by enforcing that (almost) all drugs have been evaluated at some point, which subsequently motivated the Diversity round. Thereafter, the model is free to optimise during the SMO rounds to the extend that it is able to, leveraging model predictions and model uncertainties. In fact, due to activity cliff effects \cite{stumpfe2020advances}, there are likely fundamental limits on quantifying the relationship between model uncertainty and model error; we perform a preliminary investigation of this phenomena in Appendix \ref{app:clustering_diagram}. From our prospective use of {\Recover}, we not only discover highly synergistic drug combinations, but also demonstrate that high predictive power is not strictly necessary to identify synergistic drug combinations.}

\subsection{Discovery and rediscovery of novel synergistic drug combinations}\label{sec:ct_drug}

In Figure \ref{fig:remaining_top_combos}, we plot the Bliss synergy dose-response matrices for the 12 most synergistic drug combinations (from $\sim$150 tested) with Alfacalcidol and Crizotinib achieving a max Bliss score above 90. Of note, we rapidly discover drug combinations with similar mechanisms and efficacy to those already in clinical trials. Namely, within the first SMO search round we found: (a.) Alisertib \& Pazopanib, and (b.) Flumatinib \& Mitoxantrone. The concentration intervals for the drugs used in both drug combinations that show synergy as determined from the Bliss matrices in Figure \ref{fig:combo_exemplars} are in range of therapeutically relevant plasma concentrations \cite{shah2019phase, serononovantrone}, or as observed in \textit{in vivo} animal experiments (Flumatinib) \cite{zhao2014flumatinib}.

Pazopanib inhibits angiogenesis through targeting a range of kinases including vascular endothelial growth factor receptor (VEGFR), platelet-derived growth factor receptor (PDGFR), c-KIT and fibroblast growth factor receptors (FGFR); in contrast Alisertib is a highly selective inhibitor of mitotic Aurora A kinase. Synergism between the two agents is hypothesised to be linked to the observation that mitosis-targeting agents also demonstrate antiangiogenic effects. In an independent study, the combination of Alisertib \& Pazopanib has successfully completed phase 1b clinical trials for advanced solid tumours \cite{shah2019phase}.  The combination of Flumatinib and Mitoxantrone appears to be linked to a similar mechanism, but does not seem to have been studied in the biomedical literature. Whilst Flumatinib is a tyrosine kinase inhibitor targeting Bcr-Abl, PDGFR and c-KIT, Mitoxantrone is a Type II topoisomerase inhibitor.

\subsection{{\Recover} drug embeddings capture both structural and biological information}

\begin{figure*}[ht]
\begin{center}
\hspace{0.2cm}
\begin{overpic}[trim={.01cm .01cm .01cm .01cm},clip, width=0.99\textwidth ]{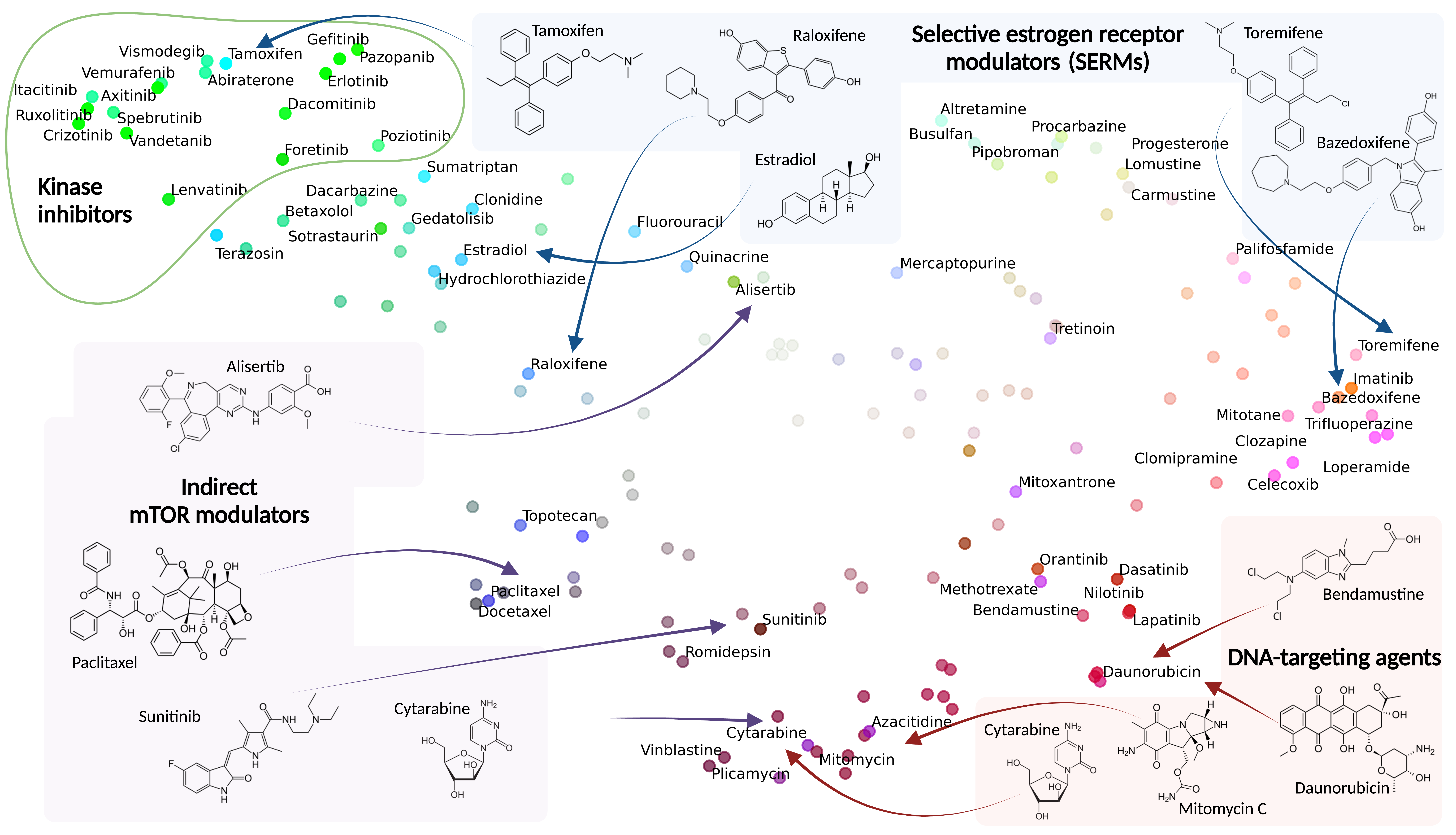}
\end{overpic}
\end{center}
\caption{UMAP of {\Recover} drug embeddings with the colour scheme generated to indicate the known target profile of the drugs.}
\label{fig:embeddings_umap}
\end{figure*}

To get a better insight into the drug embeddings learnt by {\Recover}, we report uniform manifold approximation and projection (UMAP) visualizations of the drug embeddings generated by the \textit{single drug} module in Figure~\ref{fig:embeddings_umap}. The colour of each point is chosen by applying principal component analysis (PCA) to the binary matrix of drug--targets and scaling the first 3 dimensions into an RGB triplet; high transparency indicates drugs with a PCA target profile close to the average PCA target profile (calculated over all drugs). \Review{In short, the position of the points indicates what {\Recover} has learnt about the drugs, and the colour represents information known about drug mechanisms from other databases not used in the training procedure.}

We note that the {\Recover} model does not use information on drug targets, however drugs with similar colours are located within similar areas of UMAP space. We also observe broad sensible patterns in UMAP space based on structure, for example, most kinase inhibitors (with the -\textit{nib} suffix) appear in the upper left hand of the UMAP. Moreover, drugs with similar mechanisms tend to be co-located, for example, see structurally diverse DNA-targeting agents in the bottom right of the UMAP. As a counterpoint, we observe that agents with either: mixed agonist/antagonist profiles, including selective estrogen receptor modulators (SERMS); or targeting genes through indirect mechanisms, including mammalian target of rapamycin (mTOR), lead to less structured patterns in UMAP space. We believe this is a highly novel observation, and suggests that were this screen to be scaled to a larger library of small molecules, one may be able to group diverse structures into common biological mechanisms.

\section{Discussion}

Drug combinations can achieve benefits unattainable by mono-therapies and are routinely investigated within clinical trials (e.g., PD-1/PD-L1 inhibitors combined with other agents \cite{zhu2021combination}), and utilised within clinical practice (e.g., antiretroviral treatment of HIV where between 3-4 agents may be used \cite{feng2019quadruple}). To this end, we have presented the SMO toolbox {\Recover} for drug combination identification. We showcase a general methodology, consisting of careful analysis of the properties of our machine learning pipeline --- such as its out-of-distribution generalization capacities --- to help us design key aspects of our prospective experiments, to eventually ensure a smooth and successful interaction between the SMO pipeline and the wet lab. Highly synergistic drug combinations have been identified and the resulting learnt embeddings appear to capture both structural and biological information. We provide commentary on key aspects on our approach covering datasets, computational methodology, and wet-lab techniques.

We note the considerable difficulties of working with publicly available datasets with discrepancies in the data generation process. Inconsistent media between multiple labs, the presence of spontaneous mutations within immortalised \textit{in vitro} models, and differences in experimental protocols limit ease of data integration between laboratories \cite{hirsch2019vitro}. In particular, systematic biases limit generalisability of model predictions to subsequent prospective experiments. Within oncology, protein coding mutations may drive resistance to any one chemotherapeutic agent, but also large scale gene dosing changes from non-coding mutations \cite{avsec2021effective}, copy number variation \cite{shao2019copy} and aneuploidy \cite{taylor2021rethinking}. These issues have been somewhat alleviated through careful choice of metric to optimise (max pooled Bliss synergy scores) and only using publicly available data for pretraining (when compared to using this data for prediction without adaptation). 

From a computational perspective, we experimented with a range of more complicated models. For example, we considered using graph neural networks to model biomolecular interactions \cite{gaudelet2021utilizing}, which have numerous benefits including greater biological interpretability and incorporation of prior knowledge, namely drug--target and protein--protein interactions. However, these models only resulted in marginal increases in performance whilst requiring substantially more computational resources. We believe that the limited diversity of the dataset and the simplicity of the task, a one dimensional regression, did not allow these more advanced approaches to reach their full potential. Therefore, we prioritised a framework that could be run quickly for rapid turnaround of recommendations for experimental testing.

When considering a SMO setting, we are required to collapse highly complex information into a single number to be optimised (i.e., a synergy score). Whilst there is opportunity to improve choices of metric (synergy scores may not reflect absolute cell viability), assay readouts that better characterise cell state (compared to cell viability) may provide a stronger starting point. In particular, an omic readout, through transcriptomics \cite{subramanian2017next} and/or single cell profiling \cite{chen2021high, peidli2022scperturb}; and high content imaging \cite{bray2016cell} provide a much higher dimensional measurement of cell state. Furthermore, derived properties from these readouts may be more interpretable, e.g., pathway activation \cite{nguyen2019identifying} or extracellular signalling \cite{taylor2019simulated}. Remarkably, even while only using cell viability as a readout, we achieved significant progress in identifying novel synergistic drug combinations. 

From the systematic screen by Jaak \textit{et al.} \cite{jaaks2022effective}, they conclude that: synergy between drugs is rare and highly context-dependent. {\Recover} provides a means to identify such synergies while requiring substantially less screening than an exhaustive evaluation; thus we expect {\Recover} and similar such systems may have a role to play when addressing novel emergent infectious disease such as the COVID-19 pandemic.

\section{Acknowledgements}

This work was supported, in whole or in part, by the Bill {\&} Melinda Gates Foundation [INV-019229]. Under the grant conditions of the Foundation, a Creative Commons Attribution 4.0 Generic License has already been assigned to the Author Accepted Manuscript version that might arise from this submission. The findings and conclusions contained within are those of the authors and do not necessarily reflect positions or policies of the Bill {\&} Melinda Gates Foundation. The authors would also like to thank Andrew Trister, Isabelle Lacroix, Benjamin Swerner, and Lindsay Edwards for useful discussion and support.

\bibliographystyle{unsrt}
\bibliography{references}

\begin{thebibliography}{10}

\bibitem{tyers2019drug}
Mike Tyers and Gerard~D Wright.
\newblock Drug combinations: a strategy to extend the life of antibiotics in
  the 21st century.
\newblock {\em Nature Reviews Microbiology}, 17(3):141--155, 2019.

\bibitem{mokhtari2017combination}
Reza~Bayat Mokhtari, Tina~S Homayouni, Narges Baluch, Evgeniya Morgatskaya,
  Sushil Kumar, Bikul Das, and Herman Yeger.
\newblock Combination therapy in combating cancer.
\newblock {\em Oncotarget}, 8(23):38022, 2017.

\bibitem{delou2019highlights}
Jo{\~a}o Delou, Alana~SO Souza, Leonel Souza, and Helena~L Borges.
\newblock Highlights in resistance mechanism pathways for combination therapy.
\newblock {\em Cells}, 8(9):1013, 2019.

\bibitem{al-lazikaniCombinatorialDrugTherapy2012}
Bissan {Al-Lazikani}, Udai Banerji, and Paul Workman.
\newblock Combinatorial drug therapy for cancer in the post-genomic era.
\newblock {\em Nature Biotechnology}, 30(7):679--692, July 2012.

\bibitem{janes2018reframe}
Jeff Janes, Megan~E Young, Emily Chen, Nicole~H Rogers, Sebastian
  Burgstaller-Muehlbacher, Laura~D Hughes, Melissa~S Love, Mitchell~V Hull,
  Kelli~L Kuhen, Ashley~K Woods, et~al.
\newblock The reframe library as a comprehensive drug repurposing library and
  its application to the treatment of cryptosporidiosis.
\newblock {\em Proceedings of the National Academy of Sciences},
  115(42):10750--10755, 2018.

\bibitem{clare2019industrial}
Rachel~H Clare, Catherine Bardelle, Paul Harper, W~David Hong, Ulf
  B{\"o}rjesson, Kelly~L Johnston, Matthew Collier, Laura Myhill, Andrew
  Cassidy, Darren Plant, et~al.
\newblock Industrial scale high-throughput screening delivers multiple fast
  acting macrofilaricides.
\newblock {\em Nature communications}, 10(1):1--8, 2019.

\bibitem{zhou2020artificial}
Yadi Zhou, Fei Wang, Jian Tang, Ruth Nussinov, and Feixiong Cheng.
\newblock Artificial intelligence in covid-19 drug repurposing.
\newblock {\em The Lancet Digital Health}, 2020.

\bibitem{sverchkov2017review}
Yuriy Sverchkov and Mark Craven.
\newblock A review of active learning approaches to experimental design for
  uncovering biological networks.
\newblock {\em PLoS computational biology}, 13(6):e1005466, 2017.

\bibitem{bulusuModellingCompoundCombination2016}
Krishna~C. Bulusu, Rajarshi Guha, Daniel~J. Mason, Richard P.~I. Lewis, Eugene
  Muratov, Yasaman Kalantar~Motamedi, Murat Cokol, and Andreas Bender.
\newblock Modelling of compound combination effects and applications to
  efficacy and toxicity: State-of-the-art, challenges and perspectives.
\newblock {\em Drug Discovery Today}, 21(2):225--238, February 2016.

\bibitem{cereto2015molecular}
Adri{\`a} Cereto-Massagu{\'e}, Mar{\'\i}a~Jos{\'e} Ojeda, Cristina Valls,
  Miquel Mulero, Santiago Garcia-Vallv{\'e}, and Gerard Pujadas.
\newblock Molecular fingerprint similarity search in virtual screening.
\newblock {\em Methods}, 71:58--63, 2015.

\bibitem{karczewski2018integrative}
Konrad~J Karczewski and Michael~P Snyder.
\newblock Integrative omics for health and disease.
\newblock {\em Nature Reviews Genetics}, 19(5):299, 2018.

\bibitem{wildenhain2015prediction}
Jan Wildenhain, Michaela Spitzer, Sonam Dolma, Nick Jarvik, Rachel White,
  Marcia Roy, Emma Griffiths, David~S Bellows, Gerard~D Wright, and Mike Tyers.
\newblock Prediction of synergism from chemical-genetic interactions by machine
  learning.
\newblock {\em Cell Systems}, 1(6):383--395, 2015.

\bibitem{cheng2019network}
Feixiong Cheng, Istv{\'a}n~A Kov{\'a}cs, and Albert-L{\'a}szl{\'o}
  Barab{\'a}si.
\newblock Network-based prediction of drug combinations.
\newblock {\em Nature communications}, 10(1):1--11, 2019.

\bibitem{menden2019community}
Michael~P Menden, Dennis Wang, Mike~J Mason, Bence Szalai, Krishna~C Bulusu,
  Yuanfang Guan, Thomas Yu, Jaewoo Kang, Minji Jeon, Russ Wolfinger, et~al.
\newblock Community assessment to advance computational prediction of cancer
  drug combinations in a pharmacogenomic screen.
\newblock {\em Nature communications}, 10(1):1--17, 2019.

\bibitem{lecun2015deep}
Yann LeCun, Yoshua Bengio, and Geoffrey Hinton.
\newblock Deep learning.
\newblock {\em nature}, 521(7553):436--444, 2015.

\bibitem{zitnik2018modeling}
Marinka Zitnik, Monica Agrawal, and Jure Leskovec.
\newblock Modeling polypharmacy side effects with graph convolutional networks.
\newblock {\em Bioinformatics}, 34(13):i457--i466, 2018.

\bibitem{deac2019drug}
Andreea Deac, Yu-Hsiang Huang, Petar Veli{\v{c}}kovi{\'c}, Pietro Li{\`o}, and
  Jian Tang.
\newblock Drug-drug adverse effect prediction with graph co-attention.
\newblock {\em arXiv preprint arXiv:1905.00534}, 2019.

\bibitem{preuer2018deepsynergy}
Kristina Preuer, Richard~PI Lewis, Sepp Hochreiter, Andreas Bender, Krishna~C
  Bulusu, and G{\"u}nter Klambauer.
\newblock Deepsynergy: predicting anti-cancer drug synergy with deep learning.
\newblock {\em Bioinformatics}, 34(9):1538--1546, 2018.

\bibitem{jin2021deep}
Wengong Jin, Jonathan~M Stokes, Richard~T Eastman, Zina Itkin, Alexey~V
  Zakharov, James~J Collins, Tommi~S Jaakkola, and Regina Barzilay.
\newblock Deep learning identifies synergistic drug combinations for treating
  covid-19.
\newblock {\em Proceedings of the National Academy of Sciences}, 118(39), 2021.

\bibitem{rozemberczki2021moomin}
Benedek Rozemberczki, Anna Gogleva, Sebastian Nilsson, Gavin Edwards, Andriy
  Nikolov, and Eliseo Papa.
\newblock Moomin: Deep molecular omics network for anti-cancer drug combination
  therapy.
\newblock {\em arXiv preprint arXiv:2110.15087}, 2021.

\bibitem{Kashif_Andersson_Hassan_Karlsson_Senkowski_Fryknas_Nygren_Larsson_Gustafsson_2015}
M.~Kashif, C.~Andersson, S.~Hassan, H.~Karlsson, W.~Senkowski, M.~Fryknäs,
  P.~Nygren, R.~Larsson, and M.G. Gustafsson.
\newblock In vitro discovery of promising anti-cancer drug combinations using
  iterative maximisation of a therapeutic index.
\newblock {\em Scientific Reports}, 5(1):14118, Nov 2015.

\bibitem{zagidullin2019drugcomb}
Bulat Zagidullin, Jehad Aldahdooh, Shuyu Zheng, Wenyu Wang, Yinyin Wang, Joseph
  Saad, Alina Malyutina, Mohieddin Jafari, Ziaurrehman Tanoli, Alberto Pessia,
  et~al.
\newblock Drugcomb: an integrative cancer drug combination data portal.
\newblock {\em Nucleic acids research}, 47(W1):W43--W51, 2019.

\bibitem{vzilinskas1972bayes}
A~{\v{Z}}ilinskas and J~Mockus.
\newblock On a bayes method for seeking an extremum.
\newblock {\em Automatika i vychislitelnaja tekhnika}, (3), 1972.

\bibitem{o2016unbiased}
Jennifer O'Neil, Yair Benita, Igor Feldman, Melissa Chenard, Brian Roberts,
  Yaping Liu, Jing Li, Astrid Kral, Serguei Lejnine, Andrey Loboda, et~al.
\newblock An unbiased oncology compound screen to identify novel combination
  strategies.
\newblock {\em Molecular Cancer Therapeutics}, 15(6):1155--1162, 2016.

\bibitem{holbeck2017national}
Susan~L Holbeck, Richard Camalier, James~A Crowell, Jeevan~Prasaad
  Govindharajulu, Melinda Hollingshead, Lawrence~W Anderson, Eric Polley, Larry
  Rubinstein, Apurva Srivastava, Deborah Wilsker, et~al.
\newblock {The National Cancer Institute ALMANAC: a comprehensive screening
  resource for the detection of anticancer drug pairs with enhanced therapeutic
  activity}.
\newblock {\em Cancer Research}, 77(13):3564--3576, 2017.

\bibitem{stumpfe2020advances}
Dagmar Stumpfe, Huabin Hu, and J{\"u}rgen Bajorath.
\newblock Advances in exploring activity cliffs.
\newblock {\em Journal of Computer-Aided Molecular Design}, 34(9):929--942,
  2020.

\bibitem{shah2019phase}
Hiral~A Shah, James~H Fischer, Neeta~K Venepalli, Oana~C Danciu, Sonia
  Christian, Meredith~J Russell, Li~C Liu, James~P Zacny, and Arkadiusz~Z
  Dudek.
\newblock Phase i study of aurora a kinase inhibitor alisertib (mln8237) in
  combination with selective vegfr inhibitor pazopanib for therapy of advanced
  solid tumors.
\newblock {\em American journal of clinical oncology}, 42(5):413--420, 2019.

\bibitem{serononovantrone}
EMD Serono.
\newblock Novantrone: mitoxantrone for injection concentrate, additional safety
  information, apr 2005.

\bibitem{zhao2014flumatinib}
Jie Zhao, Haitian Quan, Yongping Xu, Xiangqian Kong, Lu~Jin, and Liguang Lou.
\newblock Flumatinib, a selective inhibitor of bcr-abl/pdgfr/kit, effectively
  overcomes drug resistance of certain kit mutants.
\newblock {\em Cancer science}, 105(1):117--125, 2014.

\bibitem{zhu2021combination}
Shaoming Zhu, Tian Zhang, Lei Zheng, Hongtao Liu, Wenru Song, Delong Liu, Zihai
  Li, and Chong-xian Pan.
\newblock Combination strategies to maximize the benefits of cancer
  immunotherapy.
\newblock {\em Journal of Hematology \& Oncology}, 14(1):1--33, 2021.

\bibitem{feng2019quadruple}
Qi~Feng, Aoshuang Zhou, Huachun Zou, Suzanne Ingle, Margaret~T May, Weiping
  Cai, Chien-Yu Cheng, Zuyao Yang, and Jinling Tang.
\newblock Quadruple versus triple combination antiretroviral therapies for
  treatment naive people with hiv: systematic review and meta-analysis of
  randomised controlled trials.
\newblock {\em bmj}, 366, 2019.

\bibitem{hirsch2019vitro}
Cordula Hirsch and Stefan Schildknecht.
\newblock In vitro research reproducibility: Keeping up high standards.
\newblock {\em Frontiers in pharmacology}, 10:1484, 2019.

\bibitem{avsec2021effective}
Ziga Avsec, Vikram Agarwal, Daniel Visentin, Joseph~R Ledsam, Agnieszka
  Grabska-Barwinska, Kyle~R Taylor, Yannis Assael, John Jumper, Pushmeet Kohli,
  and David~R Kelley.
\newblock Effective gene expression prediction from sequence by integrating
  long-range interactions.
\newblock {\em bioRxiv}, 2021.

\bibitem{shao2019copy}
Xin Shao, Ning Lv, Jie Liao, Jinbo Long, Rui Xue, Ni~Ai, Donghang Xu, and
  Xiaohui Fan.
\newblock Copy number variation is highly correlated with differential gene
  expression: a pan-cancer study.
\newblock {\em BMC medical genetics}, 20(1):1--14, 2019.

\bibitem{taylor2021rethinking}
Jake~P Taylor-King.
\newblock Rethinking rare disease: longevity-enhancing drug targets through
  x-linked aneuploidy.
\newblock {\em Trends in Genetics}, 2021.

\bibitem{gaudelet2021utilizing}
Thomas Gaudelet, Ben Day, Arian~R Jamasb, Jyothish Soman, Cristian Regep,
  Gertrude Liu, Jeremy~BR Hayter, Richard Vickers, Charles Roberts, Jian Tang,
  et~al.
\newblock Utilizing graph machine learning within drug discovery and
  development.
\newblock {\em Briefings in bioinformatics}, 22(6):bbab159, 2021.

\bibitem{subramanian2017next}
Aravind Subramanian, Rajiv Narayan, Steven~M Corsello, David~D Peck, Ted~E
  Natoli, Xiaodong Lu, Joshua Gould, John~F Davis, Andrew~A Tubelli, Jacob~K
  Asiedu, et~al.
\newblock A next generation connectivity map: L1000 platform and the first
  1,000,000 profiles.
\newblock {\em Cell}, 171(6):1437--1452, 2017.

\bibitem{chen2021high}
Haide Chen, Yuan Liao, Guodong Zhang, Zhongyi Sun, Lei Yang, Xing Fang, Huiyu
  Sun, Lifeng Ma, Yuting Fu, Jingyu Li, et~al.
\newblock High-throughput microwell-seq 2.0 profiles massively multiplexed
  chemical perturbation.
\newblock {\em Cell discovery}, 7(1):1--4, 2021.

\bibitem{peidli2022scperturb}
Stefan Peidli, Tessa~Durakis Green, Ciyue Shen, Torsten Gross, Joseph Min, Jake
  Taylor-King, Debora Marks, Augustin Luna, Nils Bluthgen, and Chris Sander.
\newblock scperturb: Information resource for harmonized single-cell
  perturbation data.
\newblock {\em bioRxiv}, 2022.

\bibitem{bray2016cell}
Mark-Anthony Bray, Shantanu Singh, Han Han, Chadwick~T Davis, Blake Borgeson,
  Cathy Hartland, Maria Kost-Alimova, Sigrun~M Gustafsdottir, Christopher~C
  Gibson, and Anne~E Carpenter.
\newblock Cell painting, a high-content image-based assay for morphological
  profiling using multiplexed fluorescent dyes.
\newblock {\em Nature protocols}, 11(9):1757--1774, 2016.

\bibitem{nguyen2019identifying}
Tuan-Minh Nguyen, Adib Shafi, Tin Nguyen, and Sorin Draghici.
\newblock Identifying significantly impacted pathways: a comprehensive review
  and assessment.
\newblock {\em Genome biology}, 20(1):1--15, 2019.

\bibitem{taylor2019simulated}
Jake~P Taylor-King, Etienne Baratchart, Andrew Dhawan, Elizabeth~A Coker,
  Inga~Hansine Rye, Hege Russnes, S~Jon Chapman, David Basanta, and Andriy
  Marusyk.
\newblock Simulated ablation for detection of cells impacting paracrine
  signalling in histology analysis.
\newblock {\em Mathematical medicine and biology: a journal of the IMA},
  36(1):93--112, 2019.

\bibitem{jaaks2022effective}
Patricia Jaaks, Elizabeth~A Coker, Daniel~J Vis, Olivia Edwards, Emma~F
  Carpenter, Simonetta~M Leto, Lisa Dwane, Francesco Sassi, Howard Lightfoot,
  Syd Barthorpe, et~al.
\newblock Effective drug combinations in breast, colon and pancreatic cancer
  cells.
\newblock {\em Nature}, 603(7899):166--173, 2022.

\bibitem{perez2018film}
Ethan Perez, Florian Strub, Harm De~Vries, Vincent Dumoulin, and Aaron
  Courville.
\newblock Film: Visual reasoning with a general conditioning layer.
\newblock In {\em Proceedings of the AAAI Conference on Artificial
  Intelligence}, volume~32, 2018.

\bibitem{lakshminarayanan2017simple}
Balaji Lakshminarayanan, Alexander Pritzel, and Charles Blundell.
\newblock Simple and scalable predictive uncertainty estimation using deep
  ensembles, 2017.

\bibitem{DBLP:journals/corr/abs-2102-08501}
Moksh Jain, Salem Lahlou, Hadi Nekoei, Victor Butoi, Paul Bertin, Jarrid
  Rector{-}Brooks, Maksym Korablyov, and Yoshua Bengio.
\newblock {DEUP:} direct epistemic uncertainty prediction.
\newblock {\em CoRR}, abs/2102.08501, 2021.

\bibitem{Borkowski_Koch_Zettor_Pandi_Batista_Soudier_Faulon_2020}
Olivier Borkowski, Mathilde Koch, Agnès Zettor, Amir Pandi, Angelo~Cardoso
  Batista, Paul Soudier, and Jean-Loup Faulon.
\newblock Large scale active-learning-guided exploration for in vitro protein
  production optimization.
\newblock {\em Nature Communications}, 11(1):1872, Dec 2020.

\bibitem{King_Whelan_Jones_Reiser_Bryant_Muggleton_Kell_Oliver_2004}
Ross~D. King, Kenneth~E. Whelan, Ffion~M. Jones, Philip G.~K. Reiser,
  Christopher~H. Bryant, Stephen~H. Muggleton, Douglas~B. Kell, and Stephen~G.
  Oliver.
\newblock Functional genomic hypothesis generation and experimentation by a
  robot scientist.
\newblock {\em Nature}, 427(6971):247–252, Jan 2004.

\bibitem{Carbonell_Jervis_Robinson_Yan_Dunstan_Swainston_Vinaixa_Hollywood_Currin_Rattray_2018}
Pablo Carbonell, Adrian~J. Jervis, Christopher~J. Robinson, Cunyu Yan, Mark
  Dunstan, Neil Swainston, Maria Vinaixa, Katherine~A. Hollywood, Andrew
  Currin, Nicholas J.~W. Rattray, Sandra Taylor, Reynard Spiess, Rehana Sung,
  Alan~R. Williams, Donal Fellows, Natalie~J. Stanford, Paul Mulherin, Rosalind
  Le~Feuvre, Perdita Barran, Royston Goodacre, Nicholas~J. Turner, Carole
  Goble, George~Guoqiang Chen, Douglas~B. Kell, Jason Micklefield, Rainer
  Breitling, Eriko Takano, Jean-Loup Faulon, and Nigel~S. Scrutton.
\newblock An automated design-build-test-learn pipeline for enhanced microbial
  production of fine chemicals.
\newblock {\em Communications Biology}, 1(1):66, Dec 2018.

\bibitem{hie2020leveraging}
Brian Hie, Bryan~D Bryson, and Bonnie Berger.
\newblock Leveraging uncertainty in machine learning accelerates biological
  discovery and design.
\newblock {\em Cell Systems}, 11(5):461--477, 2020.

\bibitem{davies2015chembl}
Mark Davies, Micha{\l} Nowotka, George Papadatos, Nathan Dedman, Anna Gaulton,
  Francis Atkinson, Louisa Bellis, and John~P Overington.
\newblock {ChEMBL web services: streamlining access to drug discovery data and
  utilities}.
\newblock {\em Nucleic Acids Research}, 43(W1):W612--W620, 2015.

\bibitem{kim2019pubchem}
Sunghwan Kim, Jie Chen, Tiejun Cheng, Asta Gindulyte, Jia He, Siqian He,
  Qingliang Li, Benjamin~A Shoemaker, Paul~A Thiessen, Bo~Yu, et~al.
\newblock Pubchem 2019 update: improved access to chemical data.
\newblock {\em Nucleic acids research}, 47(D1):D1102--D1109, 2019.

\bibitem{ghandiNextgenerationCharacterizationCancer2019}
Mahmoud Ghandi, Franklin~W. Huang, Judit {Jan{\'e}-Valbuena}, Gregory~V.
  Kryukov, Christopher~C. Lo, E.~Robert McDonald, Jordi Barretina, Ellen~T.
  Gelfand, Craig~M. Bielski, Haoxin Li, Kevin Hu, Alexander~Y.
  {Andreev-Drakhlin}, Jaegil Kim, Julian~M. Hess, Brian~J. Haas, Fran{\c c}ois
  Aguet, Barbara~A. Weir, Michael~V. Rothberg, Brenton~R. Paolella, Michael~S.
  Lawrence, Rehan Akbani, Yiling Lu, Hong~L. Tiv, Prafulla~C. Gokhale, Antoine
  {de Weck}, Ali~Amin Mansour, Coyin Oh, Juliann Shih, Kevin Hadi, Yanay Rosen,
  Jonathan Bistline, Kavitha Venkatesan, Anupama Reddy, Dmitriy Sonkin, Manway
  Liu, Joseph Lehar, Joshua~M. Korn, Dale~A. Porter, Michael~D. Jones, Javad
  Golji, Giordano Caponigro, Jordan~E. Taylor, Caitlin~M. Dunning, Amanda~L.
  Creech, Allison~C. Warren, James~M. McFarland, Mahdi Zamanighomi, Audrey
  Kauffmann, Nicolas Stransky, Marcin Imielinski, Yosef~E. Maruvka, Andrew~D.
  Cherniack, Aviad Tsherniak, Francisca Vazquez, Jacob~D. Jaffe, Andrew~A.
  Lane, David~M. Weinstock, Cory~M. Johannessen, Michael~P. Morrissey, Frank
  Stegmeier, Robert Schlegel, William~C. Hahn, Gad Getz, Gordon~B. Mills,
  Jesse~S. Boehm, Todd~R. Golub, Levi~A. Garraway, and William~R. Sellers.
\newblock Next-generation characterization of the {{Cancer Cell Line
  Encyclopedia}}.
\newblock {\em Nature}, 569(7757):503--508, May 2019.

\bibitem{scikit-learn}
F.~Pedregosa, G.~Varoquaux, A.~Gramfort, V.~Michel, B.~Thirion, O.~Grisel,
  M.~Blondel, P.~Prettenhofer, R.~Weiss, V.~Dubourg, J.~Vanderplas, A.~Passos,
  D.~Cournapeau, M.~Brucher, M.~Perrot, and E.~Duchesnay.
\newblock Scikit-learn: Machine learning in {P}ython.
\newblock {\em Journal of Machine Learning Research}, 12:2825--2830, 2011.

\bibitem{rdkit}
{RDK}it: Open-source cheminformatics.
\newblock \url{http://www.rdkit.org}.

\bibitem{morgan1965generation}
Harry~L Morgan.
\newblock The generation of a unique machine description for chemical
  structures-a technique developed at chemical abstracts service.
\newblock {\em Journal of Chemical Documentation}, 5(2):107--113, 1965.

\end{thebibliography}



\appendix
\setcounter{figure}{0}
\makeatletter 
\renewcommand{\thefigure}{S\@arabic\c@figure}
\makeatother

\clearpage
\section{Methods}\label{sec:methods}

\subsection{Model description} \label{sec:model_description}

We frame the problem of pairwise drug synergy prediction as a regression task $(\{a, b\}, \hat{\textnormal{s}})$: given a pair of drugs $a, b$, we aim to predict their (pooled) level of synergy, $\hat{\textnormal{s}}$. Our proposed architecture is an end-to-end framework trained with a \textit{mean square error} (MSE) criterion. 

\begin{figure}[ht]
    \centering
    \includegraphics[width=\columnwidth]{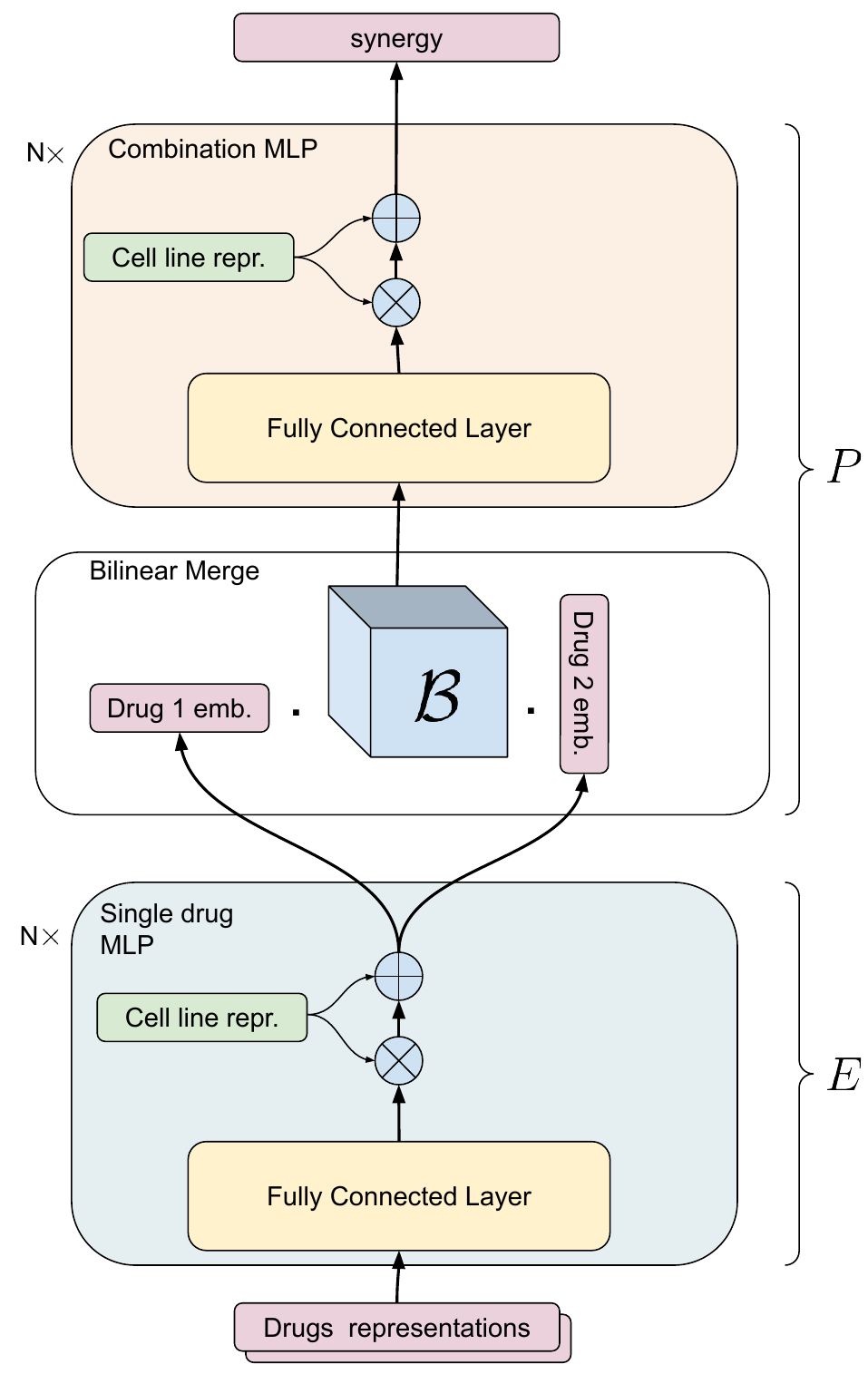}
    \caption{Overview of the {\Recover} model. Drug representations are fed into the \textit{Single drug module} which is composed of an MLP which can be conditioned on cell line. Given two drug embeddings, the synergy is predicted using the \textit{Combination module}, composed of a bilinear operation followed by an MLP. The Combination MLP can be conditioned on cell line as well.}
    \label{fig:model_overview}
\end{figure}

Our model can be decomposed into two modules. First, a \textit{single drug} module, $E$, which produces representations (or embeddings) for the drugs based on their chemical structure information. The embeddings from a pair of drugs are used as input to the \textit{combination} module $P$, which directly estimates the synergy score; see Figure \ref{fig:model_overview}. 

Further, uncertainty estimation methods are used in order to estimate the \textit{predictive distribution} of synergies $p( \hat{\textnormal{s}} | \{a, b\})$ for each drug pair $\{a, b\}$, as opposed to a point estimate. The predictive distributions of drug pairs are given as input to an acquisition function in order to decide which combinations should be tested \textit{in vitro}, balancing between combinations that are informative, \textit{i.e.} that can reduce the generalization error of the model later on, and combinations that are likely to be synergistic. 


\subsubsection{Single drug module}

Let $X_\mathcal{D}\in \mathbb{R}^{n_\mathcal{D}\times l_\mathcal{D}}$ denote the matrix of drug features, where $n_\mathcal{D}$ is the number of drugs in $\mathcal{D}$ and $l_\mathcal{D}$ corresponds to the number of raw features that describe each drug. Drug features used in this work include molecular fingerprints \cite{cereto2015molecular} and one-hot encoding of the drugs.


The single drug module can be written as a function $E: \mathcal{D} \rightarrow 
\mathbb{R}^{k_\mathcal{D}}$
where $k_\mathcal{D}$ corresponds to the dimension of the output vector representation (or embedding) of each drug. Our single drug module is a simple multi-layer perceptron (MLP) that takes raw features of drugs as input and outputs an updated vector representation. This MLP can be conditioned on cell line as described in Appendix \ref{sec:cell_line_conditioning}.


\subsubsection{Combination module}

Given a set of drugs $\mathcal{D}$, the combination module corresponds to a function $P: \binom{\mathcal{D}}{2} \mapsto \mathbb{R}$ that maps a pair of drugs to their Bliss synergy score. 
We remark first that $P$ should be agnostic to the order of the two drugs. Hence, the first operation of $P$ correspond to a permutation invariant function -- such as element-wise sum, mean, or max operations -- applied to the two vector representations corresponding to each drug. In this work, we use a bilinear operation defined by a tensor $B \in \mathbb{R}^{k_\mathcal{D}\times k_\mathcal{D} \times k}$, where $k$ is a hyperparameter corresponding to the dimension of the vector representation of a drug combination. To ensure permutation invariance, we enforce that every slice across the third dimension (denoted as $B_i$), is a symmetric matrix. Note that we do not enforce $B_i$ to be positive definite, hence $B_i$ does not necessarily define a scalar product.
The output of this permutation invariant function is fed to an MLP that outputs the predicted synergy for the pair of drugs. Again, the MLP can be conditioned on cell lines, see Appendix \ref{sec:cell_line_conditioning}.

\subsubsection{Cell line conditioning}\label{sec:cell_line_conditioning}

As a drug effect is context dependent, the synergy of a combination of two drugs can be different in experiments using different cell lines. To account for the cell line in our model we condition upon it using FiLM \cite{perez2018film}. In essence, the FiLM approach learns an affine transformation of the activation of each neuron in the MLP.

We denote the matrix of cell line features by $X_\mathcal{C} \in \mathbb{R}^{n_\mathcal{C}\times l_\mathcal{C}}$, with $\mathcal{C}$ the set of cell lines, $n_\mathcal{C}$ corresponding to the number of cell lines in $\mathcal{C}$ and $l_\mathcal{C}$ giving the number of raw features for each cell line. 

The feature representation of the cell line is either based on a one-hot encoding, or on information about mutations and basal level of gene expression. The former approach relies on having data for each cell line in the training set and cannot generalise to new cell lines. The second approach makes use of features that represent cell lines as described in Appendix ~\ref{sec:dataset_processing}.

\subsection{Searching the space of drug combinations} \label{sec:searching_space}



\subsubsection{Uncertainty estimation}
\label{subsec:uncert}

Estimating the uncertainty of the predictions is a key step towards providing reliable recommendations as well as driving the exploration with SMO. For this purpose, we use a common uncertainty estimation method: deep ensembles \cite{lakshminarayanan2017simple}. Given an ensemble of models which differ only in the initialization of the parameters, the predictions of the different models are considered as samples from the \textit{predictive} distribution. In this work, we define uncertainty as the standard deviation of the \textit{predictive} distribution, and can be estimated from the standard deviation between the predictions of the different members of the ensemble. Unless specified otherwise, we use a deep ensemble of size 5 as the uncertainty estimation method in our \textit{in silico} experiments, and of size 36 for recommendation generation, as described in Appendix \ref{sec:recommendation_generation}.

Note that for completeness, we investigated other methods for uncertainty quantification in some of the \textit{in silico} experiments, including direct estimation of the standard deviation of the predictive distribution --- in a similar fashion to Direct Epistemic Uncertainty Prediction (DEUP) \cite{DBLP:journals/corr/abs-2102-08501}, see Appendix \ref{sec:deup} for more details.

%

\subsubsection{Sequential model optimization}
\label{sec:smo}

Sequential model optimization (SMO) aims at discovering an input $x^\star \in \mathcal{X}$ maximizing an objective function $S$:
\begin{equation}
\label{eq:smo_objective}
    x^\star \in \argmax\limits_{x \in \mathcal{X}} S(x).
\end{equation}

The SMO approach consists in tackling this problem by iteratively querying the objective function $S$ in order to find a maximizer $x^\star$ in a minimal number of steps. At each step $t$, the dataset is augmented such that $\mathcal{D}_t$ contains all the inputs that have already been acquired at time $t$. The dataset $\mathcal{D}_t$ is then used to find the next query $x^{(t+1)}$. In the context of drug combinations, $x$ corresponds to a pair of drugs, and the objective function $S$ corresponds to the synergy score.

SMO has been prospectively applied to: optimize the production of proteins in cell free systems \citep{Borkowski_Koch_Zettor_Pandi_Batista_Soudier_Faulon_2020}; determine gene functions in yeast \citep{King_Whelan_Jones_Reiser_Bryant_Muggleton_Kell_Oliver_2004}; enhance the production of fine chemicals in \textit{Escherichia coli} \citep{Carbonell_Jervis_Robinson_Yan_Dunstan_Swainston_Vinaixa_Hollywood_Currin_Rattray_2018}; and to identify inhibitors of \textit{Mycobacterium tuberculosis} growth \cite{hie2020leveraging}. 

In what follows, $f$ refers to an estimator of the objective function $S$. One may notice that several properties of the potential queries $x^{(t)}$ should be taken into account. One would like to find an $x^{(t)}$ that would be informative to acquire (i.e., the uncertainty at $x^{(t)}$ is high) in order to obtain a reliable estimator of the objective function early on. On the other hand, one would like to find an $x^{(t)}$ that is a \textit{good guess} in the sense that $f(x^{(t)})$ is close to the expected maximum $\max_{x \in \mathcal{X}} f(x)$. Looking for queries which are informative is referred to as \textit{exploration} while looking for queries which are expected to maximize the objective function is called \textit{exploitation}. 

The key challenge of SMO is to balance between exploration and exploitation. This is typically achieved by designing an acquisition function (or strategy) $\alpha$ which defines a score on the space of inputs $\mathcal{X}$ and takes into account both the expected $f(x)$ and an estimate of the uncertainty at $x$. The input which maximizes the score $\alpha$ is chosen as the next query. An overview of the SMO approach is presented in Algorithm \ref{alg:smo}.

\begin{algorithm}
   \caption{Sequential model optimization}
   \label{alg:smo}
\begin{algorithmic}
  \STATE {\bfseries Input:} Initial data $\mathcal{D}_0$, objective function estimator $f$
  \FOR{$t \in \{1, 2, ...\}$}
  \STATE Select new $x^{(t + 1)}$ by optimizing an acquisition function $\alpha$
  \begin{equation*}
      x^{(t+1)} = \argmax\limits_x \alpha(x; f)
  \end{equation*}
  \STATE Query objective function $S$ to obtain $y^{t+1}$
  \STATE Augment data $\mathcal{D}_{n+1} = \mathcal{D}_n \cup \{(x^{t + 1}, y^{t+1})\}$
  \STATE Update estimator $f$
  \ENDFOR
\end{algorithmic}
\end{algorithm}

In what follows, we assume that we have access to an estimate of the mean of the predictive distribution, $\hat{\mu}(x)$, as well as an estimate of the uncertainty $\hat{\sigma}(x)$. The key acquisition functions considered are detailed below.

\paragraph{Brute-force}

$\alpha(x)$ corresponds to random noise, and therefore the drug combinations are selected at random.

\paragraph{Greedy}

$\alpha(x) = \hat{\mu}(x)$. This acquisition function corresponds to pure exploitation whereby we select drug combinations with the highest predicted synergy.

\paragraph{Pure exploration}

$\alpha(x) = \hat{\sigma}(x)$. This acquisition function corresponds to pure exploration. The strategy aims at labelling the most informative examples in order to reduce model uncertainty as fast as possible, and corresponds to the traditional strategy in \textit{Active Learning}.

\paragraph{Upper confidence bound (UCB)}%

$\alpha(x) = \mu(x) + \kappa \hat{\sigma}(x)$. This strategy balances between exploration and exploitation. $\kappa \in \mathbb{R}$ is a hyperparameter that is typically positive. Higher values of $\kappa$ give more importance to exploration.
\\

Unless specified otherwise, \textit{in silico} experiments involving SMO were performed using UCB with $\kappa = 1$.

In our synthetic SMO experiments, the model was reinitialized and trained from scratch on all visible data after each query. Whilst not designed for optimal computational efficiency, this procedure ensures that the model is not overfitting on examples that have been acquired early on.

\subsection{Recommendation generation}\label{sec:recommendation_generation}

In order to generate the recommendations for \textit{in vitro} experiments, we trained 3 models using 3 different seeds on the NCI-ALMANAC study, restricting ourselves to samples from the MCF7 cell line. We refer to these 3 models as pretrained. 

Afterwards, we \textit{fine-tune} using prospectively generated data. More precisely, the weights of one of the pretrained models were loaded, and some additional training was performed on prospectively generated data only, using early stopping. This fine-tuning process was repeated with 12 different seeds for each pretrained model. The end result being that we obtain an ensemble of 36 fine-tuned models in total.

This ensemble was used to generate predictions $(\hat{\mu}, \hat{\sigma})$ for all candidate combinations. We then use UCB with $\kappa =1$ to obtain a score according to which all candidates were ranked. 

\subsection{Dataset processing}\label{sec:dataset_processing}

Datasets have been pre-processed and normalized in a centralised data repository RESERVOIR\footnote{see \url{http:\\\\github.com\\RECOVERcoalition\\RESERVOIR}}. The repository unifies data around biologically relevant molecules and their interactions. Pre-processing and normalizing scripts are provided for traceability, and a Python API has been made available to facilitate access. Below, the major data types included are briefly described.

\subsubsection{Drugs}

Data on drugs and biologically active compounds has been extracted from Chembl \citep{davies2015chembl}, pre-processed and indexed with unique identifiers. A translation engine has been provided such that a compound can be translated to a unique identifier using generic or brand drug names, SMILES strings and Pubchem \citep{kim2019pubchem} CIDs. 

\subsubsection{Cell Line Features}

Additionally, RESERVOIR retrieved cell line features from the Cancer Dependency Map \citep{ghandiNextgenerationCharacterizationCancer2019}. These include genetic mutations, base level gene expression and metadata.

\subsubsection{Drug combinations}

Literature drug combination data was extracted from DrugComb version 1.5 \citep{zagidullin2019drugcomb}.
Quality control was applied on the experiments in DrugComb. Only blocks (i.e. combination matrices) complying to the following criteria were selected: (a.) filter out erroneous blocks that show very low variance, specifically inhibition standard deviation $\leq$0.05, (b.) filter out small blocks less than 3 $\times$ 3 dimensions, (c.) filter out blocks with extreme inhibition values, such that 5\% $<$ mean pooled growth inhibition $<$95\%.

The dataset used for model pretraining and \textit{in silico} experiments consists of 4463 data points relative to experiments on MCF7 cell line expressed as max Bliss which were reported in the Almanac study. These data correspond to 4271 unique drug combinations made up by 95 unique drugs.

The prediction set for experiment selection was built by taking 54 out of the 95 Almanac drugs for which a mode of action (MoA) was annotated in ChEMBL 25 \citep{davies2015chembl}. An additional 54 drugs were obtained by clustering 719 drugs with known MoA that are included in DrugComb but are not part of Almanac. Clustering was performed with the $k$-medoids algorithm as implemented in scikit-learn 0.24.2 \cite{scikit-learn} (n\_clusters=54, metric=Tanimoto similarity, init=k-medoids++), drugs were encoded by Morgan fingerprints with radius 2 and 1024 bits calculated with RDKit \citep{rdkit}. A representative compound for each cluster was obtained by taking the cluster centroid. 

Three of the centroid drugs were replaced due to lack of availability from commercial vendors or due to poor reported solubility. Replacements for each of the three drugs were selected by taking the nearest analogue (evaluated by Tanimoto similarity) in the same cluster. 54 Almanac and 54 non-Almanac compounds thus selected were used to build a set of 2916 binary combinations made up by one Almanac and one non-Almanac compound.

\vspace{-2mm}
\subsection{Experimental protocol}\label{sec:experimental_protocol}

\subsubsection{Cell lines}

MCF-7 cells were obtained from ATCC and maintained in DMEM (Corning) supplemented with 10\% FBS (Corning) and Antibiotic-Antimycotic (Gibco) at 37°C in 5\% $\text{CO}_2$ in a humidified incubator. Before the screens, the cell lines were passaged twice after thawing. Cultures were confirmed to be free of mycoplasma infection using the MycoAlert Mycoplasma Detection Kit (Lonza).

\vspace{-2mm}
\subsubsection{Drug Combination Assays}

A list of the tested compounds along with their concentration ranges and target mechanisms is given as a supplementary table. The compound-specific concentration ranges were selected based on their published activities. All compounds were pre-diluted in DMSO to a stock concentration that varied from 10 to 50 mM, depending on the final concentration range required for each compound. 

Briefly, compounds were plated as a 6 $\times$ 6 dose-response combination matrix in natural 384 well plates (Greiner), in a serial 1:3 dilutions of each agent (5 concentrations) and only DMSO as the lowest concentration. We used a combination plate layout where six compound pairs could be accommodated on one 384 well plate. A set of control wells with DMSO was included on all plates as negative control. To ensure reproducibility and comparability with the subsequent combination studies, the IC50 of Doxorubicin was used as reference in a 6-point dose response format in each plate as positive (total killing) control. \Review{In addition, alfacalcidol and erlotinib were evaluated in multiple rounds (and excluded from our
analysis) to ensure consistency in max Bliss synergy scores.}

Cells were seeded in white 384-well plates (Greiner) at 1000 cells/well in 50 $\mu$L of media using a multidrop dispenser and allowed to attach for 2 h. Compounds from pre-plated matrix plates were transferred to each well using a 100 nL head affixed to an Agilent Bravo automated liquid handling platform, and plates were incubated at 37°C in 5\% CO2 for an additional 72 hours. To measure the cell viability, CellTiter-Glo reagent (diluted 1:6 in water, Promega) was dispensed into the wells (30 $\mu$L), incubated for 3 minutes, and luminescence was read on a Envision plate reader (Perkin-Elmer). Final DMSO concentration in assay wells was 0.2\%. The assay was performed with 3 biological replicates.

\newpage
\onecolumn

\begin{center}
{\Huge \sc Supplementary Figures }
\end{center}

\hspace{1cm}

\begin{figure}[ht]
\centering
\begin{minipage}[t]{.27\textwidth}
\vspace{0pt}
\begin{overpic}[width=\textwidth]{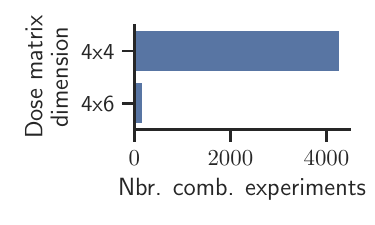}
\put(5,65){A}
\end{overpic}
\\
\begin{overpic}[width=\textwidth]{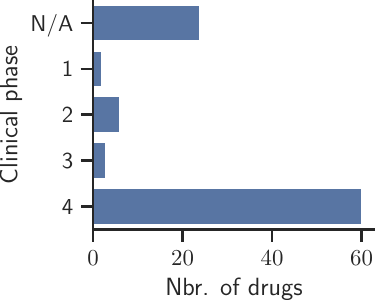}
\put(5,80){B}
\end{overpic}
\end{minipage}
\begin{minipage}[t]{.46\textwidth}
\vspace{0pt}
\begin{overpic}[width=\textwidth]{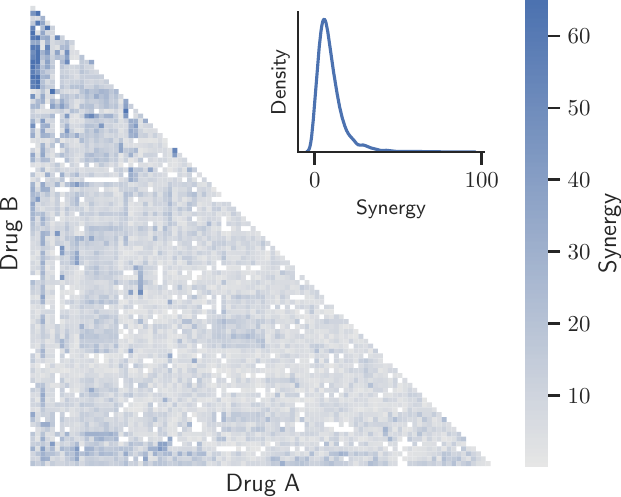}
\put(0,82){C}
\end{overpic}
\end{minipage}\hfill
\begin{minipage}[t]{.22\textwidth}
\vspace{0pt}
\begin{overpic}[width=\textwidth]{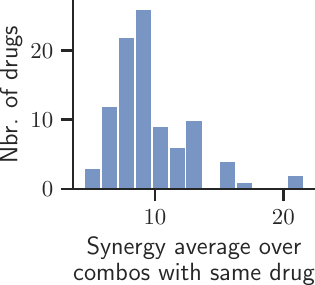}
\put(-5,93.5){D}
\end{overpic}
\\
\begin{overpic}[width=\textwidth]{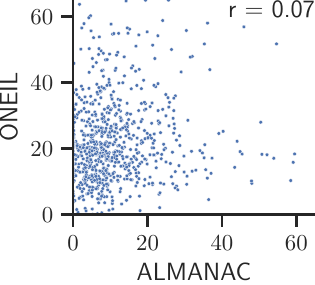}
\put(-5,90){E}
\end{overpic}
\end{minipage}
\caption{Model performance was evaluated on synergy data from the NCI-ALMANAC viability screen \cite{holbeck2017national} for the MCF7 cell line. After quality control, synergy scores could be computed for 4271 unique drug pairs. \textbf{(A.)} Distribution of dose-response matrix dimensions.
\textbf{(B.)} Distribution of clinical phases of the considered 95 distinct drugs.
\textbf{(C.)} Synergy scores for each drug-pair. White squares indicate missing or removed (low-quality) dose-response matrices. 
\textbf{(D.)} Histogram of average synergy scores for each drug computed across all drug pairs recorded. 
\textbf{(E.)} Scatter plot across 783 (drug-pair, cell line) tuples found in both NCI-ALMANAC and O'Neil 2016 \cite{o2016unbiased}.
}
\label{fig:data_summary}
\end{figure}

\begin{figure}
\begin{center}
\begin{overpic}[width=0.9\textwidth ]{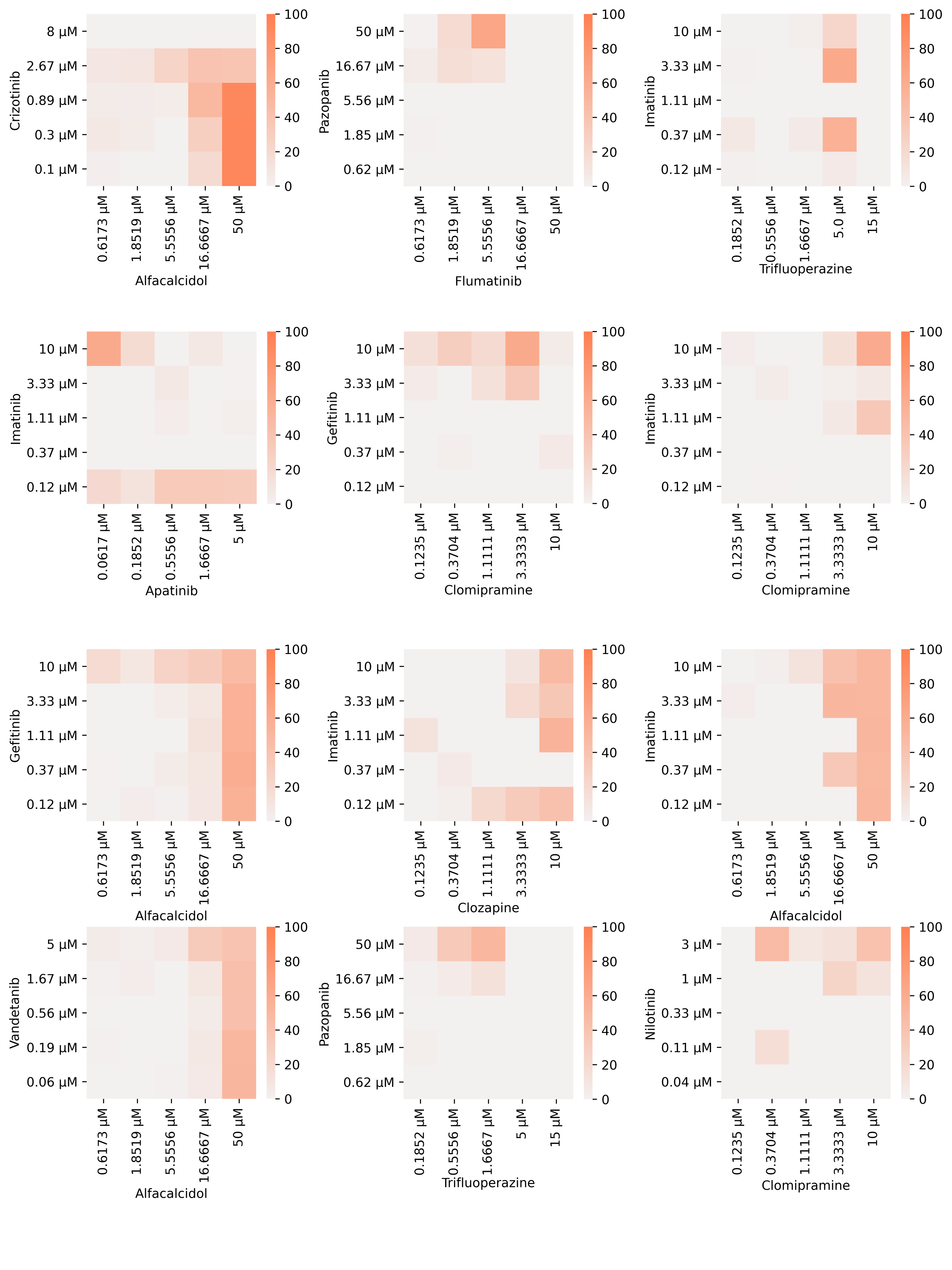}
\end{overpic}
\end{center}
\caption{Bliss synergy dose response plots the top 12 synergistic drug combinations (excluding drug combos found in Figure \ref{fig:combo_exemplars}).}
    \label{fig:remaining_top_combos}
\end{figure}

\begin{figure}
\begin{center}
\begin{overpic}[width=0.9\textwidth ]{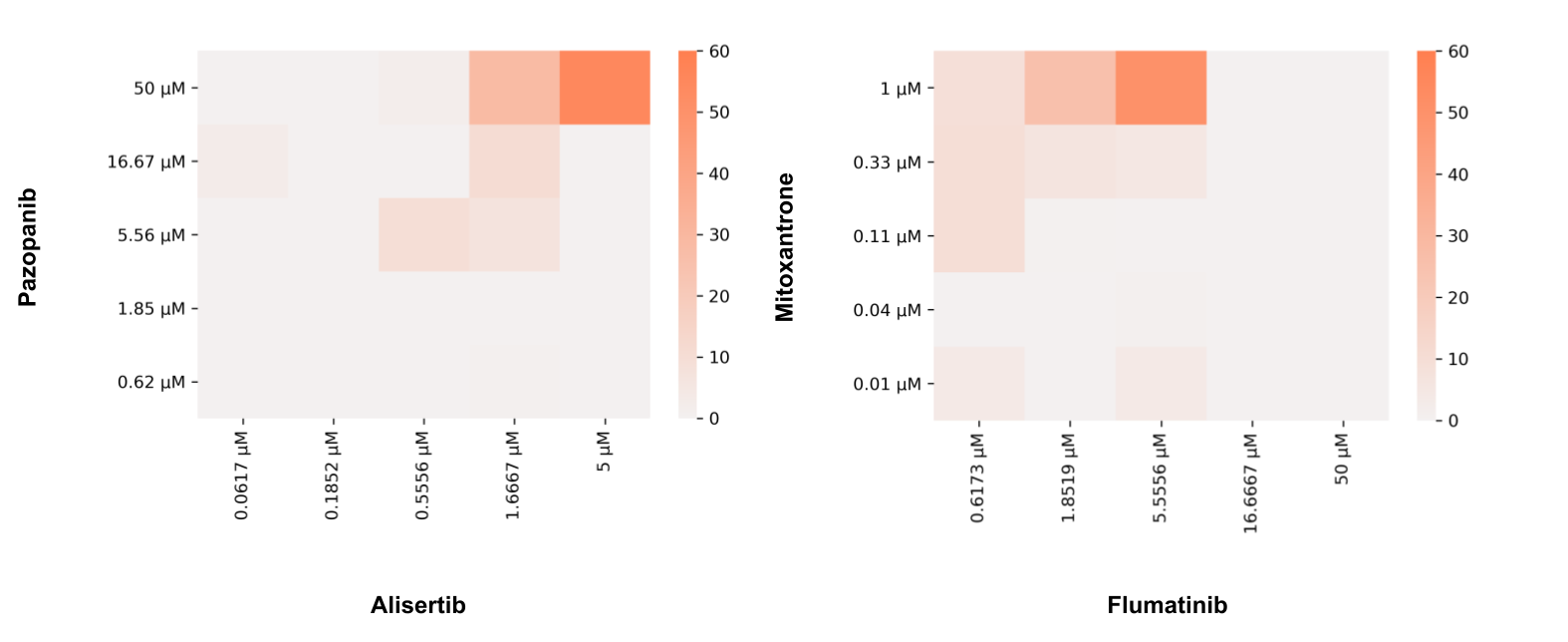}
\put(2,38){A}
\put(50.5,38){B}
\end{overpic}
\end{center}
\caption{Bliss synergy dose response plots for two drug combinations discovered within the SMO Search 1 round: \textbf{(A.)} Alisertib \& Pazopanib and \textbf{(B.)} Flumatinib \& Mitoxantrone.}
    \label{fig:combo_exemplars}
\end{figure}

\newpage

\clearpage
\begin{center}
{\Huge \sc Supplementary Information }
\end{center}
\section{Model development and evaluation}
\label{app:model_evaluation}

We investigated various aspects of the performance of {\Recover} for the prediction of Bliss synergy scores. All results presented in this section have been computed on the NCI-ALMANAC study restricted to the MCF7 cell line. Combinations are split randomly into training/validation/test (70\%/20\%/10\%). We restrict ourselves to MCF7 for consistency with prospective \textit{in vitro} experiments. 

\subsection{\Review{Benchmarking of {\Recover} on several out-of-distribution tasks}}\label{app:baseline_model_comparison}

\begin{figure*}[ht]
    \centering
    \begin{minipage}{\textwidth}
        \begin{overpic}[width=\textwidth]{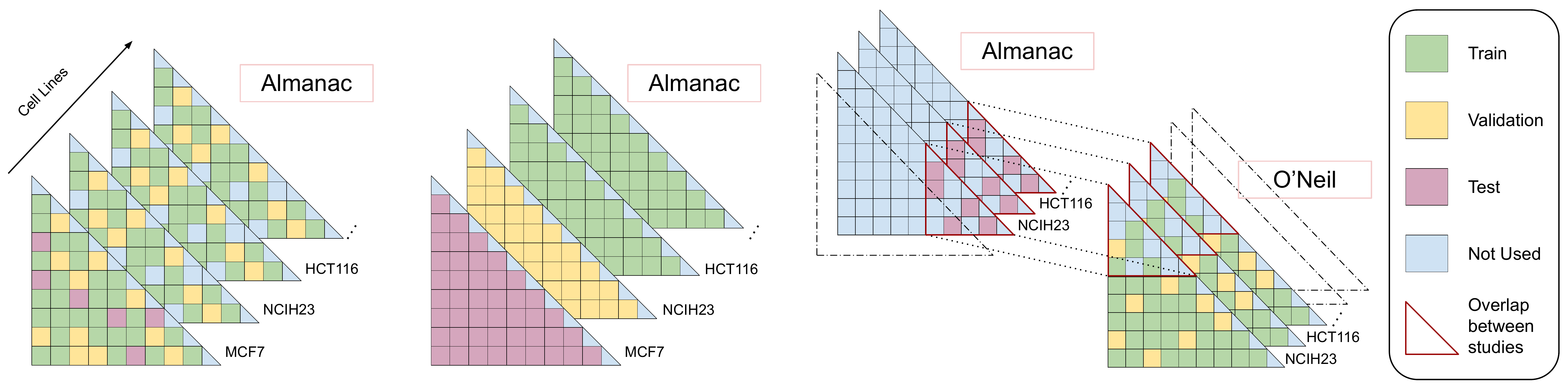}
        \put(5, -3){(iv.) Multi cell line}
        \put(30, -3){(v.) Cell line transfer}
        \put(60, -3){(vi.) Study transfer}
        \put(0, 25){A}
        \put(0, -7){B}
        \put(50, -7){C}
    \end{overpic}
    \vspace{1cm}
    \end{minipage}
    \begin{overpic}[width=0.49\textwidth]{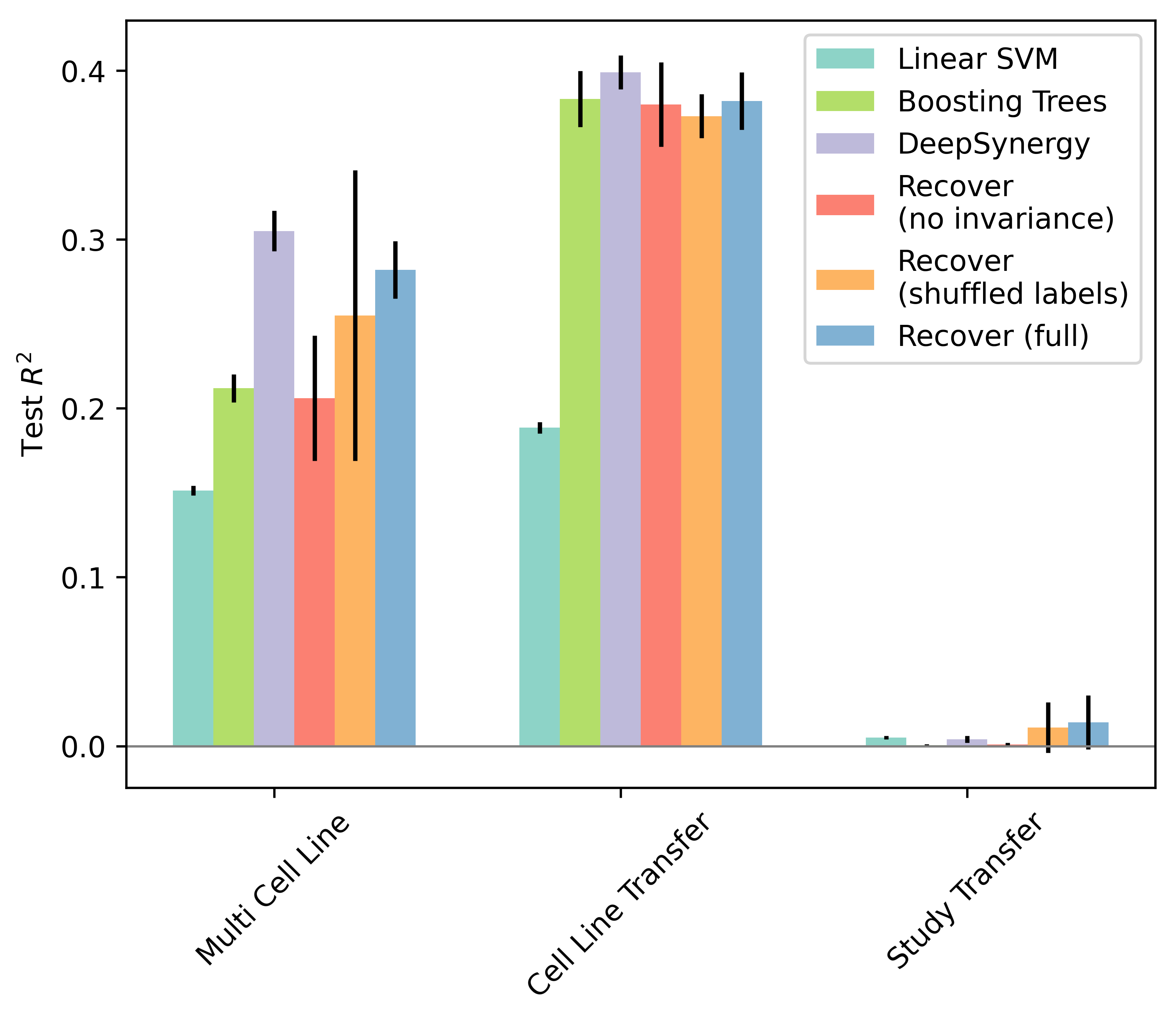}
    \end{overpic}
    \begin{overpic}[width=0.49\textwidth]{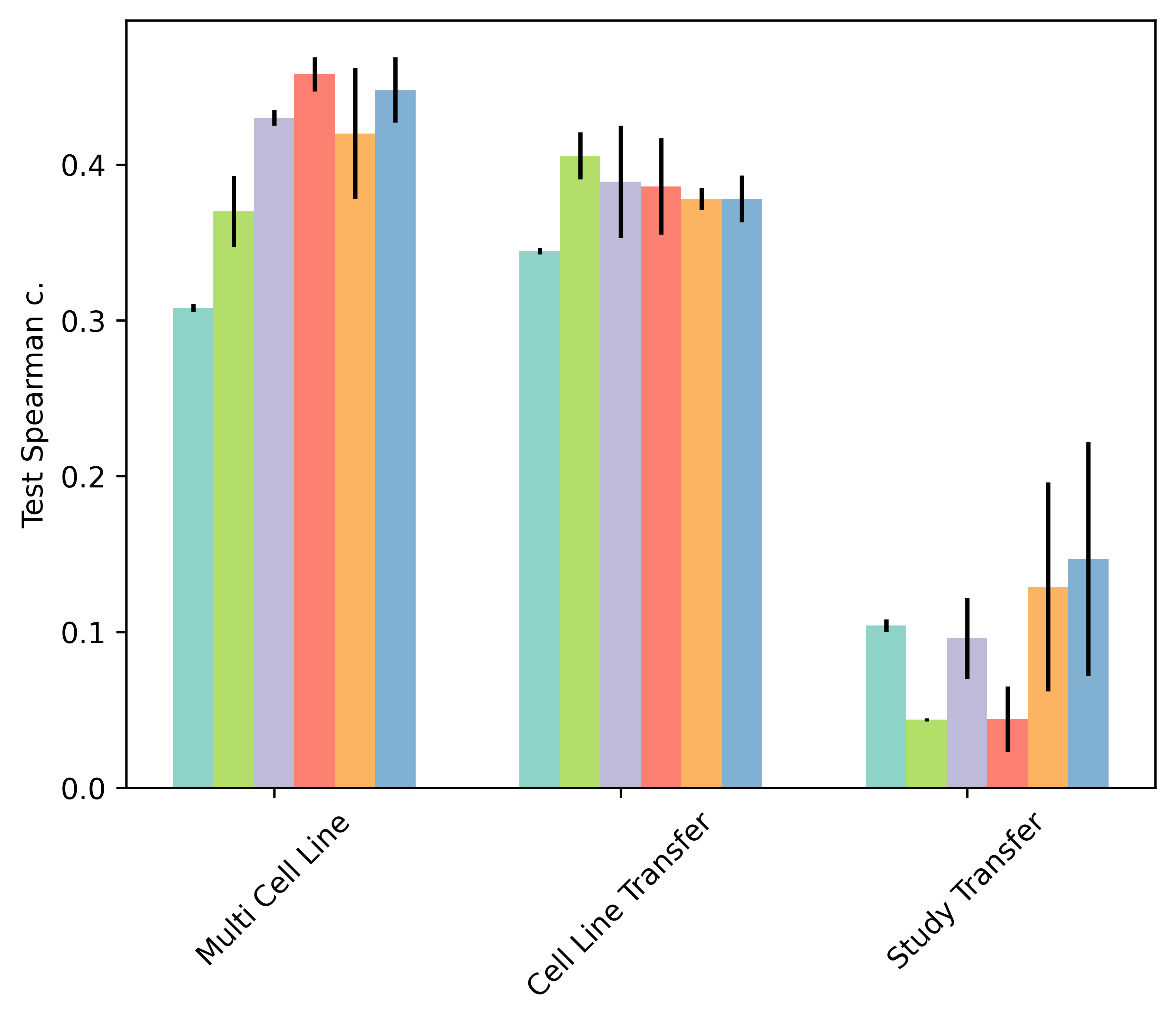}
    \end{overpic}
    \caption{\textbf{(A.)} Overview of the different tasks on which {\Recover} has been evaluated. Each task corresponds to a different way to split the training, validation and test sets, and aims at evaluating a specific generalization ability of the model. 
    For tasks (i.) to (iii.), see Figure~\ref{fig:out_of_dist_diagram}
    \textbf{(iv.)} Multi Cell Line. All cell lines are used for training and validation, but the test set is restricted to the MCF7 cell line. If a combination is part of the test set, it is never used in the training and validation sets, regardless of cell line. Training/validation split is consistent across cell lines. Proportions (70\%/20\%/10\%)
    \textbf{(v.)} Cell Line Transfer. The test set consists of all examples corresponding to the MCF7 cell line. Other cell lines are randomly assigned to training/validation (80\%/20\%). A given combination can appear in both training and test sets
    \textbf{(vi.)} Study Transfer. The O'Neil 2016 study \cite{o2016unbiased} is used to generate the training and validation sets. All overlapping cell lines are used. The test set is generated using NCI-ALMANAC and restricted to combinations for which both drugs also appear in the O'Neil study. If a combination is part of the test set, it is excluded from training and validation sets, regardless of cell line.
    \textbf{(B.) and (C.)} Performance of {\Recover} and other models for the three different tasks. Standard deviation computed over 3 seeds.}
    \label{fig:out_of_dist_diagram_appendix}
\end{figure*}

In order to understand the out-of-distribution abilities of {\Recover} as well as several other models, we evaluate a series of models on six different tasks, described in Figures~\ref{fig:out_of_dist_diagram}A and \ref{fig:out_of_dist_diagram_appendix}A. Validation and test metrics are reported in Table~\ref{tab:main_table}. Test performance can further be visualized in Figure~\ref{fig:out_of_dist_diagram}B and C, as well as Figure~\ref{fig:out_of_dist_diagram_appendix}B and C.

\begin{table*}[ht]
  \centering
  \begin{adjustbox}{width=\textwidth,center}
\begin{tabular}{lllcccccc}
\hline
& &                             & \textbf{(i.) Default} & \textbf{(ii.) One unseen drug} & \textbf{(iii.) Two unseen drugs} & \textbf{(iv.) Multi Cell Line} & \textbf{(v.) Cell Line Transfer}  & \textbf{(vi.) Study Transfer}  \\ \hline
\multirow{4}{*}{Linear SVM} 
&\multirow{2}{*}{$R^2$} 
& \textbf{Valid.}               &  $0.237 \pm 0.005$  &  $0.221 \pm 0.040$       &  $0.228 \pm 0.055$                &  $0.157 \pm 0.051$            &  $0.142 \pm 0.020$                &  $0.153 \pm 0.013$    \\ 
&& \textbf{Test}                &  $0.171 \pm 0.006$  &  $0.167 \pm 0.005$      &  $0.026 \pm 0.008$                &  $0.151 \pm 0.003$            &  $0.189 \pm 0.003$                &  $0.005 \pm 0.001$    \\ \cline{2-9}
&\multirow{2}{*}{Spearman c.} 
& \textbf{Valid.}               &  $0.493 \pm 0.014$  &  $0.440 \pm 0.024$       &  $0.455 \pm 0.045$                &  $0.371 \pm 0.011$            &  $0.373 \pm 0.007$                &  $0.421 \pm 0.009$    \\ 
&& \textbf{Test}                &  $0.453 \pm 0.004$  &  $0.333 \pm 0.002$      &  $0.106 \pm 0.015$                &  $0.308 \pm 0.003$            &  $0.344 \pm 0.002$                &  $0.104 \pm 0.004$    \\ \hline

\multirow{4}{*}{Gradient Boosting Trees} 
&\multirow{2}{*}{$R^2$} 
& \textbf{Valid.}               &  $0.276 \pm 0.022$  &  $0.320 \pm 0.084$       &  $0.273 \pm 0.065$                &  $0.294 \pm 0.056$            &  $0.310 \pm 0.014$                &  $0.311 \pm 0.006$    \\ 
&& \textbf{Test}                &  $0.172 \pm 0.002$  &  $0.184 \pm 0.011$      &  $0.024 \pm 0.002$                &  $0.212 \pm 0.008$            &  $0.383 \pm 0.017$                &  $0.001 \pm 0.001$    \\ \cline{2-9}
&\multirow{2}{*}{Spearman c.} 
& \textbf{Valid.}               &  $0.484 \pm 0.033$  &  $0.475 \pm 0.042$      &  $0.430 \pm 0.022$                &  $0.474 \pm 0.010$            &  $0.507 \pm 0.008$                &  $0.580 \pm 0.010$    \\ 
&& \textbf{Test}                &  $0.459 \pm 0.001$  &  $\mathbf{0.362 \pm 0.002}$      &  $0.137 \pm 0.015$                &  $0.370 \pm 0.023$            &  $0.406 \pm 0.015$                &  $0.044 \pm 0.001$    \\ \hline

\multirow{4}{*}{DeepSynergy} 
&\multirow{2}{*}{$R^2$} 
& \textbf{Valid.}               &  $0.329 \pm 0.034$  & $0.283 \pm 0.079$       &  $0.356 \pm 0.111$                &  $0.393 \pm 0.015$            &  $0.366 \pm 0.011$                &  $0.378 \pm 0.021$    \\ 
&& \textbf{Test}                &  $0.144 \pm 0.025$  &  $0.170 \pm 0.012$       &  $\mathbf{0.059 \pm 0.003}$       &  $\mathbf{0.305 \pm 0.012}$   &  $\mathbf{0.399 \pm 0.010}$       &  $0.004 \pm 0.002$    \\ \cline{2-9}
&\multirow{2}{*}{Spearman c.} 
& \textbf{Valid.}               &  $0.453 \pm 0.011$  &  $0.424 \pm 0.056$      &  $0.401 \pm 0.032$                &  $0.520 \pm 0.006$            &  $0.350 \pm 0.045$                &  $0.640 \pm 0.009$    \\ 
&& \textbf{Test}                &  $0.454 \pm 0.008$  &  $0.302 \pm 0.018$      &  $\mathbf{0.176 \pm 0.003}$       &  $0.430 \pm 0.005$            &  $\mathbf{0.389 \pm 0.036}$       &  $0.096 \pm 0.026$    \\ \hline

\multirow{4}{*}{\Recover~(no invariance)} 
&\multirow{2}{*}{$R^2$} 
& \textbf{Valid.}               &  $0.332 \pm 0.074$ &  $0.275 \pm 0.064$       &  $0.330 \pm 0.123$                &  $0.361 \pm 0.057$            &  $0.301 \pm 0.017$                &  $0.358 \pm 0.032$    \\ 
&& \textbf{Test}                &  $0.208 \pm 0.001$ & $\mathbf{0.195 \pm 0.008}$        &  $0.043 \pm 0.005$                &  $0.206 \pm 0.037$            &  $0.380 \pm 0.025$                &  $0.001 \pm 0.001$    \\ \cline{2-9}
&\multirow{2}{*}{Spearman c.} 
& \textbf{Valid.}               &  $0.457 \pm 0.036$ &  $0.471 \pm 0.051$       &  $0.455 \pm 0.073$                &  $0.504 \pm 0.007$            &  $0.294 \pm 0.023$                &  $0.625 \pm 0.008$    \\ 
&& \textbf{Test}                &  $0.437 \pm 0.003$ &  $0.329 \pm 0.006$       &  $0.145 \pm 0.023$                &  $\mathbf{0.458 \pm 0.011}$   &  $\mathbf{0.386 \pm 0.031}$       &  $0.044 \pm 0.021$    \\ \hline

\multirow{4}{*}{\Recover~(shuffled labels)} 
&\multirow{2}{*}{$R^2$} 
& \textbf{Valid.}               &  $0.325 \pm 0.058$  & $0.298 \pm 0.086$    &  $0.384 \pm 0.161$                &  $0.391 \pm 0.030$            &  $0.271 \pm 0.028$                &  $0.320 \pm 0.020$      \\ 
&& \textbf{Test}                &  $0.237 \pm 0.014$  & $0.086 \pm 0.015$    &  $0.003 \pm 0.003$                &  $0.255 \pm 0.086$            &  $0.373 \pm 0.013$                &  $0.011 \pm 0.015$       \\ \cline{2-9}
&\multirow{2}{*}{Spearman c.} 
& \textbf{Valid.}               &  $0.460 \pm 0.009$  & $0.466 \pm 0.053$    &  $0.439 \pm 0.070$                &  $0.510 \pm 0.019$            &  $0.268 \pm 0.063$                &  $0.598 \pm 0.032$       \\ 
&& \textbf{Test}                &  $0.449 \pm 0.016$  &  $0.228 \pm 0.016$   &  $-0.050 \pm 0.04$               &  $0.420 \pm 0.042$            &  $0.378 \pm 0.007$                &  $0.129 \pm 0.067$       \\ \hline

\multirow{4}{*}{\Recover (full)} 
&\multirow{2}{*}{$R^2$} 
& \textbf{Valid.}               &  $0.343 \pm 0.053$  &  $0.275 \pm 0.067$      &   $0.401 \pm 0.147$              &  $0.387 \pm 0.032$            &  $0.278 \pm 0.027$                &  $0.304 \pm 0.021$      \\ 
&& \textbf{Test}                & $\mathbf{0.242 \pm 0.006}$ & $\mathbf{0.196 \pm 0.008}$&   $0.038 \pm 0.002$              &  $0.282 \pm 0.017$            &  $0.382 \pm 0.017$                &  $\mathbf{0.014 \pm 0.016}$      \\ \cline{2-9}
&\multirow{2}{*}{Spearman c.} 
& \textbf{Valid.}               &  $0.474 \pm 0.021$  &  $0.433 \pm 0.023$     &  $0.459 \pm 0.069$               &  $0.518 \pm 0.021$            &  $0.299 \pm 0.047$                &  $0.589 \pm 0.032$      \\ 
&& \textbf{Test}                & $\mathbf{0.466 \pm 0.007}$ & $0.326 \pm 0.025$&  $0.157 \pm 0.012$               &  $\mathbf{0.448 \pm 0.021}$   &  $\mathbf{0.378 \pm 0.015}$       &  $\mathbf{0.147 \pm 0.075}$      \\ \hline

\end{tabular}
\end{adjustbox}
\caption{Performance of {\Recover} and baselines for the different tasks, as detailed in Figure \ref{fig:out_of_dist_diagram} and Figure \ref{fig:out_of_dist_diagram_appendix} . Standard deviation computed over 3 seeds.}
\label{tab:main_table}
\end{table*}

For this evaluation, the hyperparameters of {\Recover} have been optimized (on the validation set of the default task) within the following set of values. Because it was not tractable to perform a grid search over all possible values at once, hyperparameters have been optimized one at a time in an iterative way. The set of parameters that yielded best performance is highlighted, and  used for all following experiments (both \textit{in silico} and \textit{in vitro} experiments):
\begin{itemize}
    \item Learning rate: [$1  \times 10^{-1}$, $1  \times 10^{-2}$, $1  \times 10^{-3}$, $\mathbf{1 \times 10^{-4}}$, $1 \times 10^{-5}$, $1 \times 10^{-6}$]
    \item Batch size: [16, 32, 64, \textbf{128}, 256]
    \item Weight decay: [$1$, $1  \times 10^{-1}$, $\mathbf{1  \times 10^{-2}}$, $1  \times 10^{-3}$, $1 \times 10^{-4}$, $1 \times 10^{-5}$, $1 \times 10^{-6}$, $0$]
    \item Morgan fingerprint radius: [\textbf{2}, 3, 4, 5]
    \item Morgan fingerprint dimension: [\textbf{1024}, 2048]
    \item Output dimension of the single drug MLP: [16, 32, 64, \textbf{128}, 256]
    \item Dimension(s) of the hidden layer(s) of the single drug MLP:  [[512], [256], \textbf{[1024]}, [2048], [4096], [1024, 1024], [1024, 1024, 1024], [1024, 512], [1024, 512, 256]]
    \item Dimension(s) of the hidden layer(s) of the combination MLP: [[32], \textbf{[64]}, [128], [256], [64, 16], [64, 32], [64, 64]]
\end{itemize}

Depending on the task at hand, the model configuration will differ slightly. For all tasks (excluding task (iii.)), the drug feature representations consists of the Morgan fingerprints \citep{morgan1965generation} concatenated with a \textit{one-hot} encoding specifying the identity of drug. For task (iii.), only the Morgan fingerprint is used. When several cell lines are available, the {\Recover} model is conditioned on cell lines using feature-wise linear modulation (FiLM) \cite{perez2018film}, see Appendix \ref{sec:cell_line_conditioning}. The cell line features are either a one-hot encoding of the cell line for tasks (iv.) and (vi.), or some information about mutations and basal level of mRNA gene expression for task (v.).

We will now describe a few baseline models and how their hyperparameters have been optimized. A grid search has been performed to optimize the hyperparameters of the \textit{Gradient Boosting Trees} baseline model. The number of trees was set to $100$. The set of parameters that yielded best performance is highlighted:
\begin{itemize}
    \item Maximum tree depth: [2, 5, 10, \textbf{20}]
    \item Minimum number of samples to split a node: [2, 5, 10, \textbf{20}, 50]
    \item Learning rate: [0.0001, 0.001, 0.01, \textbf{0.1}, 1]
    \item Maximum feature: [all, \textbf{$\sqrt{\text{total number of features}}$},  $\log_2[\text{total number of features}]$]
\end{itemize}
Similarly, a grid search has been performed for the  \textit{Linear SVM} baseline model:
\begin{itemize}
    \item Tolerance for stopping criterion: [$1  \times 10^{-1}$, $1  \times 10^{-2}$, $\mathbf{1  \times 10^{-3}}$, $1 \times 10^{-4}$, $1 \times 10^{-5}$]
    \item Regularization C: [0.0001, 0.001, 0.01, \textbf{0.1}, 1., 10]
\end{itemize}

Moreover, we compare against DeepSynergy \cite{preuer2018deepsynergy}, which is a deep learning based approach. We replicated the original model, more precisely:
\begin{itemize}
    \item  Cell line features and drug features (described in Subsection~\ref{sec:dataset_processing}) were given as input (for all tasks). 
    \item Normalization: Input features' mean and standard deviation were set to $(0, 1)$, followed by a $\tanh$ normalization. 
    \item The architecture from the original paper was used (two hidden layers of dimension $8182$ and $4096$), with input dropout ($p=0.2$) and layer dropout ($p=0.5$).
    \item The final prediction is the average of two predictions made from $(d_1, d_2)$ and $(d_2, d_1)$ where $d_1$ and $d_2$ are the two drugs in the combination.
\end{itemize}

Finally, we evaluate two variants of {\Recover}. In {\Recover} (no invariance), the two drug embeddings (outputs of the single drug MLP) are concatenated and directly fed into the combination MLP, instead of first being fed into the invariance module. In {\Recover} (shuffled labels), prior to training, drug features are randomly permuted such that each drug gets represented by the features of another drug. A similar procedure is applied to cell line features when they are used, \textit{c.f.} tasks (iv.) to (vi.).

We will now briefly comment the results of the benchmarking study. {\Recover} outperforms baseline models in terms of $R^2$ and Spearman correlation metrics on the Default task (i.). {\Recover} (shuffled labels) performs well compared to other models on the default task, multi cell line task, cell line transfer task and study transfer task. In these cases, the information contained in drug fingerprints and cell line features only provides a limited gain in performance, thus merely knowing the identity of the drugs is sufficient. We further confirmed in Appendix~\ref{app:feature_importance} that drug structure information only provided a minimal increase in performance on the default task.

In task (iii.) we note a considerable drop in performance when compared to task (i.) for all models alike, demonstrating that {\Recover} will have markedly reduced performance when attempting to predict the synergy of drug combinations in which both drugs have not been observed in earlier experiments. The results pertaining to tasks (iv.) and (v.) demonstrate that leveraging experiments from other cell lines does provide a benefit when compared to the performance from task (i.), although the effect is most significant when the specific drug combination in question has been seen in other cell lines, i.e., task (v.). For completion, we confirm the significant batch effects between the NCI-ALMANAC and the O'Neil 2016 studies render using the same model parameters for both studies impossible --- notice task (vi.) performing at the level of randomness and Figure \ref{fig:data_summary}E.

\subsection{Feature importance study}
\label{app:feature_importance}

Through investigation of different drug features, we find a large proportion of the performance of {\Recover} can be achieved given the identity of the drugs alone, and that structural information allows for a slight increase in performance. As shown in Table \ref{tab:comparison_input_features}, the performance of the model is similar whether the one hot encoding of the drug or its Morgan fingerprint is used as input. We notice a slight improvement when using both feature types together. Note that the number of parameters of the model is always the same regardless of the type of feature provided as input. When a feature type is not used, the corresponding part of the drug feature vector is set to zero without changing the underlying dimension.

\begin{table}[H]
\centering
\begin{tabular}{lcc}
\hline
                                                & \textbf{$R^2$}                & \textbf{Spearman corr.}     \\ \hline
{\Recover} (fingerprint + one hot)         &  $\mathbf{0.242 \pm 0.006}$             &  $\mathbf{0.466 \pm 0.007}$           \\
{\Recover} (fingerprint)                             &  $0.232 \pm 0.007$                      &  $0.458 \pm 0.010$                    \\
{\Recover} (one hot)                    &  $0.230 \pm 0.029$                 &  $0.449 \pm 0.017$                    \\ \hline
\end{tabular}
\caption{Feature importance study for the prediction of max Bliss score on the MCF7 cell line. Standard deviation computed over 3 seeds.}
\label{tab:comparison_input_features}
\end{table}

\subsection{Upper bounds on model performance}
\label{app:subsection_theoretical_upper_bounds}

We investigate {\Recover} performance with regards to Spearman correlation and $R^2$. Whilst predictive power appears modest, we are still able to identify highly synergistic drug combinations in simulated SMO experiments, see Appendix \ref{app:tail_dist}. Several aspects that may limit predictive power: experimental noise, and nonuniformity of maximum Bliss synergy scores. 

In Figure \ref{fig:data_summary}C, we note most data points are close to zero, with some examples very far from the mean, i.e., the examples of interest. As an example, let us consider the case of Spearman correlation. Given that the observations are noisy, the observed rank among synergies might be corrupted compared to the true ordering --- especially in the region close to zero where the density of examples is very high. 

The non-uniformity of synergy scores leads to some difficulties in evaluating fairly the performance of {\Recover}. For example, the positive tail of the distribution, which is the region of interest, represent a very small percentage of the total number of examples and thus have a little effect on the value of the aggregated statistic.

In order to get a better understanding of the performance of our model, we compare the reported aggregated statistics to an upper bound which takes into account the presence of noise in the observations in addition to the distribution of synergy scores.

\paragraph{Evaluating the level of noise}

Consider all replicates from the NCI-ALMANAC study. Two examples are considered replicates when the same pair of drugs has been tested on the same cell line. We found $1960$ triplets (\{drug1, drug2\}, cell line) that had been tested several times. For each triplet, we computed the standard deviation of the maximum Bliss score across the replicates. We refer to this as the level of noise for a given triplet. We then computed the average level of noise $\Bar{\eta}$ across all triplets.

\paragraph{Estimating upper bounds}

Given an average level of noise $\eta$ and the distribution of synergy scores in NCI-ALMANAC, we simulated a noisy acquisition process as follows: the synergies from NCI-ALMANAC were considered as the \textit{true} synergies, and noisy observations were obtained by corrupting the \textit{true} synergies with some Gaussian noise $\mathcal{N}(0, \eta^2 )$. We then considered a \textit{perfect} regression model which fits the noisy observations exactly, and evaluated its performance on the \textit{true} synergies. Upper bounds are defined by the performance of this \textit{perfect} regression model.

\begin{figure}[h]
\centering
\begin{overpic}[width=0.45\textwidth]{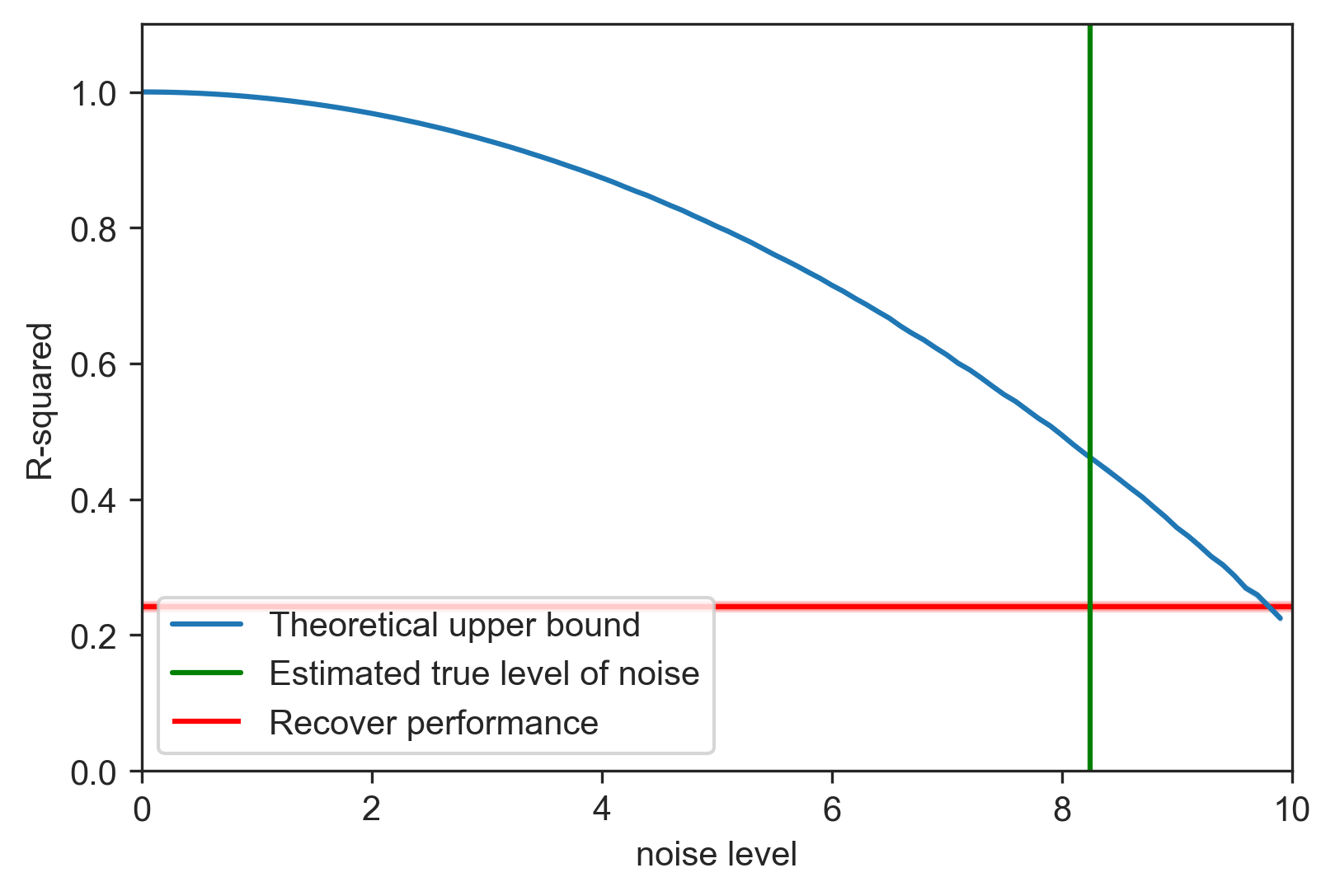}
\put(2,65){A}
\end{overpic}
\hspace{0.5cm}
\begin{overpic}[width=0.45\textwidth]{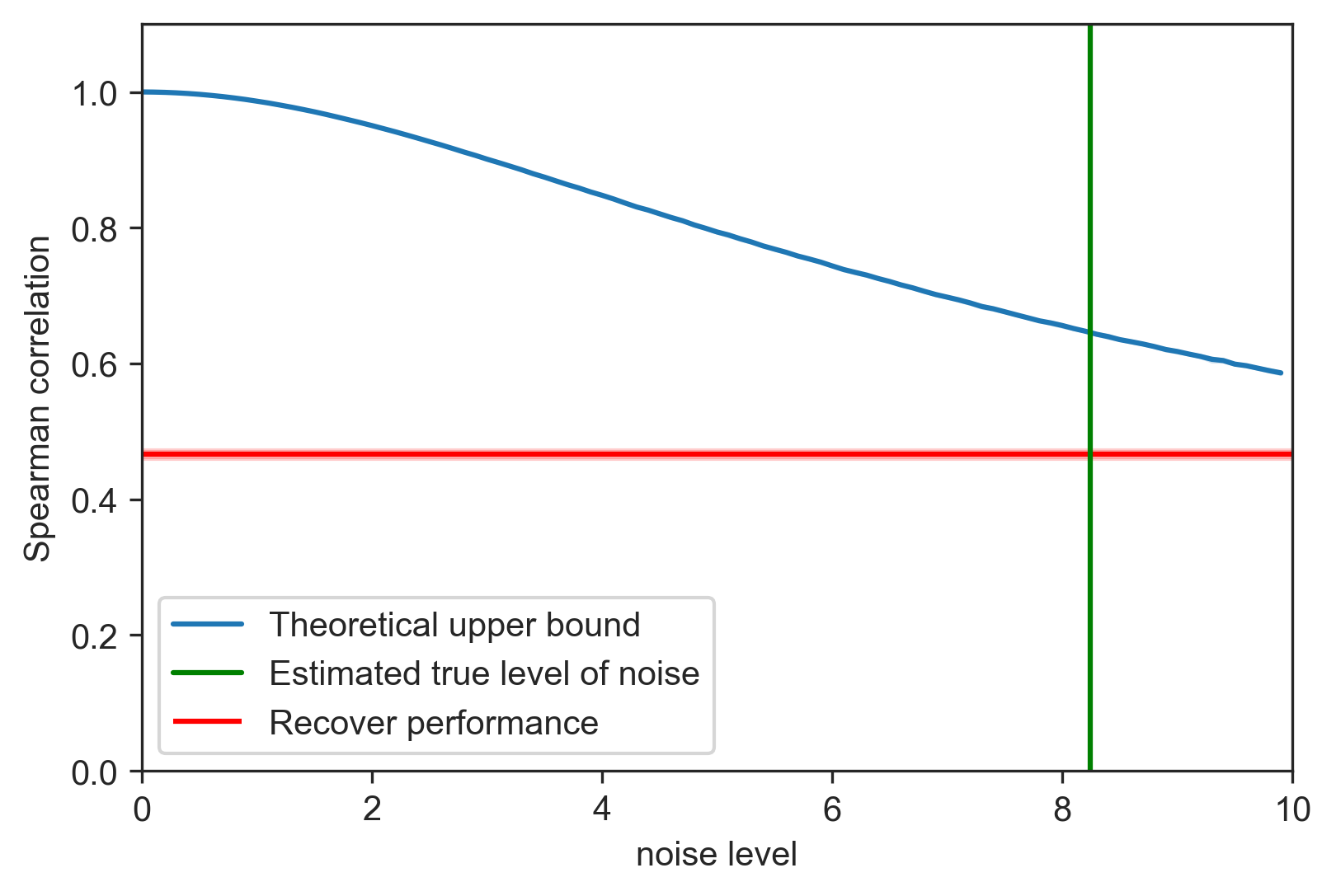}
\put(2,65){B}
\end{overpic}
\caption{Comparison of the performance of {\Recover} with upper bounds that take into account the level of noise in the data, and the non uniform distribution of synergies: \textbf{(A.)} $R^2$ value, \textbf{(B.)} Spearman correlation. The upper bound, shown in blue, is a function of the level of noise. The standard deviation for the upper bound is lower than $10^{-3}$ for both statistics (estimated over 3 seeds). The estimated level of noise in our dataset is shown in green. The performance of {\Recover} 
    as well as its standard error are shown in red.}
    \label{fig:upper_bounds}
\end{figure}

Upper bounds have been computed for $R^2$ and Spearman correlation using various levels of noise and are reported in Figure \ref{fig:upper_bounds}. We see that the noisy acquisition process alone leads to significant limitations in the performance that can be reached. While there is still room for improvement, the performance of {\Recover} is reasonably close to the hypothetical maximum. For example, {\Recover} achieves 0.47 Spearman correlation, while the highest achievable Spearman correlation is estimated to be 0.64.

\subsection{Performance analysis on the tail of the distribution of synergies}\label{app:tail_dist}

We now show that {\Recover} achieves good performance on the positive tail of the distribution of synergies, which is necessary to successfully identify highly synergistic combinations within a SMO setting.

\begin{figure}[h]
\centering
\includegraphics[width=0.45\linewidth]{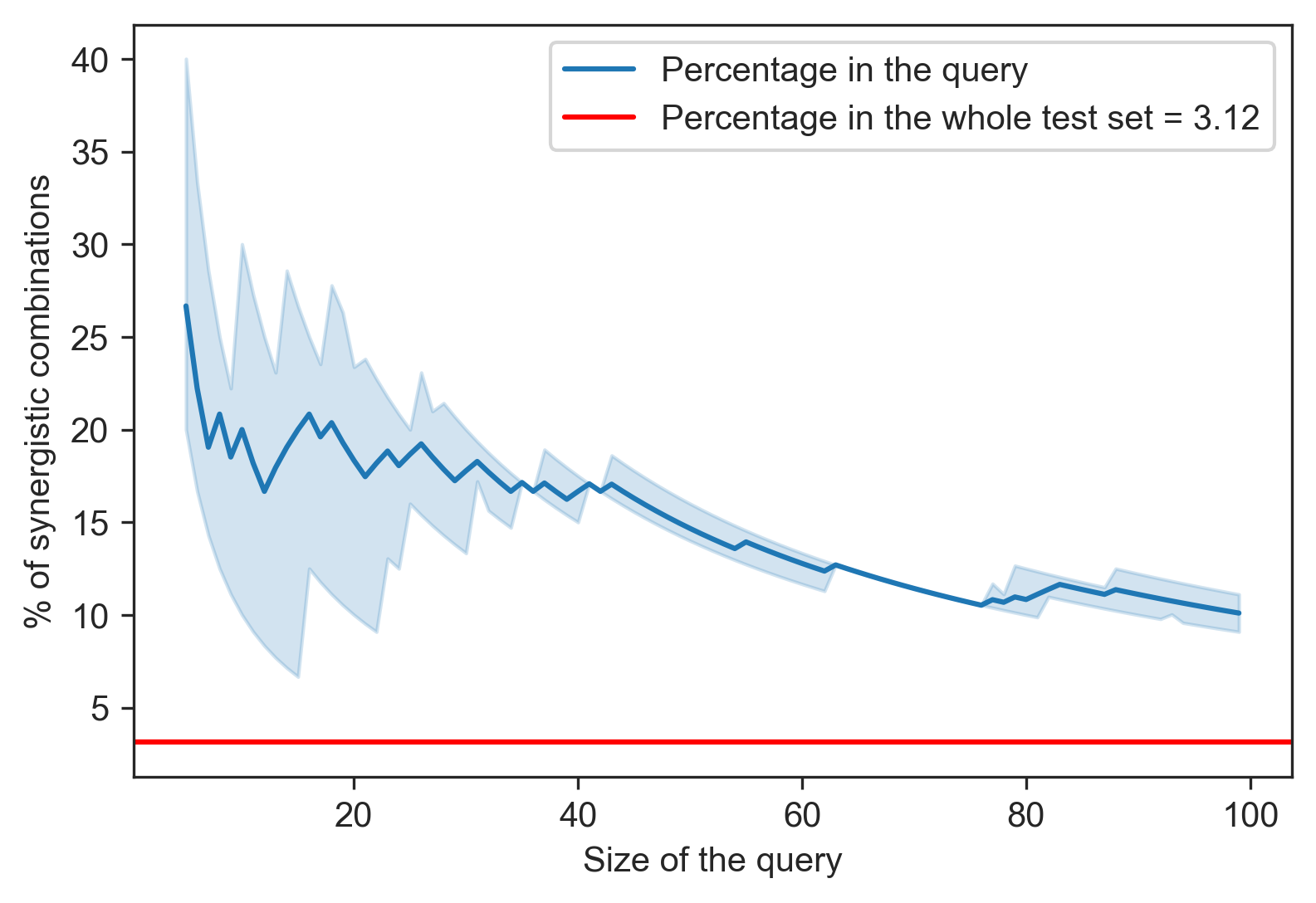}
\caption{Percentage of highly synergistic combinations (synergy score $> 30$) in queried combinations using {\Recover} as a function of the size of the query. Shaded area corresponds to the standard deviation computed over 3 seeds.}
\label{fig:detailed_eval_app}
\end{figure}

Given a model trained on NCI-ALMANAC, we query the top $k$ combinations with highest predicted synergy within the test set, and compute the percentage of combinations which are truly synergistic (arbitrarily defined by a maximum Bliss synergy score of above $30$) within queried examples. We refer to $k$ as the size of the query. In Figure \ref{fig:detailed_eval_app}, we report the percentage of synergistic combinations as a function of the size of the query. The percentage of synergistic combinations is superior to the proportion in the whole test set, meaning that the model performs far beyond the level of randomness. For instance, with a query size of $30$, we observe a $\sim$5-fold enrichment in synergistic combinations. However, this enrichment decreases with the size of the query.

\subsection{\Review{Analysis of the relationship between drug similarity, model uncertainty, and model error}}
\label{app:clustering_diagram}

\begin{figure}[H]
\centering
\begin{overpic}[width=\textwidth]{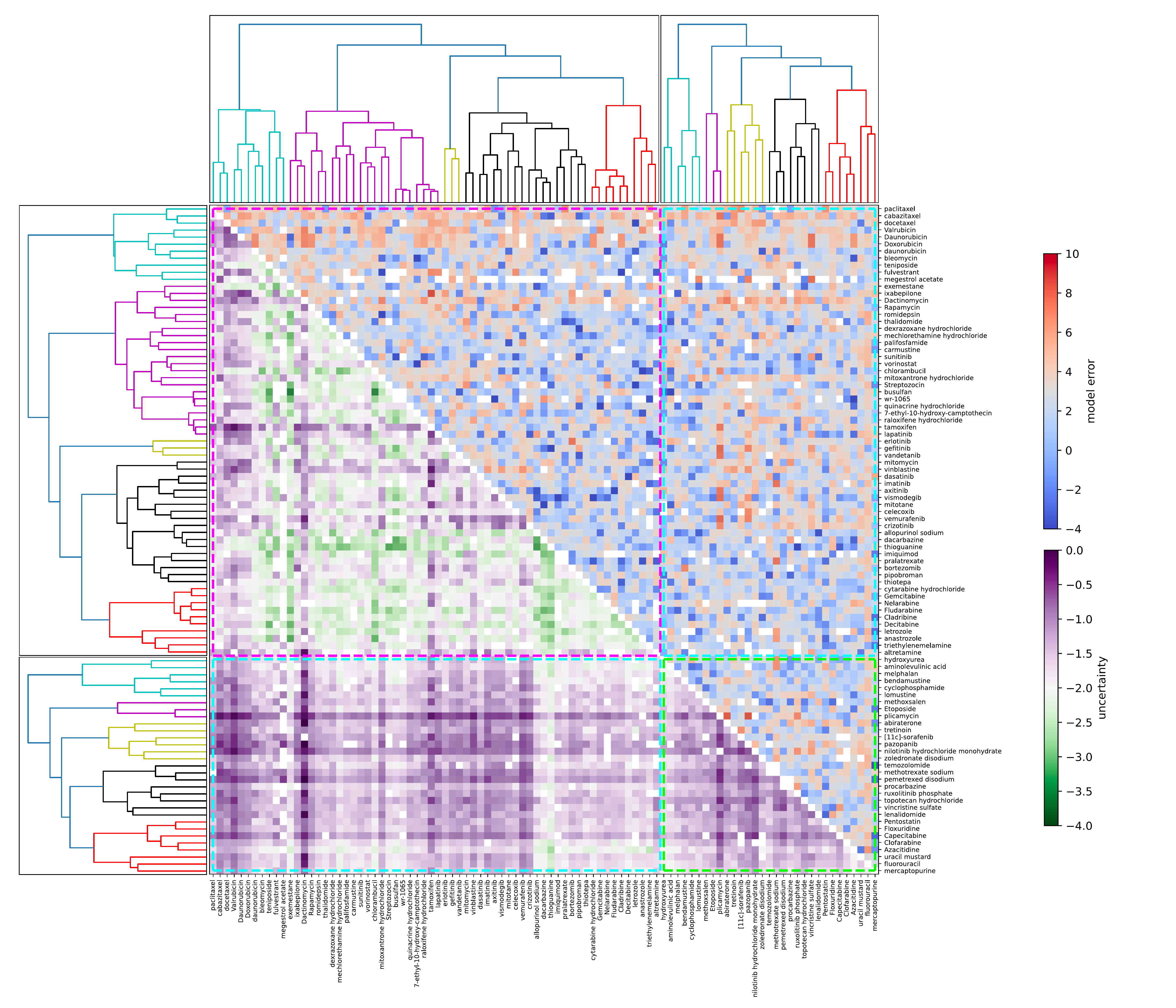}
\end{overpic}
\caption{Uncertainty of {\Recover} (lower left triangle) and mean square error (upper right triangle). The magenta square corresponds to the training-validation set. The part of the matrix outside of the magenta square corresponds to the test set, which can be subdivided into two parts: combinations where one of the drugs has been seen during training (cyan rectangles), and combinations where none of the drugs have been seen during training (light green square). The ordering of the drugs is based on a hierarchical clustering wherein distances between drugs are derived from their Tanimoto similarities. The color map is on a logarithmic scale, both for model error and for model uncertainty. White entries correspond to combinations absent from the dataset.}
\label{fig:matrix_uncertainty_model_error}
\end{figure}

\begin{figure}[H]
\centering
\begin{overpic}[width=0.49\textwidth]{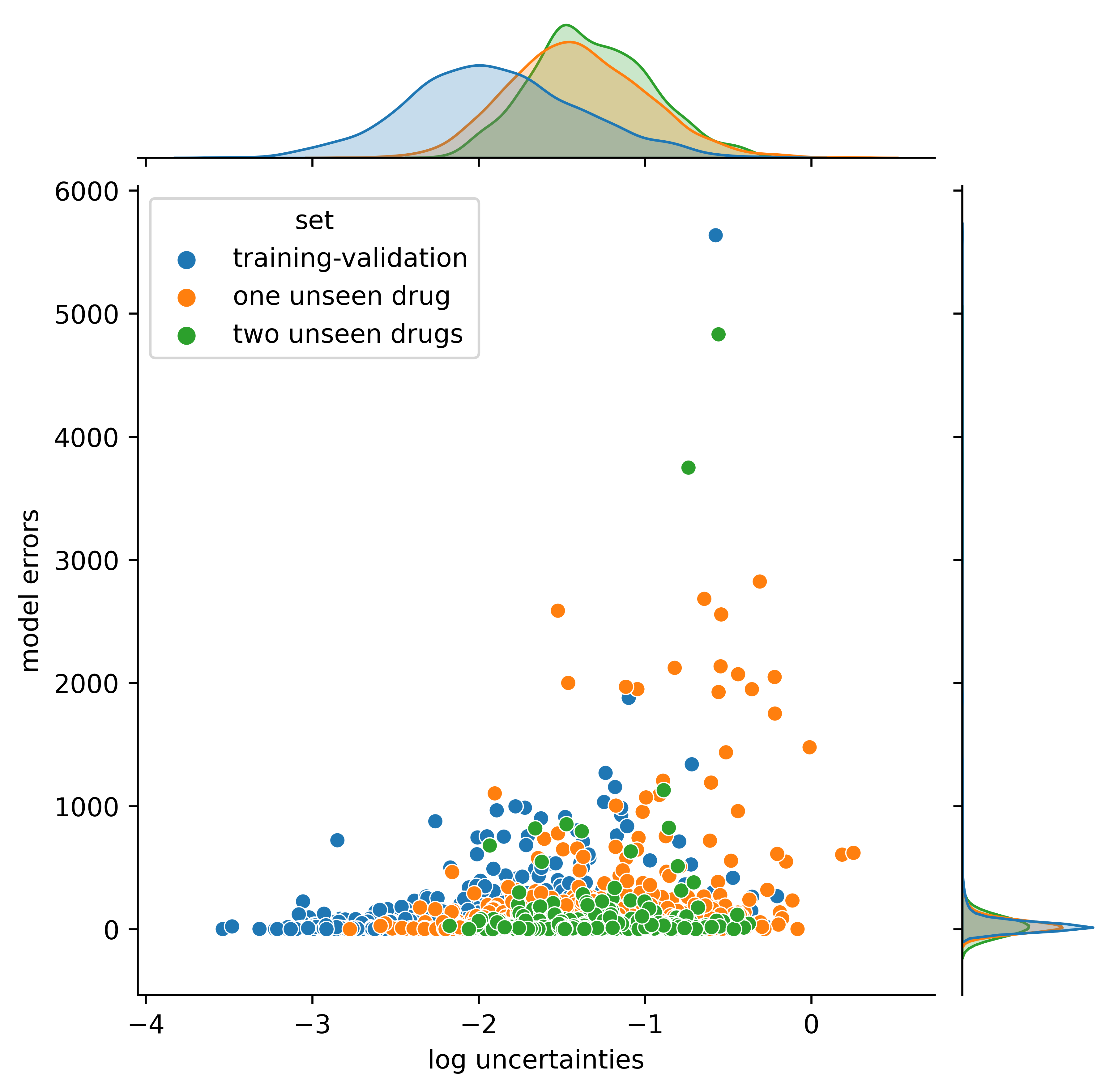}
\end{overpic}
\caption{Errors made by the {\Recover} model as a function of its log-uncertainty. Color corresponds to the dataset split to which the combination belongs.}
\label{fig:jointplot_uncert_model_error}
\end{figure}

\Review{To help build intuition for the relationship between drug similarity, model uncertainty and model accuracy, we provide some additional results using {\Recover} trained on combinations from the ALMANAC database (data pertaining to the MCF7 cell line).}

\Review{In Figure \ref{fig:matrix_uncertainty_model_error}, we report both model uncertainty (lower left triangle) and mean square error (upper right triangle). Uncertainty is estimated using an ensemble of 20 models. The ordering of the drugs is based on a hierarchical clustering wherein distances between drugs are derived from their Tanimoto similarities. Therefore, the ordering of lines and columns is based on drug structures. Clustering is performed separately for drugs contained within the training/validation set and drugs contained within the test set. }

\Review{We observe that there is some local consistency in the values of model uncertainty, suggesting that the model is confident in some regions of chemical space, and less confident in other regions of the chemical space. Moreover, we observe columns and rows of high uncertainty, meaning that the model is less confident about specific drugs (regardless of the other drug found in the combination). On average, uncertainty is lowest on the training-validation set, and highest on combinations where neither of the drugs have been seen, \textit{c.f.} marginal distributions of log-uncertainties in Figure~\ref{fig:jointplot_uncert_model_error} (top). Moreover, large model errors seem to preferentially occur in regions of high uncertainty – as shown in Figure~\ref{fig:jointplot_uncert_model_error}. Most importantly, large errors never occur when model is confident (low uncertainty), \textit{c.f.} absence of points in the upper left part of Figure~\ref{fig:jointplot_uncert_model_error}. Furthermore, logged model errors and logged uncertainties are weakly positively correlated ($0.22$ on the \textit{Training+Validation} split, $0.29$ on the \textit{One unseen drug} split, and $0.14$ on the \textit{Two unseen drugs} split).}

\Review{The modeling of structure-activity relationships (SAR) is an extremely difficult task when considering emergent complex phenotypes (e.g., cell death) due to the presence of cliff effects, i.e. there are superficially similar molecules with similar properties, and others with divergent properties. We note that this is such a challenging problem for machine learning that it undoubtedly deserves standalone research without the complicating factor of studying pairs of drugs.}

\section{\textit{In silico} Sequential Model optimization}
\label{app:smo_evaluation}

We benchmarked the SMO pipelines, whereby the model is shown a fraction of the full data set and can choose sample points to unblind. Our \textit{in silico} experiments try to mirror as closely as possible the setting of the \textit{in vitro} experiments. Therefore, unless specified otherwise, experiments are restricted to the MCF7 cell line and 30 combinations were acquired at a time using the same model as the one used to generate recommendations for the \textit{in vitro} experiments. Uncertainty is estimated using a deep ensemble of size 5, unless stated otherwise.

\subsection{Alternative uncertainty estimators: Direct Estimation}\label{sec:deup}

As deep ensembles are only one method for uncertainty quantification, we tested another approach: directly estimating the uncertainty in the style of DEUP \cite{DBLP:journals/corr/abs-2102-08501}. Here, two models are initialized: the first one, denoted as the \textit{mean predictor}, predicts the expected synergy $\hat{\mu}$ and is trained with Mean Square Error (MSE). The second model, denoted as the \textit{uncertainty predictor}, outputs an estimate of the standard deviation of the predictive distribution $\hat{\sigma}$ and is trained to minimize the following negative log-likelihood criterion:
\begin{equation}
\label{eq:nll}
NLL = \frac{\log(\hat{\sigma}^2)}{2} + \frac{(y - \hat{\mu})^2}{2\hat{\sigma}^2},
\end{equation}
where $y$ is the ground truth variable of interest. This criterion allows us to get an estimator $\hat{\sigma}$ of the standard deviation of the \textit{predictive} distribution for each combination: given the expected synergy $\hat{\mu}$ predicted by the mean predictor, and the actual observation $y$, we wish to find $\hat{\sigma}$ that maximizes the probability of $y$ assuming a predictive distribution of the form $\mathcal{N}(\hat{\mu}, \hat{\sigma})$.

\begin{align}
\log p(y | \hat{\mu}, \hat{\sigma}) &= \log \left[ \frac{1}{\sqrt{2 \pi  \hat{\sigma}^2}} e^{-\frac{1}{2}(\frac{y - \hat{\mu}}{\hat{\sigma}})^2} \right] \\
&= -\frac{1}{2} \log(2 \pi) - \frac{1}{2}\log(\hat{\sigma}^2) - \frac{1}{2} \frac{(y - \hat{\mu})^2}{\hat{\sigma}^2}
\end{align}

Therefore, maximizing $\log p(y | \hat{\mu}, \hat{\sigma})$ w.r.t. $\hat{\sigma}$ is equivalent to minimizing the $NLL$ criterion presented in Equation \ref{eq:nll}. Note that only the uncertainty predictor is trained using this criterion. The mean predictor is trained using Mean Square Error (MSE) as we found experimentally that it was more stable. When fixing $\hat{\sigma}=1$, $NLL$ corresponds to the MSE criterion.

\subsection{Backtesting {\Recover} demonstrates efficient exploration of drug combination space}\label{sec:smo_results}

In order to simulate real-world interactions with the wet lab, we start with a set of 30 randomly chosen drug pairs that the pipeline is initially trained on, while the rest of the data is hidden from the pipeline. Hence, these combinations are always part of the visible set. We now perform an \textit{iteration}: we split the visible set into a training and validation set (80/20), train the model using early stopping and then acquire 30 additional drug pairs from the hidden set. The entire procedure is repeated again with the new training set of $60$ drug pairs (i.e., $30 + 30$), and then with $90$, and so on, until no more drug pairs are left to acquire from the hidden set. A range of methods can be used to generate recommendations for new drug pairs, typically as specified by an \textit{acquisition function} \citep{vzilinskas1972bayes}. The acquisition functions that we investigate are detailed in Appendix~\ref{sec:smo}.

\begin{figure}
\begin{overpic}[width=0.45\textwidth]{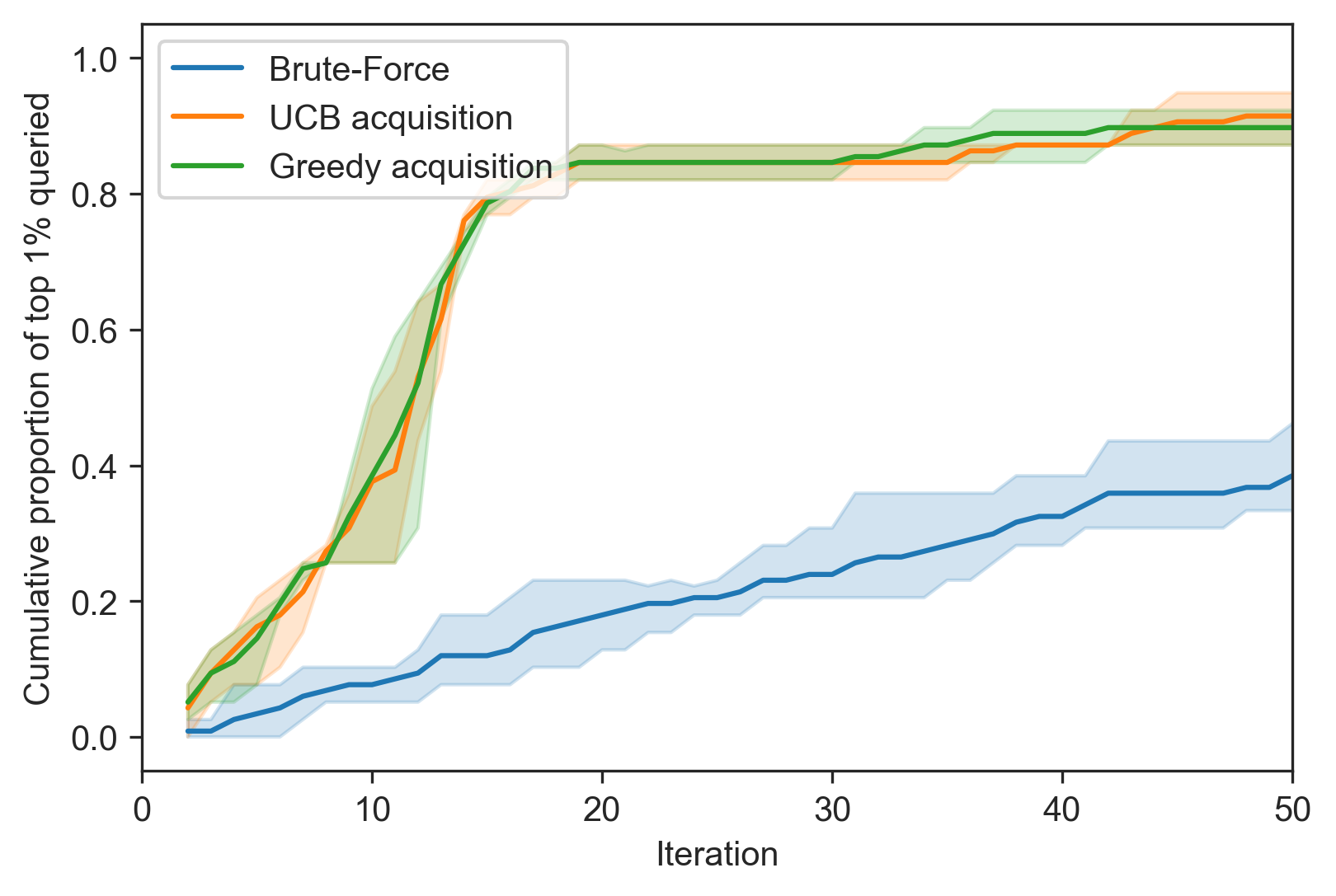}
\put(2,65){A}
\end{overpic}
\hspace{0.5cm}
\begin{overpic}[width=0.45\textwidth]{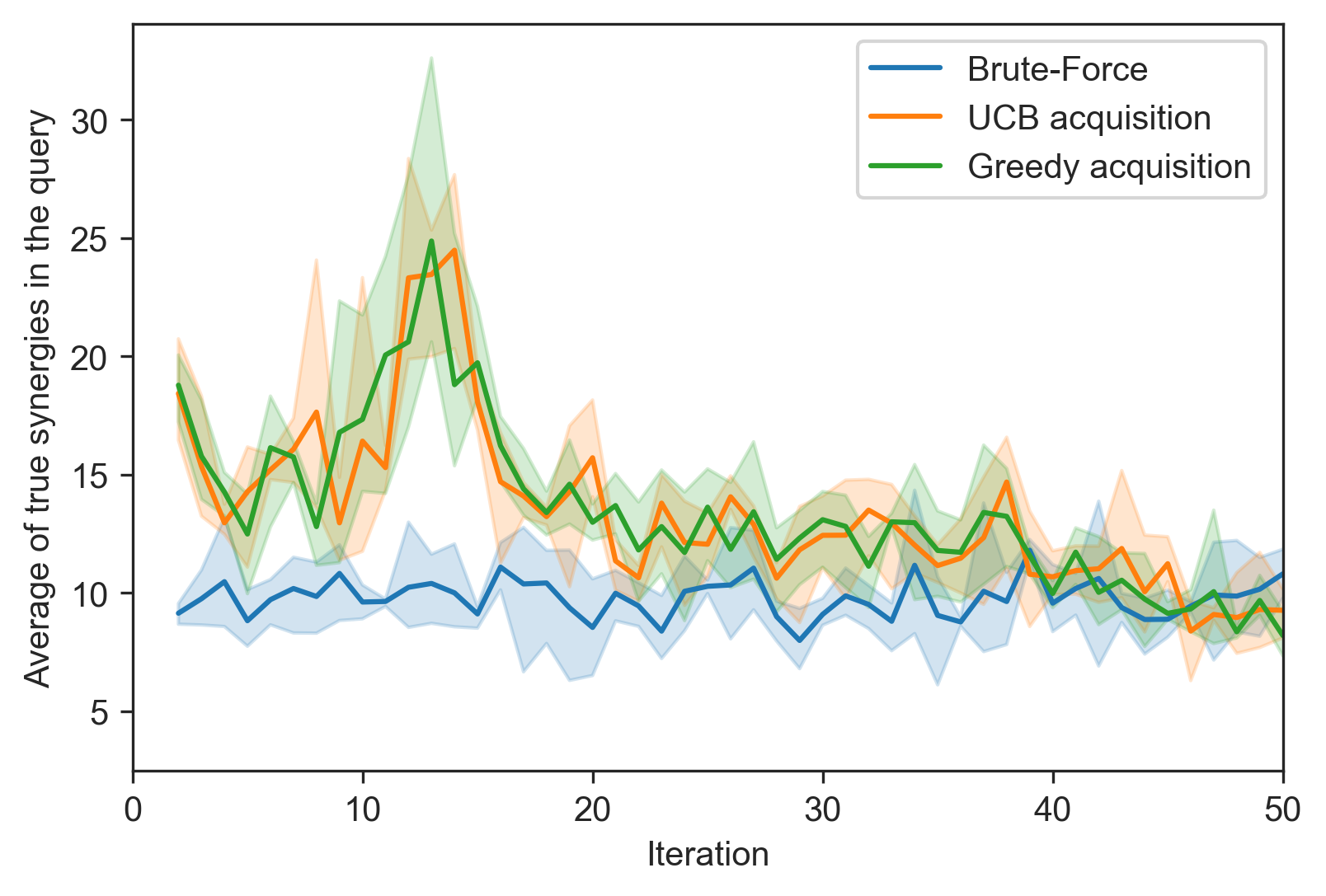}
\put(45,10){\includegraphics[clip,width=4cm]{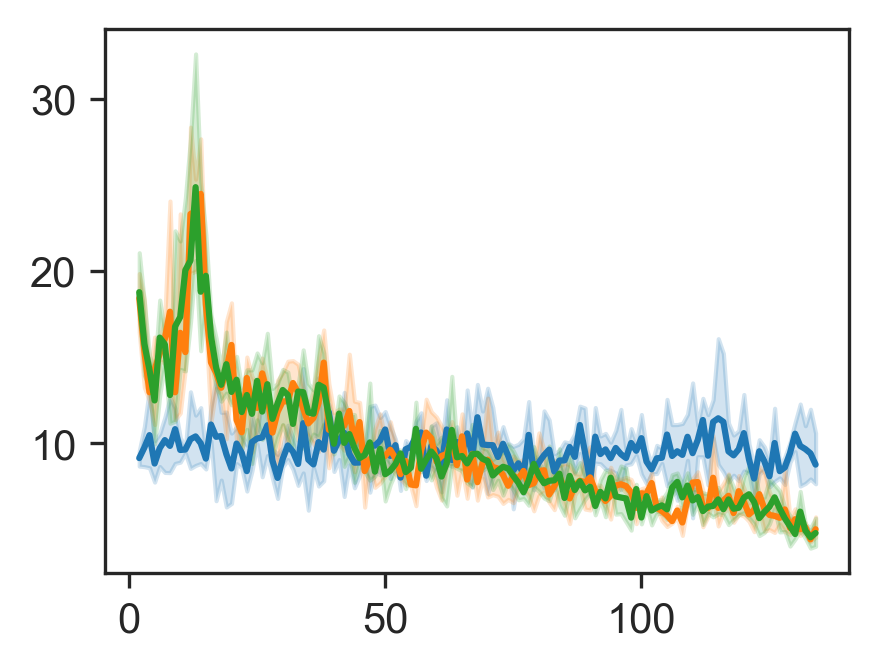}}
\put(2,65){B}
\end{overpic}
\caption{Comparison of acquisition functions through \textit{in silico} SMO experiments: \textbf{(A.)} Cumulative proportion of the top 1\% synergistic combinations that have been rediscovered by {\Recover}, \textbf{(B.)} Average of the true synergies of the drug pairs that have just been queried at each iteration of the SMO pipeline, \textbf{(B, inset)} Zoomed out view. Uncertainty estimated via deep ensembles. Shaded areas correspond to the standard deviation computed over 3 seeds.}
\label{fig:active_learning_average_syn}
\end{figure}

We compare acquisition functions by the rate at which they unblind the set of top 1\% of synergistic combinations in the NCI-ALMANAC dataset within each synthetic experimental round (or \textit{iteration}). As shown in Figure~\ref{fig:active_learning_average_syn}A, Greedy acquisition and Upper Confidence Bound (UCB) acquisition, two model-based strategies, perform on par with each other and outperform the random strategy by a large margin. Both approaches discover 80\% of the top 1\% of synergistic drug pairs in approximately 15 iterations whereas the random strategy discovers less than 20\% in the same number of iterations.

Figure~\ref{fig:active_learning_average_syn}B presents the average synergy among queried combinations at each iteration of the SMO pipeline. While the average synergy using random strategy is always approximately $10$, model-based strategies query batches for which the average synergy can be up to 25. After approximately 15 iterations, the average synergy in the queries starts to decrease, as there are fewer highly synergistic combinations left to query.

\subsection{Transfer learning: O'Neil 2016 study to NCI-ALMANAC study}
\label{app:effect_pretraining}

An important aspect of the SMO pipeline is its ability to leverage publicly available data in order to improve performance in a new experimental setting. To this end, we analyzed the impact of pretraining the model on the O'Neil database before simulating SMO experiments on a subset of the NCI-ALMANAC database. 

While the out-of-distribution analysis presented in Table~\ref{tab:main_table} showed that {\Recover} does not generalize well to new experimental settings \textit{without adaptation}, these experiments demonstrate, quite remarkably, that some latent knowledge can still be transferred from one experimental setting to another, resulting in increased performance in the latter setting.

\begin{figure}[H]
\centering
\begin{overpic}[width=0.45\textwidth]{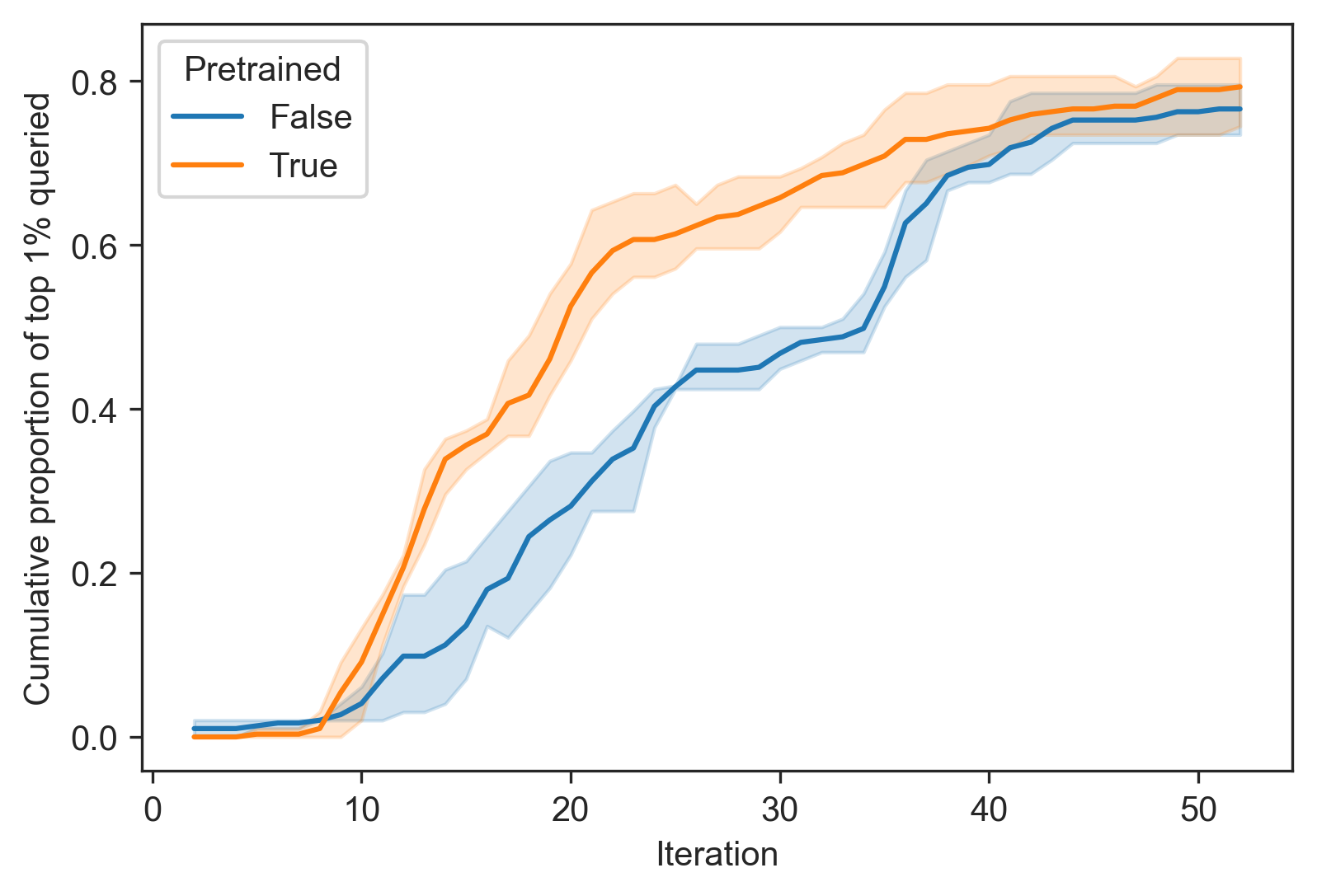}
\put(2,65){A}
\end{overpic}
\hspace{0.5cm}
\begin{overpic}[width=0.45\textwidth]{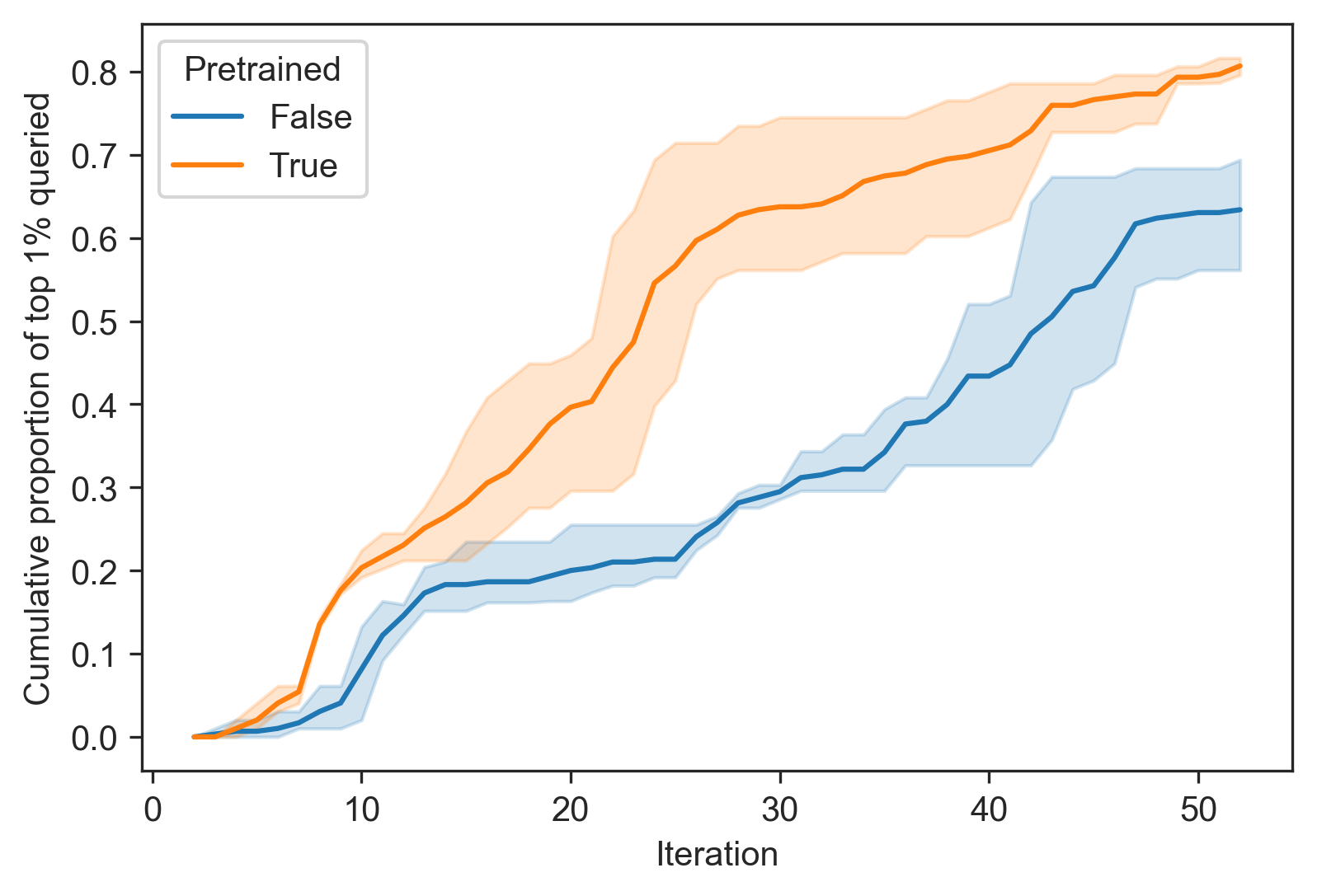}
\put(2,65){B}
\end{overpic}
\caption{Effect of pretraining on the rate of discovery of highly synergistic combinations via a UCB acquisition function: \textbf{(A.)} Deep Ensemble, \textbf{(B.)} Direct Uncertainty Estimation. Evaluation performed on the subset of NCI-ALMANAC consisting of drug pairs for which at least one of the drugs is included in O'Neil. All overlapping cell lines were included. Shaded areas correspond to the standard deviation computed over 3 seeds.}
\label{fig:active_learning_pretraining}
\end{figure}

As shown in Figure~\ref{fig:active_learning_pretraining}, the model pretrained on O'Neil outperforms the other model, initialized randomly. The model was first pretrained on the O'Neil study, thereafter, we simulated the SMO process on the subset of NCI-ALMANAC consisting of drug pairs for which at least one of the drugs was included in the O'Neil study. We compared against a model that had not been pretrained. All models are conditioned on cell line, and two different uncertainty estimation methods were tested. 

In these experiments, we restricted ourselves to cell lines which were covered in both the O'Neil and NCI-ALMANAC studies. As the MCF7 cell line is not included in the O'Neil study, we could not restrict ourselves to MCF7 as usual. Instead, we performed SMO on all overlapping cell lines, resulting in a space of possible queries which is bigger than in other experiments where only MCF7 was used. This explains why the rate of discovery is slower in this case.

\subsection{Demonstration that an uncertainty driven strategy can outperform a greedy strategy}
\label{app:comparison_different_acquisition_functions}

When trying to mirror as closely as possible the setting of the \textit{in vitro} experiments, we did not notice any significant difference in performance between the two Model-based acquisition functions Greedy and UCB, as shown in Figure \ref{fig:active_learning_average_syn}A. However, depending on the model set up, we can demonstrate that taking uncertainty into account to guide experiments can increase the performance of the pipeline over a naive Greedy acquisition strategy.

In the following, uncertainty was directly estimated using an DEUP-style \textit{uncertainty predictor}, and the task is the prediction of the average Bliss synergy score (instead of maximum Bliss), which corresponds to the average over the dose-response matrix of the concentration specific Bliss scores. In the following, 5 combinations were acquired at a time, instead of 30.

\begin{figure}[H]
\begin{center}
\includegraphics[width=0.45\textwidth]{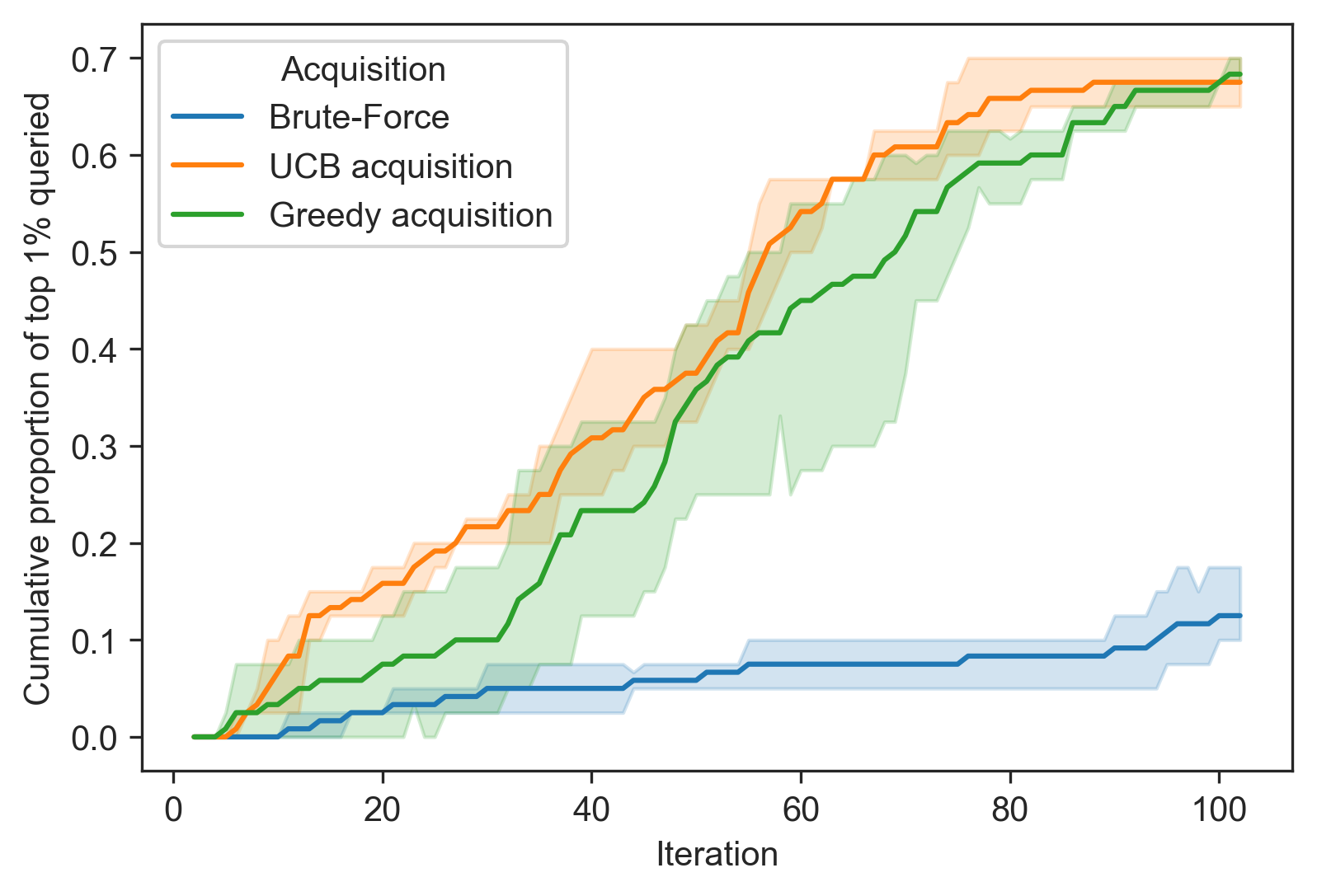}
\caption{Comparison of acquisition functions through in silico SMO experiments. Prediction of the average Bliss synergy score. 5 combinations acquired at a time. Uncertainty estimated using direct uncertainty estimation. Shaded areas correspond to the standard deviation computed over 3 seeds.}
\label{fig:active_learning_bliss_average_app}
\end{center}
\end{figure}

As shown in Figure~\ref{fig:active_learning_bliss_average_app}, UCB can outperform Greedy acquisition, demonstrating the value of taking uncertainty into account in the exploration strategy.

\end{document}